\documentclass[hidelinks]{article}
\usepackage[utf8]{inputenc}
\usepackage[a4paper, total={6.5in, 10in}]{geometry}

\usepackage[authoryear, square]{natbib}
\setcitestyle{numbers}

\usepackage{times, amssymb, amsmath, hyperref, amsthm}

\newtheorem*{definition}{Definition}

\usepackage{tensor}
\usepackage{eucal}
\usepackage{verbatim}

\usepackage{authblk}
\usepackage{braket}
\usepackage{url}
\usepackage{xargs}                      
\usepackage{graphicx}
\usepackage[table]{xcolor}  
\usepackage[disable, colorinlistoftodos,prependcaption,textsize=tiny]{todonotes}
\usepackage{dirtytalk}

\usepackage{tikz}
\tikzset{
  treenode/.style = {shape=rectangle, rounded corners,
                     draw, align=center,
                     top color=white, bottom color=blue!20},
  root/.style     = {treenode, font=\Large, bottom color=red!30},
  env/.style      = {treenode, font=\ttfamily\normalsize},
  dummy/.style    = {circle,draw}
}

\definecolor{cblue}{RGB}{100,5,255}    
\definecolor{cred}{RGB}{255,10,10} 
\definecolor{cgreen}{RGB}{5,165,20}  
\definecolor{corange}{rgb}{1.0,0.49,0.0}

\title{The general set of Noetherian energy-momentum tensors in linearized gravity: mathematical framework}

\author[1,2]{Lydia Beth Taylor\thanks{ltayl59@uwo.ca}}
\author[1,2]{Mark Robert Baker\thanks{mbaker66@uwo.ca}}

\affil[1]{Department of Physics and Optical Engineering, Rose-Hulman Institute of Technology, 5500 Wabash Ave, Terre Haute, IN, USA-47803}

\affil[2]{Department of Physics, St. Francis Xavier University, 4130 University Ave, Antigonish, NS, CA-B2G2W5}

\date{October 3, 2023}

\begin{document}

\maketitle

\large

\vspace{0.5cm}

\begin{abstract}
Energy-momentum tensors are foundational objects which are uniquely defined in standard physical field theories such as electrodynamics and Yang-Mills theory. In general relativity, and in particular linearized gravity where symmetries required for an energy-momentum tensor derived from Noether's first theorem are well defined, there exists a long standing non-uniqueness problem; numerous distinct energy-momentum expressions exist in the literature, and there is not consensus which, if any, is the unique expressions for the theory. Recently, the viability of the superpotential `improvement' method was shown to be insufficient for addressing the non-uniqueness problem of energy-momentum tensors in linearized gravity. In the present article, the mathematical framework for the general set of Noetherian energy-momentum tensors in linearized gravity is derived using Noether's first theorem, which consists of all possible energy-momentum tensors from the Noether current which yield the linearized Einstein field equations in the corresponding Euler-Lagrange equation of the Noether identity without introducing any `improvement' terms. This result has several advantages in addition to not requiring `improvements', such as the ability to impact the Lagrangian proportional piece of the energy-momentum tensor. Numerous common published gravitational energy-momentum expressions are then compared to these general results to assess which can be classified as Noetherian, and which cannot. Standard physical criteria such as symmetry and tracelessness are then used to prove that it is possible to directly obtain an energy-momentum tensor which is simultaneously symmetric and traceless from the Noether current; such an expression is derived from the general results. Consequences of these results and their relation to the aforementioned non-uniqueness problem are discussed.
\end{abstract}

\vspace{0.5cm}

\section{Introduction}

In a recent article \cite{Baker2021b}, the complete set of energy-momentum tensors in linearized gravity that follow from the canonical Noether energy-momentum tensor and the addition of superpotential \say{improvement} terms was determined. This result showed that there are infinitely many such energy-momentum tensors, and infinitely many off-shell even if a single superpotential is fixed (such as the Belinfante superpotential \cite{belinfante1940}). A consequence of \cite{Baker2021b} is that the superpotential \say{improvement} method cannot be used to define uniqueness criteria for energy-momentum tensors in linearized gravity, which is highly desirable for a number of reasons, such as in the Padmanabhan-Deser debate \cite{padmanabhan2008,deser2010,butcher2009,barcelo2014}, as well as uniqueness of the energy-momentum tensor being a standard feature of foundational physical theories such as electrodynamics and Yang-Mills theory \cite{Baker2021PHD}. There exist numerous gravitational energy-momentum expressions in the literature, and attempts at determining uniqueness for linearized gravity has a long history \cite{szabados1992,babak1999,magnano2002,bicak2016,Toth2022}. We refer the reader to \cite{Baker2021b} for significant discussion on the topic of energy-momentum in linearized gravity, the non-uniqueness problem, and the issues that this raises with the superpotential \say{improvement} method.\footnote{The superpotential \say{improvement} method has long been used to add desired terms ad-hoc to an energy-momentum tensor derived from Noether’s first theorem \cite{belinfante1940,forger2004,blaschke2016}; these \say{improvements} are terms which do not exist in the Noether identity \cite{noether1918,kosmann2011}. Typically this is performed after only the 4-parameter Poincar{\'e} translation symmetry (the symmetry associated with deriving an energy-momentum tensor from the Noether identity) is used without the complete set of symmetries of the action, leading to the so-called canonical Noether energy-momentum tensor, which often does not correspond to the known physical energy-momentum tensor \cite{forger2004,baker2021noether}; therefore \say{improvement} terms are added via superpotentials to obtain the desired result. A major issue with this approach is that if the desired result is not known as in linearized gravity (unlike in the case of e.g. electrodynamics) it is not clear which of infinitely many possible \say{improvements} should be added ad-hoc to the Noether identity. We also note that energy-momentum tensors are sometimes referred to as stress-energy tensors; we will use the former terminology in this article.}\\

There is another distinct approach to Noether’s first theorem that does not require the ad-hoc introduction of any \say{improvements}, known as the Bessel-Hagen method \cite{baker2021noether,besselhagen1921,besseltranslation2006}: by considering only Noether’s first theorem \cite{noether1918,kosmann2011} and the complete set of action symmetries, one can directly obtain the correct, unique, physical energy-momentum tensor directly from Noether’s first theorem in cases such as electrodynamics, Yang-Mills theory, and linearized Gauss-Bonnet gravity \cite{Baker2021PHD}. The Bessel-Hagen method uses Noether’s first theorem directly, therefore results from this method are completely results of Noether’s first theorem, as no additional ad-hoc terms must be added to the Noether identity to obtain known physical expressions \cite{baker2021noether}. For this reason, alongside the results in \cite{Baker2021b}, a clear question arises: which gravitational energy-momentum expressions (if any) can be derived directly from Noether’s first theorem without the introduction of ad-hoc terms not present in the Noether identity? Such energy-momentum tensors are referred to as Noetherian energy-momentum tensors \cite{baker2021}:\\

\begin{definition}
Noetherian energy-momentum tensor $T^{\mu \nu}_N$: an energy-momentum tensor which is derived directly from Noether’s first theorem and satisfies the Noether identity without the introduction of any ad-hoc correction (or \say{improvement}) terms, which are terms that do not exist explicitly in the Noether identity \cite{noether1918,kosmann2011}\\
\end{definition}

Deriving the mathematical framework for the general set of such Noetherian energy-momentum tensors in linearized gravity is a major focus of this article. How this will be accomplished is to consider the most general possible set of the two internal freedoms in Noether's first theorem available if one restricts themselves to expressions only present in the Noether identity \cite{noether1918,kosmann2011}: i) all Lagrangian densities which yield the linearized Einstein field equations, and ii) all field transformations of the spin-2 tensor potential $h_{\mu\nu}$ which can be inserted into the Noether identity. We will then use the resulting set of energy-momentum tensors derived from the Noether current to determine which of the linearized well-known gravitational energy-momentum expressions are derivable directly from Noether’s first theorem, and whether or not uniqueness criteria can be determined based on standard physical requirements of energy-momentum tensors such as symmetry and tracelessness.\\

The reason linearized gravity is essential for asking these questions is because in general relativity, a finite group of global coordinate symmetries $\delta x_\alpha$ which can be inserted into the Noether identity are not available. In Minkowski spacetime however, the 10-parameter Poincar{\'e} group provides such global coordinate symmetries (which include the 4-parameter Poincar{\'e} translation) that can be inserted into the Noether identity; the Minkowski metric yields the 10-parameter Poincar{\'e} group of symmetries when substituted into Killing’s equation. In particular, the 4-parameter Poincar{\'e} translation is the coordinate symmetry associated with the definition of an energy-momentum tensor in the case of Noether’s first theorem. For this reason linearized gravity allows for the strict application of Noether's first theorem for deriving an energy-momentum tensor. We note that when we refer to Minkwoski spacetime we are considering the 4D Minkowski metric with signature ($+,-,-,-$). \\

Linearized gravity is conventionally defined by the linearization of the metric tensor of general relativity about a Minkowski background ($g_{\mu\nu} \approx \eta_{\mu\nu} + \epsilon h_{\mu\nu}$); when linearizing the Einstein field equations, terms of order $\epsilon$ are kept and higher order terms discarded. From this linearization technique one obtains the linearized Einstein field equations (sometimes referred to as the spin-2 equation of motion),

\begin{equation}
    E^{\mu \nu} = \frac{1}{2}
    [- \eta^{\mu \nu} \square h^\alpha_\alpha 
    + \square h^{\mu \nu} 
    + \partial^\mu \partial^\nu h^\alpha_\alpha 
    - \partial_\alpha \partial^\nu h^{\mu \alpha} 
    - \partial_\alpha \partial^\mu h^{\nu \alpha} 
    + \eta^{\mu \nu} \partial^\alpha \partial^\beta h_{\alpha \beta}]  , 
 \label{linEFE}
\end{equation}

\noindent where $h_{\alpha\beta} = h_{\beta\alpha}$ is the symmetric spin-2 potential, analogous to the role played by the vector potential $A_\mu$ in classical electrodynamics. This equation is invariant under the spin-2 gauge transformation (linearized diffeomorphism) $h_{\mu\nu} \to h_{\mu\nu} + \partial_\mu \xi_\nu + \partial_\nu \xi_\mu$.\\

However, this is not the only possible Lagrangian density which yields the linearized Einstein field equations (\ref{linEFE}); there is a general system of possible Lagrangians which will yield this in the Euler-Lagrange equation. As discussed above, this is one of the two possible freedoms that exist explicitly in the Noether identity, and using this freedom will change the resulting energy-momentum tensor \cite{gieres2022,Butcher2012PHD,kuzmin2001}.\footnote{The ability to modify an energy-momentum tensor by modifying the Lagrangian density compared to adding ad-hoc \say{improvement} terms has been discussed extensively in the general case in \cite{gieres2022}; the former is a freedom that exists within the Noether identity, the latter is a method that exists external to the Noether identity. This approach has been used in the past for example in \cite{kuzmin2001}, and discussed in \cite{Butcher2012PHD} for the case of linearized gravity. The basic idea is that any two Lagrangians which yield the same equation of motion can be related by a boundary term, which will not effect the equation of motion, but will effect the resulting energy-momentum tensor. Therefore by adding boundary terms to the action one can modify the energy-momentum tensor without effecting the equation of motion, and in some cases \cite{kuzmin2001}, avoid the need for \say{improvement} terms all together; it is this freedom that we will use in Section \ref{SectionAllLag} as one of our aforementioned two possible freedoms in the Noether identity. } We will begin Section \ref{SectionGenNoetherT} by determining the general system of equations associated to the free coefficients of all possible Lagrangian densities in Section \ref{SectionAllLag}, which will serve as the first set of restrictions on a general set of energy-momentum tensors that we will derive from the linearized gravity Noether current (which is presented in Section \ref{SectionLinGravNoetherIdentity}). After which we will write down the most general set of field transformations in Section \ref{SectionAllSymmTrans}, and together with the freedom from Section \ref{SectionAllLag}, obtain the most general set of energy-momentum tensors in linearized gravity that can be derived from the Noether current in Section \ref{SectionGeneralResultSetFock}. We will use this general set of energy-momentum tensors in Section \ref{LiteratureComparison} to determine which of the various published gravitational energy-momentum expressions are Noetherian, and which are non-Noetherian. In Section \ref{ResultsRestriction} we will apply possible uniqueness criteria to the general set of energy-momentum tensors derived from the Noether current based on standard physical energy-momentum tensor requirements such as symmetry and tracelessness; properties which are features of energy-momentum tensors in foundational physical theories such as electrodynamics and Yang-Mills theory. Finally we will summarize and discuss our results in Section \ref{SectionSummaryDiscussion}.

\section{Mathematical framework for the general set of Noetherian energy-momentum tensors in linearized gravity} \label{SectionGenNoetherT}

\subsection{Determining all possible Lagrangian densities in linearized gravity} \label{SectionAllLag}

If we want to consider the most general set of Noetherian energy-momentum tensors in linearized gravity, we must consider all possible Lagrangian densities that yields the linearized Einstein field equations (\ref{linEFE}). This is because numerous Lagrangian densities can yield (\ref{linEFE}), however each different Lagrangian density will yield a different energy-momentum tensor. Therefore terms of the form $\partial h \partial h$ and $h \partial \partial h$ must be considered in the action when deriving the Noether identity. This is one of the two freedoms that we have in deriving an energy-momentum tensor from the Noether current. Writing down the general set of possible terms in the Lagrangian density,

\begin{multline}
    \mathcal{L} = 
    C_1 \partial_\beta h_{\mu \nu} \partial^\beta h^{\mu \nu} 
    + C_2 \partial_\beta h^\mu_\mu \partial^\beta h^\nu_\nu 
    + C_3 \partial_\beta h^{\beta \mu} \partial_\mu h^\nu_\nu 
    + C_4 \partial_\beta h^{\beta \mu} \partial^\nu h_{\mu \nu} 
    + C_5 \partial_\mu h_{\nu \beta} \partial^\nu h^{\mu \beta} 
    \\ 
    + D_1 h_{\mu \nu} \partial^\mu \partial^\nu h^\beta_\beta 
    + D_2 h_{\mu \nu} \partial^\mu \partial_\beta h^{\nu \beta} 
    + D_3 h_{\mu \nu} \partial^\beta \partial_\beta h^{\mu \nu} 
    + D_4 h^\mu_\mu \partial^\beta \partial_\beta h^\nu_\nu 
    + D_5 h^\beta_\beta \partial^\mu \partial^\nu h_{\mu \nu} , \label{genlag}
\end{multline}

\noindent and inserting into the Euler-Lagrange equation,

\begin{equation}
    E^{\omega \sigma} = 
    \frac{\partial \mathcal{L}}{\partial h_{\omega \sigma}} 
    - \partial_\rho \frac{\partial \mathcal{L}}{\partial(\partial_\rho h_{\omega \sigma})} 
    + \partial_\rho \partial_\zeta \frac{\partial \mathcal{L}}{\partial(\partial_\rho \partial_\zeta h_{\omega \sigma})}  ,
\end{equation}

\noindent we obtain the terms associated to linearized gravity equation of motion (linearized Einstein field equations),

\begin{multline}
    E^{\omega\sigma} = 
    (2D_4 - 2C_2)\eta^{\omega \sigma} \square h^\mu_\mu 
    + (2D_3 - 2C_1)\square h^{\omega \sigma} 
    + (D_1 - C_3 + D_5)\partial^\omega \partial^\sigma h^\mu_\mu 
    \\
    + (D_2 - C_4 - C_5)\partial_\mu \partial^\sigma h^{\omega \mu} 
    + (D_2 - C_4 - C_5)\partial_\mu \partial^\omega h^{\sigma \mu} 
    + (D_1 - C_3 + D_5)\eta^{\omega \sigma} \partial^\mu \partial^\nu h_{\mu \nu} .
\end{multline}

\noindent  If the above coefficients are fixed as as any multiple $n$ of the linearized Einstein's field equations (\ref{linEFE}), then the above equation of motion must have a particular system of equations satisfied; we have included the resulting system of equations (\ref{eomcondition1})-(\ref{eomcondition4}) in Appendix \ref{Restriction1} in order to be in the same location as the other required equations for the general set of energy-momentum tensors derived from the Noether current in Appendix \ref{Fock to Noether Coefficient Comparison}. Any solution to this system of equations (\ref{eomcondition1})-(\ref{eomcondition4}) will give an equation of motion which is equivalent to the linearized Einstein's field equations (\ref{linEFE}) up to some numerical coefficient $n$. Therefore this system of equations puts a restriction on all possible Lagrangian densities in equation (\ref{genlag}) that can be used to obtain (\ref{linEFE}), and only solutions to this system of equations can be considered as possible Noetherian energy-momentum tensors, provided they satisfy the remaining conditions in Appendix \ref{Fock to Noether Coefficient Comparison}.

\subsection{Noether identity for linearized gravity} \label{SectionLinGravNoetherIdentity}

Different solutions to the system of equations (\ref{eomcondition1})-(\ref{eomcondition4}) may yield the same equation of motion, but they will yield different energy-momentum tensors in the Noether current. Since we have both terms of the form $\partial h \partial h$ and $h \partial \partial h$ in the Lagrangian $\mathcal{L}(h, \partial h, \partial \partial h)$, we must take into account both of these when deriving the Noether identity \cite{gelfand2012}. Noether’s first theorem in the case of this general linearized gravity Lagrangian yields the following Noether identity \cite{Baker2021b},\\

\begin{multline}
    \left( \frac{\partial \mathcal{L}}{\partial h_{\omega \sigma}} 
    - \partial_\rho \frac{\partial \mathcal{L}}{\partial(\partial_\rho h_{\omega \sigma})} 
    + \partial_\rho \partial_\zeta \frac{\partial \mathcal{L}}{\partial(\partial_\rho \partial_\zeta h_{\omega \sigma})} 
     \right) \delta h_{\omega \sigma} 
    \\
    + \partial_\rho \left( \eta^{\rho \lambda} \mathcal{L} \delta x_\lambda 
    + \frac{\partial \mathcal{L}}{\partial(\partial_\rho h_{\omega \sigma})} \delta h_{\omega \sigma} 
    + \frac{\partial \mathcal{L}}{\partial(\partial_\rho \partial_\zeta h_{\omega \sigma})} \partial_\zeta \delta h_{\omega \sigma} 
    - [\partial_\zeta \frac{\partial \mathcal{L}}{\partial(\partial_\rho \partial_\zeta h_{\omega \sigma})}]\delta h_{\omega \sigma} 
    \right)
    = 0 . \label{s2NoetherIdentity}
\end{multline}

\noindent It is from the second line in the above Noether identity that we can derive an energy-momentum tensor for a linearized gravity Lagrangian density of the form $\mathcal{L}(h, \partial h, \partial \partial h)$; see \cite{Baker2021b,jackiw1994,leclerc2006,leclerc2006b} for additional discussion on Noether's theorem and its relation to gravitational energy-momentum expressions. The Noether identity (\ref{s2NoetherIdentity}) can be expressed compactly in terms of the Euler-Lagrange equation $E^{\omega\sigma}$ and the Noether current $J^\rho$ as,

\begin{equation}
E^{\omega\sigma} \delta h_{\omega \sigma} + \partial_\rho J^\rho = 0 , \label{CompactNoetherIdentity}
\end{equation}

\noindent where the Noether current is,

\begin{equation}
J^\rho =  \eta^{\rho \lambda} \mathcal{L} \delta x_\lambda 
    + \frac{\partial \mathcal{L}}{\partial(\partial_\rho h_{\omega \sigma})} \delta h_{\omega \sigma} 
    + \frac{\partial \mathcal{L}}{\partial(\partial_\rho \partial_\zeta h_{\omega \sigma})} \partial_\zeta \delta h_{\omega \sigma} 
    - [\partial_\zeta \frac{\partial \mathcal{L}}{\partial(\partial_\rho \partial_\zeta h_{\omega \sigma})}]\delta h_{\omega \sigma} . \label{NoetherCurrent}
\end{equation}

\noindent An energy-momentum tensor is derived from Noether's first theorem by factoring the 4-parameter Poincar{\'e} translation $a_\lambda$ out of the Noether current (\ref{NoetherCurrent}), yielding the Noether identity,

\begin{equation}
E^{\omega\sigma} \delta h_{\omega \sigma} + a_\lambda \partial_\rho T^{\rho\lambda} = 0 . \label{CompactNoetherIdentityT}
\end{equation}

\noindent It is from (\ref{NoetherCurrent}) that we will obtain the general set of energy-momentum tensors in linearized gravity which can be derived from the Noether current.

\subsection{Determining all possible field transformations $\delta h_{\rho\sigma}$ in linearized gravity}  \label{SectionAllSymmTrans}

The freedom in the Lagrangian density is one of only two freedoms that we have to work with to derive a Noetherian energy-momentum tensor; the Noether current (\ref{NoetherCurrent}) contains field transformations $\delta h_{\omega \sigma}$ which must be specified in order to derive a specific energy-momentum tensor. Therefore the second freedom is to consider the most general set of field transformations possible in this identity. If one wishes to derive the canonical energy-momentum tensor for example, they would use only the so-called canonical transformation,

\begin{equation}
\delta h_{\rho\sigma} = - \partial_\beta h_{\rho\sigma} \delta x^\beta . \label{canontrans}
\end{equation}

\noindent However, in the case of the Bessel-Hagen method, for gravity models such as linearized Gauss-Bonnet gravity, these transformations have been shown to be \cite{baker2019},

\begin{equation}
\delta h_{\rho\sigma} = - 2 \bar{\Gamma}^\nu_{\ \rho\sigma} \delta x_v ,
\end{equation}

\noindent which are proportional to the linearized Christoffel symbol of the second kind $\bar{\Gamma}^\nu_{\ \rho\sigma} = \frac{1}{2} (\partial^\nu h_{\rho\sigma} - \partial_\rho h^\nu_{\ \sigma} - \partial_\sigma h^\nu_{\ \rho})$, similar to standard gauge theories where the transformations are proportional to the field strength tensor \cite{Baker2021PHD}. This transformation follows from the fact that the spin-2 gauge symmetry (linearized diffeomorphisms) are required to yield the known gauge invariant energy-momentum tensor of the theory. In the case of linearized gravity however, with terms proportional to $\partial h \partial h$ and $h \partial \partial h$, it is not possible to construct a gauge invariant energy-momentum tensor \cite{magnano2002}. For this reason straightforward derivation of a unique energy-momentum tensor is problematic, as we discussed in \cite{Baker2021b}. We will consider the most general mathematical set of possible field transformations, to derive the most general set of possible linearized gravity energy-momentum tensors from the Noether current. We will then restrict our set of solutions based on known expressions in the literature, and various physical arguments, after this derivation is complete.\\
 
The field transformations which yield an energy-momentum tensor must be proportional to the 4-parameter Poincar{\'e} translation coordinate symmetry $\delta x_\lambda = a_\lambda$, and to the the derivative spin-2 potential $h_{\mu\nu}$. Considering all such possible terms with free coefficients, we are left with,

\begin{multline}
    \delta h_{\omega \sigma} 
    = B_1 \partial^\alpha h_{\omega \sigma} a_\alpha 
    + B_2 \partial^\alpha h_{\alpha \sigma} a_\omega 
    + B_3 \partial^\alpha h_{\omega \alpha} a_\sigma 
    + B_4 \partial_\omega h_{\alpha \sigma} a^\alpha 
    \\
    + B_5 \partial_\sigma h_{\omega \alpha} a^\alpha 
    + B_6 \partial_\omega h^\alpha_\alpha a_\sigma 
    + B_7 \partial_\sigma h^\alpha_\alpha a_\omega 
    + B_8 \eta_{\omega \sigma} \partial^\alpha h_{\alpha \gamma} a^\gamma 
    + B_9 \eta_{\omega \sigma} \partial^\alpha h^\gamma_\gamma a_\alpha .  \label{GenSymTransform}
\end{multline}

\noindent There is a secondary restriction on these coefficients, in order to preserve the symmetry $\delta h_{\omega \sigma} = \delta h_{\sigma \omega } $ one requires solutions of the form $B_2 = B_3$, $B_4 = B_5$ and $B_6 = B_7$. We have included these conditions in Appendix \ref{Restriction2} as they are part of the general set of conditions for an energy-momentum tensor derived from the Noether current, which are all included in Appendix \ref{Fock to Noether Coefficient Comparison}.\\

This general set of transformations (\ref{GenSymTransform}) is our second freedom in the Noether identity, which are the most general set of possible field transformations which could yield an energy-momentum tensor in linearized gravity. We can now combine these with the freedom in the Lagrangian density to obtain the general set of linearized gravity energy-momentum tensors which can be derived from the Noether current.

\subsection{Mathematical framework for the general set of Noetherian energy-momentum tensors in linearized gravity}  \label{SectionGeneralResultSetFock}

Starting from the Noether identity (\ref{s2NoetherIdentity}), we can insert the general Lagrangian density (\ref{genlag}) and the general set of field transformations (\ref{GenSymTransform}) into the Noether current (\ref{NoetherCurrent}),

\begin{equation}
J^\rho =  \eta^{\rho \lambda} \mathcal{L} \delta x_\lambda 
    + \frac{\partial \mathcal{L}}{\partial(\partial_\rho h_{\omega \sigma})} \delta h_{\omega \sigma} 
    + \frac{\partial \mathcal{L}}{\partial(\partial_\rho \partial_\zeta h_{\omega \sigma})} \partial_\zeta \delta h_{\omega \sigma} 
    - [\partial_\zeta \frac{\partial \mathcal{L}}{\partial(\partial_\rho \partial_\zeta h_{\omega \sigma})}]\delta h_{\omega \sigma} . \label{NoetherCurrent2}
\end{equation}

\noindent For brevity we have calculated all four of the terms in (\ref{NoetherCurrent2}) and included them in the Appendix \ref{AppendixFourTerms}. We note that the required derivatives can be expressed in the form $\frac{\partial (\partial_\alpha h_{\sigma \beta})}{\partial(\partial_\rho h_{\mu \nu})} = 
    \delta^\rho_\alpha \Delta^{\mu \nu}_{\sigma \beta}$ and $\frac{\partial (\partial_\alpha \partial_\gamma h_{\sigma \beta})}{\partial(\partial_\rho \partial_\lambda h_{\mu \nu})} = 
    \Delta^{\rho \lambda}_{\alpha \gamma} \Delta^{\mu \nu}_{\sigma \beta}$ where $\Delta^{\mu \nu}_{\sigma \beta} = 
    \frac{1}{2}(\delta^\mu_\sigma \delta^\nu_\beta 
    + \delta^\nu_\sigma \delta^\mu_\beta)$. In our resulting energy-momentum tensor (\ref{FockEMT}) we will use the trace abbreviation $h = h^\alpha_\alpha$.\\

The next step is to factor out the 4-parameter Poincar{\'e} translation from the Noether current (\ref{NoetherCurrent2}) according to (\ref{CompactNoetherIdentityT}) as $E^{\omega \sigma} \delta h_{\omega \sigma} + a_\lambda \partial_\rho T^{\rho \lambda} = 0$. Collecting like terms yields the general set of energy-momentum tensors in linearized gravity which can be derived from the Noether current,

\begin{multline}
    T^{\rho \lambda} = 
    b_1 \partial_\alpha h^{\rho \lambda} \partial_\beta h^{\alpha \beta} 
    + b_2 \partial_\alpha h^{\rho \lambda} \partial^\alpha h
    + b_3 \partial_\alpha h^{\rho \alpha} \partial_\beta h^{\lambda \beta} 
    + b_4 \partial_\alpha h^\rho_{\ \beta} \partial^\alpha h^{\lambda \beta} 
    + b_5 \partial_\alpha h^{\rho \beta} \partial_\beta h^{\lambda \alpha} 
    \\
    + b_6 \partial^\rho h \partial^\lambda h 
    + b_7 \partial^\rho h_{\alpha \beta} \partial^\lambda h^{\alpha \beta}
    + b_{8_i} \partial^\rho h^{\lambda \alpha} \partial_\alpha h
    + b_{8_{ii}} \partial^\lambda h^{\rho \alpha} \partial_\alpha h
    + b_{9_i} \partial^\rho h \partial_\alpha h^{\lambda \alpha} 
    + b_{9_{ii}} \partial^\lambda h \partial_\alpha h^{\rho \alpha} 
    \\
    + b_{10_i} \partial^\rho h^{\lambda \alpha} \partial^\beta h_{\alpha \beta} 
    + b_{10_{ii}} \partial^\lambda h^{\rho \alpha} \partial^\beta h_{\alpha \beta}
    + b_{11_i} \partial^\rho h_{\alpha \beta} \partial^\alpha h^{\lambda \beta} 
    + b_{11_{ii}} \partial^\lambda h_{\alpha \beta} \partial^\alpha h^{\rho \beta}
    + c_1 \eta^{\rho \lambda} \partial_\alpha h \partial^\alpha h
    \\
    + c_2 \eta^{\rho \lambda} \partial_\alpha h_{\beta \sigma} \partial^\alpha h^{\beta \sigma} 
    + c_3 \eta^{\rho \lambda} \partial_\alpha h^{\alpha \beta} \partial_\sigma h^\sigma_{\ \beta}
    + c_4 \eta^{\rho \lambda} \partial_\alpha h_{\sigma \beta} \partial^\sigma h^{\alpha \beta}
    + c_5 \eta^{\rho \lambda} \partial_\alpha h^{\alpha \beta} \partial_\beta h
    + d_1 h^{\rho \lambda} \partial_\alpha \partial^\alpha h 
    \\
    + d_2 h^{\rho \lambda} \partial_\alpha \partial_\beta h^{\alpha \beta}
    + d_3 h \partial_\alpha \partial^\alpha h^{\rho \lambda} 
    + d_4 h^{\alpha \beta} \partial_\alpha \partial_\beta h^{\rho \lambda} 
    + d_{5_i} h^{\rho \alpha} \partial^\beta \partial_\beta h^\lambda_{\ \alpha} 
    + d_{5_{ii}} h^{\lambda \alpha} \partial^\beta \partial_\beta h^\rho_{\ \alpha}
    + d_{6_i} h^{\rho \alpha} \partial_\alpha \partial_\beta h^{\lambda \beta}
    \\
    + d_{6_{ii}} h^{\lambda \alpha} \partial_\alpha \partial_\beta h^{\rho \beta} 
    + d_7 h \partial^\rho \partial^\lambda h 
    + d_8 h_{\alpha \beta} \partial^\rho \partial^\lambda h^{\alpha \beta}
    + d_{9_i} h^{\rho \alpha} \partial^\lambda \partial_\alpha h 
    + d_{9_{ii}} h^{\lambda \alpha} \partial^\rho \partial_\alpha h 
    + d_{10_{i}} h^{\rho \alpha} \partial^\lambda \partial^\beta h_{\alpha \beta}
    \\
    + d_{10_{ii}} h^{\lambda \alpha} \partial^\rho \partial^\beta h_{\alpha \beta}
    + d_{11_{i}} h \partial^\rho \partial_\alpha h^{\lambda \alpha} 
    + d_{11_{ii}} h \partial^\lambda \partial_\alpha h^{\rho \alpha} 
    + d_{12_{i}} h_{\alpha \beta} \partial^\rho \partial^\alpha h^{\lambda \beta}
    + d_{12_{ii}} h_{\alpha \beta} \partial^\lambda \partial^\alpha h^{\rho \beta}
    \\
    + a_1 \eta^{\rho \lambda} h_{\alpha \beta} \partial^\alpha \partial^\beta h
    + a_2 \eta^{\rho \lambda} h \partial_\alpha \partial_\beta h^{\alpha \beta}
    + a_3 \eta^{\rho \lambda} h_{\alpha \beta} \partial^\alpha \partial_\sigma h^{\sigma \beta}
    + a_4 \eta^{\rho \lambda} h \partial_\alpha \partial^\alpha h 
    + a_5 \eta^{\rho \lambda} h_{\alpha \beta} \partial_\sigma \partial^\sigma h^{\alpha \beta} , \label{FockEMT}
\end{multline}

\noindent where for brevity each of the coefficients in (\ref{FockEMT}) can be found in terms of the general symmetry coefficients $B_m$ and general Lagrangian coefficients $C_m$ and $D_m$ in Appendix \ref{Restriction3}. Together with conditions in Appendix \ref{Restriction1} and \ref{Restriction2} they form the complete set of conditions in Appendix \ref{Fock to Noether Coefficient Comparison}. This calculation is relatively straightforward but lengthy; the four terms in the Noether current in Appendix \ref{AppendixFourTerms} must be inserted into (\ref{NoetherCurrent2}) and all like terms must be collected after factoring out the 4-parameter Poincar{\'e} translation $a_\lambda$ --- the resulting coefficients form the system of equations in \ref{Restriction3}. In order for a linearized gravity energy-momentum tensor to be derivable from the Noether current it must satisfy all conditions in Appendix \ref{Fock to Noether Coefficient Comparison}: the equation of motion conditions (\ref{eomcondition1})-(\ref{eomcondition4}) in Appendix \ref{Restriction1}, the symmetry conditions on (\ref{GenSymTransform}) in Appendix \ref{Restriction2}, and the system of equations in Appendix \ref{Restriction3}. Therefore as long as one knows the appropriate Lagrangian density, and field transformations, they can immediately derive an energy-momentum tensor. Conversely, if one knows an energy-momentum tensor, they can immediately determine if it satisfies these equations, hence immediately determine if it is non-Noetherian; it is by this process which we will check numerous gravitational energy-momentum expressions in Section \ref{LiteratureComparison}. We note that the only term in (\ref{FockEMT}) that cannot be obtained in a Noetherian energy-momentum tensor is the $d_{5_{ii}} = h^{\lambda \alpha} \partial^\beta \partial_\beta h^\rho_{\ \alpha}$ contribution which is why $d_{5_{ii}} = 0$ in Appendix \ref{Restriction3}.\footnote{This term cannot be introduced by the product rule because this would require contributions from the corresponding expansion $\partial_\alpha \partial^\alpha [h^{\lambda \beta} h^\rho_{\ \beta}] = 
    2 \partial_\alpha h^{\lambda \beta} \partial^\alpha h^\rho_{\ \beta} 
    + \underline{h^{\lambda \beta} \partial_\alpha \partial^\alpha h^\rho_{\ \beta}} 
    + h^\rho_{\ \beta} \partial_\alpha \partial^\alpha h^{\lambda \beta} $ which are the $b_4 \partial_\alpha h^{\lambda \beta} \partial^\alpha h^\rho_{\ \beta}$ and $d_{5_i} h^\rho_{\ \beta} \partial_\alpha \partial^\alpha h^{\lambda \beta}$ terms (the $d_{5_{ii}}$ term is underlined). This would require is to write the solution as $T^{\rho \lambda} = 
    (\frac{b_4}{2} + d_{5_i})\partial_\alpha \partial^\alpha [h^{\lambda \beta} h^\rho_{\ \beta}] 
    - \frac{b_4}{2} h^\rho_{\ \beta} \partial_\alpha \partial^\alpha h^{\lambda \beta}
    - 2 d_{5_i} \partial_\alpha h^{\lambda \beta} \partial^\alpha h^\rho_{\ \beta} 
    - \underline{(\frac{b_4}{2} + d_{5_i}) h^{\lambda \beta} \partial_\alpha \partial^\alpha h^\rho_{\ \beta}}
    + ...$ which requires $\frac{b_4}{2} + d_{5_i} = 0$ to eliminate the $\partial_\alpha \partial^\alpha [h^{\lambda \beta} h^\rho_{\ \beta}] $ contribution. From this solution we have $T^{\rho \lambda} = 
    (0)\partial_\alpha \partial^\alpha [h^{\lambda \beta} h^\rho_{\ \beta}] 
    + d_{5_i} h^\rho_{\ \beta} \partial_\alpha \partial^\alpha h^{\lambda \beta}
    + b_4 \partial_\alpha h^{\lambda \beta} \partial^\alpha h^\rho_{\ \beta} 
    - \underline{(0) h^{\lambda \beta} \partial_\alpha \partial^\alpha h^\rho_{\ \beta}}
    + ...$, which again eliminates the $d_{5_{ii}}$ from the general expression and returns us to the solution in Appendix \ref{Fock to Noether Coefficient Comparison}.}\\

We have chosen the coefficients in (\ref{FockEMT}) to match exactly the Fock coefficients in \cite{Baker2021b} so that comparison between the results of the two studies is straightforward; subscripts i and ii indicate terms which form a symmetric pair, e.g. $b_{8i}$ and $b_{8ii}$. All such coefficients must be equal to form a symmetric energy-momentum tensor; such restrictions are discussed in Section \ref{SectionSymmetric}. 

\subsection{Important note on the mathematical framework developed in Section \ref{SectionGeneralResultSetFock}} \label{SectionImportantNote}

We make an important note here regarding the relationship of the results in this article which are a) the mathematical framework for the general set of Noetherian energy-momentum tensors in linearized gravity, and the ultimate goal of determining b) the general set of Noetherian energy-momentum tensors in linearized gravity. We are being clear in this distinction because we are considering the most general freedoms in the Noether identity i) all Lagrangian densities which yield the linearized Einstein field equations, and ii) all field transformations ($\delta h_{\mu\nu}$) which can be inserted into the Noether current. Solutions of the form a) are more general than b), but they do not guarantee Noetherian classification or satisfaction of the Noether identity. In other words, just because an energy-momentum tensor is derived from the Noether current, does not guarantee that Noether identity itself is satisfied. This means that all possible {\it{field transformations}} transformations must be further restricted to those which satisfy the Noether identity, yielding all possible {\it{symmetry transformations}} of the fields that will lead to result b); these two types of transformations are at the center of the distinction between a) and b). For this reason our results of type a) can be thought of as the general set of linearized gravity energy-momentum tensors which can be derived from the Noether current. Results of the type b) are the general set of Noetherian energy-momentum tensors in linearized gravity, which consist of all Noetherian energy-momentum tensors which satisfy the Noether identity. The derivation of a) in this article develops the mathematical framework for b) which is the subject of future work.\\

We would like to explain why we have calculated a) in this article and left b) for future work. The Noether identity is derived under simultaneous variation of coordinates ($\delta x_\alpha$) and fields ($\delta h_{\mu\nu}$), however these transformations must be symmetries of the action itself in order to satisfy the Noether identity. The converse is also true, if a given Noetherian energy-momentum tensor is conserved, then the corresponding transformations of coordinates and field are symmetry transformations. Deriving the general set of symmetry transformation restrictions for our broad study of linearized gravity is a highly lengthy and nontrivial calculation for reasons discussed in Section \ref{SectionConservedIdentity}. If these restrictions are calculated and applied, combined with our results in (\ref{FockEMT}) and Appendix \ref{Fock to Noether Coefficient Comparison}, one would have the result b). However, since b) is a subset of a), any results that do not satisfy a) will also not satisfy b) and therefore be non-Noetherian. Therefore our extensive study of the Noetherian nature of various published gravitational energy-momentum tensors in Section \ref{LiteratureComparison} is valid independent of the symmetry transformation calculations required for b), as the only possible Noetherian expression (Einstein's) is a known Noetherian result. The main reason to emphasize the distinction between a) and b) in this article is because in Section \ref{ResultsRestriction} we restrict the results of a) based on common physical criteria of an energy-momentum tensor, such as symmetry, tracelessness, and specific symmetry transformations which are published in the literature, however these results do not guarantee the simultaneous satisfaction of the criteria for b). We have stopped at a) because the numerous results of this article are valid for a), and extension to b) would involve considerably more restrictions, calculations and corresponding results that would dramatically increase the length of an already long study and article, and make the standalone results of the current article unnecessarily more cumbersome for the reader. Therefore we will leave the study of b) to future work where we will more broadly study the general set of action symmetries in linearized gravity and their relationship to the Noether identity as discussed in Section \ref{SectionConservedIdentity}.

\section{Comparison to published gravitational energy-momentum expressions in the literature} \label{LiteratureComparison}

Numerous gravitational energy-momentum expressions exist in the literature; an ultimate goal of such programs is to have an energy-momentum tensor for the gravitational field as in the case of physical field theories such as classical electrodynamics and Yang-Mills theory. Such attempts at uniqueness in general relativity have failed for a number of reasons, such as the lack of a finite global symmetry group which includes the 4-parameter Poincar{\'e} translation that an energy-momentum tensor derived from Noether's first theorem is defined with; see \say{{\it{Section 6 An Assertion of Hilbert}}} in Noether's article for related discussion to the lack of a finite symmetry group \say{{\it{concerning the relationship between the lack of a proper law of energy and general relativity}}} \cite{noether1918,kosmann2011}. Lack of uniqueness in the case of linearized gravity is primarily attributed to the lack of a possible gauge invariant energy-momentum tensor \cite{magnano2002}, a feature essential for uniqueness in foundational gauge theories such as classical electrodynamics and Yang-Mills theory \cite{Baker2021PHD}. These uniqueness issues have resulted in a large volume of published energy-momentum expressions in the literature that are primarily in two different categories: a) gravitational energy-momentum pseudotensors in general relativity and b) energy-momentum tensors defined in the area of linearized gravity. A distinction is made within a) due to the technical origin of pseudotensors and "complexes", where complexes include additional terms added to the traditional pseudotensor from the matter tensor \cite{moller1961}; however, sometimes both are referred to as gravitational energy-momentum pseudotensors. For our purpose, this distinction is irrelevant as we are interested in the precise mathematical form of each energy-momentum expression, therefore we will refer to both as pseudotensors. Expressions of the form a) can be linearized following the standard linearization method to determine an energy-momentum tensor for linearized gravity \cite{favata2001,bicak2016}; this process has allowed for uniqueness criteria to be explored, such as the Landau-Lifshitz energy-momentum expression being the unique conserved and symmetric expression contained only terms of the form $\partial h \partial h$ \cite{bicak2016}. The linearization of pseudotensors is nothing new as this is how Landau and Lifshitz themselves complete their famous quadrupole radiation formula \cite{landau1971}. Therefore, it is possible to determine whether or not any published gravitational energy-momentum expression is Noetherian in the context of linearized gravity, and by determining this it may be possible to narrow down the scope of published expressions by those which can be derived from the Noether identity, and those which can not. In this section we will determine this for many of the standard gravitational energy-momentum expressions which exist in the literature.\\

Several gravitational energy-momentum pseudotensors exist in the literature, such as Einstein \cite{Einstein1916}, Landau-Lifshitz \cite{landau1971}, Goldberg \cite{Goldberg1958} and Weinberg \cite{weinberg1972}. There are also several gravitational energy-momentum complexes, such as Papapetrou \cite{Papapetrou1948}, Bergmann-Thomson \cite{Bergmann1953}, Imitation Einstein \cite{Moller1958} and M{\o}ller \cite{Moller1958}. All of these expressions we will linearize and compare to the general set of energy-momentum tensors in linearized gravity which can be derived from the Noether current. There are also numerous energy-momentum tensors which have been defined in linearized gravity, such as Hilbert \cite{babak1999}, Fierz \cite{Toth2022}, Butcher \cite{Butcher2012PHD} and Padmanabhan \cite{padmanabhan2008}; all of these will also be compared to our general result (\ref{FockEMT}). As discussed, there is a long history of attempts to derive gravitational energy-momentum expressions and questions about their uniqueness --- we refer the reader to \cite{szabados1992,favata2001,tolman1930,cooperstock2000,xulu2003phd,sharif2005,padmanabhan2010,duerr2019,deHaro2022} for more discussion about some of the motivation behind and problems with the various energy-momentum expressions in general relativity and \cite{Baker2021b,padmanabhan2008,barcelo2014,magnano2002,bicak2016,Butcher2012PHD} for more discussion of the non-uniqueness problem in linearized gravity. Depending on the research article, different arguments have been presented in favour of different linearized gravity energy-momentum tensors. Some authors have argued for uniqueness of the Hilbert tensor in linearized gravity \cite{babak1999}, the Landau-Lifshitz expression due to its use in the quadrupole radiation formula \cite{landau1971} which is related after an approximation to the results of Hulse and Taylor \cite{hulse1975} (recipients of the 1993 Nobel Prize in Physics for linking the period decay of a binary pulsar system to energy loss due to gravitational radiation) or the more recent Fierz formalism of linearized gravity \cite{Toth2022}.  Other authors have argued against the notion of a physical linearized gravity energy-momentum tensor altogether, for example due to the aforementioned lack of gauge invariance \cite{magnano2002}. While there are numerous attempts to bypass some of the issues of defining an energy-momentum expression in general relativity, perhaps the most notable attempts are associated to quasi-local energy methods \cite{chen1999,chang1999,szabados2009}. The basic idea behind these methods is to assert that each pseudotensor is a Hamiltonian boundary term for a particular physical situation, and for a quasi-local region with given boundary conditions, the appropriate pseudotensor must be selected. This approach is perhaps the best available if one wishes to use the energy-momentum expressions directly in general relativity, thereby avoiding the issue of requiring a finite set of global symmetry transformations for Noether's first theorem. However, because our focus is on the Noetherian nature of linearized gravity energy-momentum tensors due to a set of globally defined symmetry transformations (e.g., the 10-parameter Poincar{\'e} group), such quasi-local methods will not be applicable to our study. We are interested in linearized gravity energy-momentum tensors which are globally defined as in the case of e.g. classical electrodynamics. Our focus in this section is specifically to determine which of the various published expressions for gravitational energy-momentum can be derived directly from Noether's first theorem, and which cannot.

\subsection{Linearization method} 

Before linearizing the various gravitational energy-momentum pseudotensors for comparison with the general set of energy-momentum tensors in linearized gravity which were derived from the Noether current in (\ref{FockEMT}) and Appendix \ref{Fock to Noether Coefficient Comparison}, we will briefly overview the standard linearization procedure, because some of these details are scattered throughout the literature. \\

The linearization of general relativity was proposed by Einstein who linearized his field equations as part of determining the linearized gravitational wave equation \cite{einstein1916gw}. The linearization of nonlinear differential equations via perturbation methods is attributed the work of Poincar{\'e} \cite{poincare1893} where the solution of a given nonlinear differential equation is expanded in terms of a small parameter $\epsilon$ such that at each order of the expansion a linear equation can be solved for. In this case of general relativity the solution is the metric tensor, and it is expanded as $g_{\mu \nu} = \eta_{\mu \nu} + \epsilon h_{\mu \nu} + \mathcal{O}(\epsilon^2)$ where all terms of order $\epsilon^2$ and higher represented by $\mathcal{O}(\epsilon^2)$ are neglected due to the small parameter $\epsilon$. The difference with this approach and that of Poincar{\'e} is that the leading term is taken to be the Minkowski metric, which takes on a different role than that of the subsequent terms as the background spacetime. Commonly the $\mathcal{O}(\epsilon^2)$ terms are omitted in the literature and the linearization about the Minkowski background is given simply as \cite{bicak2016,Papapetrou1948},

\begin{tabular}{@{}p{.5\linewidth}@{}p{.5\linewidth}@{}}
\begin{equation}
g_{\mu \nu} = \eta_{\mu \nu} + \epsilon h_{\mu \nu} ,   \label{eqn:cov. metric}
\end{equation}
  &
\begin{equation}
g^{\mu \nu} = \eta^{\mu \nu} - \epsilon h^{\mu \nu}  .  \label{eqn:con. metric}
\end{equation}
\end{tabular}

\noindent The reason for the sign change is to preserve the Kronecker delta definition in terms of the metric tensor to order $\epsilon$ as $\delta_\alpha^\beta = g_{\alpha\gamma} g^{\beta\gamma} = (\eta_{\alpha\gamma} + \epsilon h_{\alpha\gamma})(\eta^{\beta\gamma} - \epsilon h^{\beta\gamma}) = \eta_{\alpha\gamma} \eta^{\beta\gamma} = \delta_\alpha^\beta$. We note that some of the articles we refer to use abbreviations for the derivative of the metric such as $\partial_\sigma g^{\mu \nu} = g^{\mu \nu}_{\ \ \ , \sigma} = g^{\mu \nu}_\sigma$. Following the linearization procedure we can linearize the metric determinant $g = det(g_{\alpha\beta})$, as well as $\frac{1}{-g}$, $\sqrt{-g}$ and $\frac{1}{\sqrt{-g}}$ which all appear in some of the pseudotensors we will consider, 

\begin{tabular}{@{}p{.5\linewidth}@{}p{.5\linewidth}@{}}
\begin{gather}
    g = -1 - \epsilon h^\sigma_\sigma ,   \label{eqn:det g}
\\
 \frac{1}{-g} = 1 -  \epsilon h^\sigma_\sigma , \label{eqn:1/negg}
\end{gather}
  &
\begin{gather}
    \sqrt{-g} = 1 + \frac{1}{2} \epsilon h^\sigma_\sigma  ,  \label{eqn:sqrtg}
\\
    \frac{1}{\sqrt{-g}} = 1 - \frac{1}{2} \epsilon h^\sigma_\sigma .\label{eqn:1/sqrtg}
\end{gather}
\end{tabular}

\noindent The way in which these expressions are obtained is to calculate the linearization of the metric determinant $g$, up to order $\epsilon$, then use the binomial expansion to calculate the remaining expressions. The determinant linearization $g = det(g_{\mu \nu}) =  det(\eta_{\mu \nu} + \epsilon h_{\mu \nu})$ is calculated as,

\begin{equation}
    g = det
    \begin{bmatrix}
    1 + \epsilon h_{11} & h_{12} & h_{13} & h_{14} \\
    h_{21} & -1 + \epsilon h_{22} & h_{23} & h_{24} \\
    h_{31} & h_{32} & -1 + \epsilon h_{33} & h_{34} \\
    h_{41} & h_{42} & h_{43} & -1 + \epsilon h_{44}
    \end{bmatrix} .
\end{equation}

\noindent Taking this determinant we have $g = (1 + \epsilon h_{11})(-1 + \epsilon h_{22})(-1 + \epsilon h_{33})(-1 + \epsilon h_{44}) +\dots$. Keeping terms at most order $\epsilon$ we are left with $g \approx -1 - \epsilon h_{11} + \epsilon h_{22} + \epsilon h_{33} + \epsilon h_{44}$. Noting that $\epsilon \eta^{\mu\nu} h_{\mu\nu} = \epsilon h_{11} - \epsilon h_{22} - \epsilon h_{33} - \epsilon h_{44} $ and $h^\sigma_\sigma = \eta^{\mu\nu} h_{\mu\nu}$ we are left with the result $g \approx -1 - \epsilon \eta^{\mu\nu} h_{\mu\nu}$ which we have in (\ref{eqn:det g}).\\

From here we can use the binomial expansion ($[1 + x]^n = 1 + n x + \frac{n(n-1)}{2!} x^2 + \frac{n(n-1)(n-2)}{3!} x^3 + ...$) on the remaining expressions (\ref{eqn:1/negg})-(\ref{eqn:1/sqrtg}). Since $g = -1 - \epsilon h^\sigma_\sigma$ we have to expand $\frac{1}{-g} = \frac{1}{1 + \epsilon \eta^{\alpha \beta} h_{\alpha \beta}}$, $\sqrt{-g} = \sqrt{1 + \epsilon \eta^{\alpha \beta} h_{\alpha \beta}}$ and $\frac{1}{\sqrt{-g}} = \frac{1}{\sqrt{1 + \epsilon \eta^{\alpha \beta} h_{\alpha \beta}}}$: these expand to $\frac{1}{-g} = 1 - \epsilon \eta^{\alpha \beta} h_{\alpha \beta} + O(\epsilon^2)$, $\sqrt{-g} = 1 + \frac{1}{2} \epsilon \eta^{\alpha \beta} h_{\alpha \beta} + O(\epsilon^2)$ and $\frac{1}{\sqrt{-g}} = 1 - \frac{1}{2} \epsilon \eta^{\alpha \beta} h_{\alpha \beta} + O(\epsilon^2)$, which to $\epsilon$ order are what we have in (\ref{eqn:1/negg})-(\ref{eqn:1/sqrtg}).\\

There are also various abbreviations of the metric combined with $\sqrt{-g}$ which can be found in the literature such as \cite{bicak2016,Freud1939},

\begin{equation}
  \sqrt{-g} g^{\mu \nu}  = \hat{g}^{\mu \nu} = \mathfrak{g}^{\mu \nu} ,  \label{eqn:g hat}
\end{equation} 

\noindent which is sometimes referred to as the gothic metric. The Christoffel symbols of the first ($\Gamma_{\alpha \gamma \mu}$) and second ($\Gamma^\lambda_{\gamma \mu}$) kind are also often required,

\begin{equation}
    \label{eqn: Christoffel Def.}
    \Gamma^\lambda_{\gamma \mu} = 
    g^{\lambda \alpha} \Gamma_{\alpha \gamma \mu} = 
    \frac{1}{2} g^{\lambda \alpha} 
    (-\partial_\alpha g_{\gamma \mu} 
    + \partial_\mu g_{\gamma \alpha} 
    + \partial_{\gamma} g_{\alpha \mu}) .
\end{equation}

In order to linearize a pseudotensor, this standard linearization method is followed, keeping the $\epsilon^2$ (lowest order) terms \cite{bicak2016}, as in the case of linearizing the Einstein-Hilbert action --- both the Einstein-Hilbert action and the various pseudotensors have two $h$ contributions per term upon linearization (either of the form $\partial h \partial h$ or $h \partial \partial h$). The resulting expressions are linearized gravity energy-momentum tensors consisting of these $\partial h \partial h$ and $h \partial \partial h$ terms, all of which can be found in the general expression (\ref{FockEMT}); this equation includes all possible terms. We will start by linearizing the pseudotensors of general relativity and comparing them to the general set of energy-momentum tensors from (\ref{FockEMT}) and Appendix \ref{Fock to Noether Coefficient Comparison}.

\subsection{Pseudotensors, linearization and comparison}

We will now consider various published expressions for gravitational energy-momentum pseudotensors in the literature, their linearization, and whether or not the resulting expression can be derived from Noether's first theorem using the general set of energy-momentum tensors (\ref{FockEMT}) which were derived from the Noether current. If a given energy-momentum expression cannot be derived from the Noether current, then it follows that it is non-Noetherian. While some of these expressions have been linearized in the past, such as the Landau Lifshitz expression in \cite{bicak2016}, many others have not, therefore for completeness we will calculate all linearized expressions that we use in this article. \\

\subsubsection{Einstein pseudotensor}

In 1916 Einstein published the first of many gravitational pseudotensors, which is most often referred to as the Einstein pseudotensor \cite{Einstein1916}. If one linearizes the Einstein pseudotensor, they obtain the canonical Noether energy-momentum tensor \cite{Baker2021b,Chen2015}, which is the energy-momentum tensor derived directly from Noether's first theorem if one only uses the canonical transformations (\ref{canontrans}). For an article that discusses some of the history of the Einstein pseudotensor, see \cite{Chen2015}. Einstein's pseudotensor, like the canonical Noether tensor, is not symmetric. This lack of symmetry was a large part of the motivation to introduce some of the additional pseudotensors which we will linearize and compare in this section; a symmetric energy-momentum tensor is often viewed as a necessity in a physical field theory for reasons such as angular momentum conservation. The Einstein pseudotensor \cite{Einstein1916} is defined as,

\begin{equation}
    t^\nu_\sigma = 
    \frac{1}{2} (\mathcal{L} \delta^\nu_\sigma 
    - \frac{\partial \mathcal{L}}{\partial g^{\mu \rho}_\nu} g^{\mu \rho}_\sigma) , \label{Einsteinpseudo}
\end{equation}

\noindent where the Lagrangian originally used for this purpose comes from the so called Gamma-Gamma (or $\Gamma\Gamma$) part of the Einstein-Hilbert action,

\begin{equation}
    \mathcal{L} = 
    \sqrt{-g} g^{\mu \nu} 
    [\Gamma^\beta_{\mu \alpha} \Gamma^\alpha_{\nu \beta} 
    - \Gamma^\alpha_{\mu \nu} \Gamma^\beta_{\alpha \beta}] .
\end{equation}

\noindent Linearization of this Lagrangian yields the linearized Einstein-Hilbert action, which is equivalent (up to a negative sign) to the Fierz-Pauli Lagrangian density. Therefore one can use this linearized Lagrangian in the case of Noether's first theorem to yield the canonical Noether energy-momentum tensor directly from Noether's first theorem \cite{Baker2021b}. It is important to note that in his original article \cite{Einstein1915}, Einstein uses notation for the Christoffel symbol which is related to (\ref{eqn: Christoffel Def.}) as $\Gamma^\nu_{\alpha \beta} = - \{^{\alpha \beta}_{\ \nu} \}$.\\

Due to an ambiguity in the order of indices in (\ref{Einsteinpseudo}), we will consider both cases $t^{\rho \lambda}_L$ (the bottom index is raised to the left, becoming the first index) and $t^{\rho \lambda}_R$ (the bottom index raised to the right, becoming the second index). This is important because this expression is not symmetric; from Noether's theorem only one of these possibilities ($t^{\rho \lambda}_R$) is related to the canonical Noether tensor, and the other ($t^{\rho \lambda}_L$) is related to one of the Goldberg pseudotensors. Linearizing both possibilities we have,

\begin{multline}
    2 t^{\rho \lambda}_L = 
    \frac{1}{2} \eta^{\lambda \rho} \partial_\alpha  h_{\beta \mu} \partial^\beta  h^{\alpha \mu} 
    - \frac{1}{4} \eta^{\lambda \rho} \partial_\mu  h_{\alpha \beta} \partial^\mu  h^{\alpha \beta} 
    + \frac{1}{4} \eta^{\lambda \rho} \partial_\mu  h^\alpha_\alpha \partial^\mu  h^\beta_\beta 
    - \frac{1}{2} \eta^{\lambda \rho} \partial_\alpha  h^{\alpha \beta} \partial_\beta  h^\mu_\mu
    \\
    - \partial^\rho  h_{\alpha \beta} \partial^\alpha  h^{\beta \lambda} 
    + \frac{1}{2} \partial^\rho  h^{\alpha \beta} \partial^\lambda  h_{\alpha \beta} 
    - \frac{1}{2} \partial^\rho  h^\alpha_\alpha \partial^\lambda  h^\beta_\beta 
    + \frac{1}{2} \partial^\rho  h^\alpha_\alpha \partial_\beta  h^{\beta \lambda}
    + \frac{1}{2} \partial^\rho  h^{\alpha \lambda} \partial_\alpha  h^\beta_\beta , \label{EinsteinLeftRaise}
\end{multline} 

\begin{multline}
    2 t^{\rho \lambda}_R = 
    \frac{1}{2} \eta^{\rho \lambda} \partial_\alpha  h_{\beta \mu} \partial^\beta  h^{\alpha \mu} 
    - \frac{1}{4} \eta^{\rho \lambda} \partial_\mu  h_{\alpha \beta} \partial^\mu  h^{\alpha \beta} 
    + \frac{1}{4} \eta^{\rho \lambda} \partial_\mu  h^\alpha_\alpha \partial^\mu  h^\beta_\beta 
    - \frac{1}{2} \eta^{\rho \lambda} \partial_\alpha  h^{\alpha \beta} \partial_\beta  h^\mu_\mu
    \\
    - \partial^\lambda  h_{\alpha \beta} \partial^\alpha  h^{\beta \rho} 
    + \frac{1}{2} \partial^\lambda  h^{\alpha \beta} \partial^\rho  h_{\alpha \beta} 
    - \frac{1}{2} \partial^\lambda  h^\alpha_\alpha \partial^\rho  h^\beta_\beta 
    + \frac{1}{2} \partial^\lambda  h^\alpha_\alpha \partial_\beta  h^{\beta \rho}
    + \frac{1}{2} \partial^\lambda  h^{\alpha \rho} \partial_\alpha  h^\beta_\beta . \label{EinsteinRightRaise}
\end{multline}

\noindent We note that the terms proportional to the Minkowski metric in both cases are identical and proportional to the Fierz-Pauli Lagrangian density \cite{Baker2021b}.\\ 

We can now compare the two expressions to the general set of energy-momentum tensors in (\ref{FockEMT}). For $t^{\rho \lambda}_L$, the coefficients are: $b_6 = -\frac{1}{2}, b_7 = \frac{1}{2}, b_{8_i} = \frac{1}{2}, b_{9_i} = \frac{1}{2}, b_{11_i} = -1, c_1 = \frac{1}{4}, c_2 = -\frac{1}{4}, c_4 = \frac{1}{2}$ and $c_5 = -\frac{1}{2}$, the rest are zero. For $t^{\rho \lambda}_R$, the coefficients are: $b_6 = -\frac{1}{2}, b_7 = \frac{1}{2}, b_{8_{ii}} = \frac{1}{2}, b_{9_{ii}} = \frac{1}{2}, b_{11_{ii}} = -1, c_1 = \frac{1}{4}, c_2 = -\frac{1}{4}, c_4 = \frac{1}{2}$ and $c_5 = -\frac{1}{2}$, the rest are zero. \\

When solving $t^{\rho \lambda}_L$ we get $C_1 = -\frac{1}{4}, C_5 = \frac{1}{2}, D_3 = 0$ from (\ref{eqn:c_2 Coefficient}), (\ref{eqn:c_4 Coefficient}) and (\ref{eqn:a_5 Coefficient}). The equations (\ref{eqn:b_7 Coefficient}) and (\ref{eqn:b_11_ii Coefficient}) conflict once those coefficients are used. These conflicting equations are $\frac{1}{2} = b_7 = 2B_1 C_1 - B_1 D_3 $ and $0 = b_{11_{ii}} = 2 B_1 C_5 - \frac{1}{2} B_1 D_2 $ because $b_7$ requires $ B_1 = -1$ and $b_{11_{ii}}$ requires $D_2 = 2$, which conflicts with $0 = d_{12_{ii}} = \frac{1}{2} B_1 D_2$ which requires $D_2 = 0$. Therefore the linearized Einstein pseudotensor $t^{\rho \lambda}_L$ in (\ref{EinsteinLeftRaise}) is non-Noetherian.\\

When solving $t^{\rho \lambda}_R$ the non-Minkowski terms are swapped as compared to $t^{\rho \lambda}_L$. The above issue disappears as $b_{11_{ii}} = -1$. The system can be solved fully and the coefficients can be determined. For the non-zero symmetry coefficients (\ref{GenSymTransform}) we have only $B_1 = -1$. The non-zero Lagrangian coefficients (\ref{genlag}) are $C_1 = -\frac{1}{4}$, $C_2 = \frac{1}{4}$, $C_3 = - \frac{1}{2}$ and $C_5 = \frac{1}{2}$. The free parameter $n$ in Appendix \ref{Restriction1} is $n=1$. This result corresponds to the canonical Noether energy-momentum tensor that follows from the linearized Einstein-Hilbert action \cite{Baker2021b}. Therefore the linearized Einstein pseudotensor $t^{\rho \lambda}_R$ in (\ref{EinsteinRightRaise}) is Noetherian.

\subsubsection{Landau-Lifshitz pseudotensor} \label{SectionLL}

Landau and Lifshitz developed a symmetric pseudotensor in \cite{landau1971}, which involves a lengthy derivation not present in their book. Perhaps the only complete and explicit derivation of this expression can be found in \cite{Antonov2023}. In their book \cite{landau1971} they give their tensor in two forms,

\begin{multline}
    \tilde{t}^{\alpha \beta} = 
    \frac{c^4}{16 \pi k} [(g^{\alpha \lambda} g^{\beta \mu} 
    - g^{\alpha \beta} g^{\lambda \mu}) 
    (2 \Gamma^\nu_{\lambda \mu} \Gamma^\rho_{\nu \rho} 
    - \Gamma^\nu_{\lambda \rho} \Gamma^\rho_{\mu \nu} 
    - \Gamma^\nu_{\lambda \nu} \Gamma^\rho_{\mu \rho})
    + g^{\alpha \lambda} g^{\mu \nu} 
    (\Gamma^\beta_{\lambda \rho} \Gamma^\rho_{\mu \nu} 
    + \Gamma^\beta_{\mu \nu} \Gamma^\rho_{\lambda \rho} 
    - \Gamma^\beta_{\nu \rho} \Gamma^\rho_{\lambda \mu} 
    - \Gamma^\beta_{\lambda \mu} \Gamma^\rho_{\nu \rho})
    \\
    + g^{\beta \lambda} g^{\mu \nu} 
    (\Gamma^\alpha_{\lambda \rho} \Gamma^\rho_{\mu \nu} 
    + \Gamma^\alpha_{\mu \nu} \Gamma^\rho_{\lambda \rho} 
    - \Gamma^\alpha_{\nu \rho} \Gamma^\rho_{\lambda \mu} 
    - \Gamma^\alpha_{\lambda \mu} \Gamma^\rho_{\nu \rho})
    + g^{\lambda \mu} g^{\nu \rho} 
    (\Gamma^\alpha_{\lambda \nu} \Gamma^\beta_{\mu \rho} 
    - \Gamma^\alpha_{\lambda \mu} \Gamma^\beta_{\nu \rho})] ,
\end{multline}

\begin{multline}
    16 \pi (-g) \bar{t}^{\alpha \beta} = 
    \hat{g}^{\alpha \beta}_{\ \ ,\mu} \hat{g}^{\mu \nu}_{\ \ ,\nu} 
    - \hat{g}^{\alpha \mu}_{\ \ ,\mu} \hat{g}^{\beta \nu}_{\ \ ,\nu} 
    + \frac{1}{2} g^{\alpha \beta} g_{\mu \nu} \hat{g}^{\mu \lambda}_{\ \ ,\gamma} \hat{g}^{\gamma \nu}_{\ \ ,\lambda}
    - g_{\mu \nu} \hat{g}^{\mu \lambda}_{\ \ ,\gamma} (g^{\alpha \gamma} \hat{g}^{\beta \nu}_{\ \ ,\lambda} + g^{\beta \gamma} \hat{g}^{\alpha \nu}_{\ \ ,\lambda})
    \\
    + g_{\mu \nu} g^{\lambda \gamma} \hat{g}^{\alpha \mu}_{\ \ ,\lambda} \hat{g}^{\beta \nu}_{\ \ ,\gamma} 
    + \frac{1}{8} (2 g^{\alpha \mu} g^{\beta \nu} - g^{\alpha \beta} g^{\mu \nu})
    (2 g_{\lambda \gamma} g_{\rho \sigma} - g_{\gamma \rho} g_{\lambda \sigma}) \hat{g}^{\lambda \sigma}_{\ \ ,\mu} \hat{g}^{\gamma \rho}_{\ \ ,\nu} ,
\end{multline}

\noindent the latter is what is linearized in \cite{bicak2016}. We linearized both and confirmed that they yield the same equivalent linearization,

\begin{multline}
    \frac{64 \pi k}{c^4} \tilde{t}^{\alpha \beta} =  16 \pi \bar{t}^{\alpha \beta} =
    3 \eta^{\alpha \beta} \partial_\nu  h^\rho_\rho \partial^\nu  h^\mu_\mu 
    - 4 \eta^{\alpha \beta} \partial_\nu  h^\rho_\rho \partial_\mu  h^{\mu \nu}
    + 2 \eta^{\alpha \beta} \partial_\rho  h_{\mu \nu} \partial^\nu  h^{\mu \rho} 
    - \eta^{\alpha \beta} \partial_\rho  h_{\mu \nu} \partial^\rho  h^{\mu \nu}
    \\
    - 6 \partial_\nu  h^\rho_\rho \partial^\nu  h^{\alpha \beta} 
    + 2 \partial_\nu  h^\rho_\rho \partial^\beta  h^{\alpha \nu} 
    + 2 \partial_\nu  h^\rho_\rho \partial^\alpha  h^{\beta \nu} 
    + 4 \partial_\rho  h^{\beta \nu} \partial^\rho  h^\alpha_{\ \nu} 
    - 4 \partial^\beta  h^\rho_\rho \partial^\alpha  h^\nu_\nu 
    + 4 \partial_\nu  h^{\nu \rho} \partial_\rho  h^{\alpha \beta} 
    \\
    + 4 \partial^\alpha  h^\rho_\rho \partial_\nu  h^{\beta \nu} 
    + 2 \partial^\alpha  h^{\nu \rho} \partial^\beta  h_{\nu \rho} 
    - 4 \partial^\alpha  h_{\nu \rho} \partial^\nu  h^{\beta \rho} 
    + 4 \partial^\beta  h^\rho_\rho \partial_\nu  h^{\alpha \nu} 
    - 4 \partial^\beta  h_{\nu \rho} \partial^\nu  h^{\alpha \rho} 
    - 4 \partial_\nu  h^{\beta \nu} \partial_\rho  h^{\alpha \rho} . \label{LLLin}
\end{multline}

The non-zero coefficients in (\ref{FockEMT}) are $b_1 = 4, b_2 = -6, b_3 = -4, b_4 = 4, b_6 = -4, b_7 = 2, b_{8_i} = 2, b_{8_{ii}} = 2, b_{9_i} = 4, b_{9_{ii}} = 4, b_{11_i} = -4, b_{11_{ii}} = -4, c_1 = 3, c_2 = -1, c_4 = 2$ and $c_5 = -4$. We immediately have conflicting equations (\ref{eqn:b_4 Coefficient}) and (\ref{eqn:b_5 Coefficient}), which are $4 = b_4 = B_4 C_5 + B_5 C_5 - \frac{1}{4} B_4 D_2 - \frac{1}{4} B_5 D_2 $ and $0 = b_5 = B_4 C_5 + B_5 C_5 - \frac{1}{4} B_4 D_2 - \frac{1}{4} B_5 D_2 $, therefore the linearized Landau-Lifshitz pseudotensor (\ref{LLLin}) is non-Noetherian.

\subsubsection{Goldberg mixed index pseudotensor}

Goldberg published a generalized set of pseudotensors to an arbitrary weight $n$ in 1958 \cite{Goldberg1958}; one non-symmetric mixed index set of expressions, and one symmetric set of expressions. We will first consider his mixed index pseudotensor for weight $n=0$. For Goldberg's expressions we are using the gothic metric $\mathfrak{g}_{\mu \nu} = \sqrt{-g} g_{\mu \nu}$ from (\ref{eqn:g hat}); however we note that the reader has to be careful when reading the pseudotensor literature as some authors \cite{Notte2009, Bhattacharyya2021, Kopeikin2013} define instead $\mathfrak{g}_{\mu \nu} = \frac{1}{\sqrt{-g}} g_{\mu \nu}$ whose linearization (\ref{eqn:1/sqrtg}) differs from the more standard version in (\ref{eqn:sqrtg}). In this case one will yield equivalent results to order $\epsilon^2$ regardless of which $\mathfrak{g}_{\mu \nu}$ definition is used, but in scenarios where this is not the case one would yield conflicting results. The mixed index Goldberg pseudotensor is,

\begin{equation}
    t_{(n) \mu}^{\ \ \ \ \ \nu} = (-g)^{\frac{n}{2}} [t_{\mu}^{\ \nu} + \frac{1}{2} n U_{\mu}^{\ [\nu \sigma]} (\ln(|g|))_{, \sigma}] ,
\end{equation}

\noindent where $U_{\mu}^{\ [\nu \rho]} = \mathfrak{g}_{\mu \lambda} H^{[\lambda \sigma] [\nu \rho]}_{\ \ \ \ \ \ \ \ , \sigma} $ and $H^{[\lambda \sigma] [\nu \rho]}  = \mathfrak{g}^{\lambda \nu} \mathfrak{g}^{\rho \sigma} - \mathfrak{g}^{\lambda \rho} \mathfrak{g}^{\nu \sigma}$. When $n=0$, this Goldberg pseudotensor is related to Einstein's pseudotensor, with particular index order,

\begin{multline}
    t_{\mu}^{\ \nu} = 
    -\frac{1}{8} \delta_{\mu}^{\ \nu} 
    [2 \mathfrak{g}^{\rho \sigma} \mathfrak{g}_{\lambda \iota} \mathfrak{g}_{\kappa \tau} 
    - \mathfrak{g}^{\rho \sigma} \mathfrak{g}_{\iota \kappa} \mathfrak{g}_{\lambda \tau} 
    - 4 \delta_{\kappa}^{\ \sigma} \delta_{\lambda}^{\ \rho} \mathfrak{g}_{\iota \tau}] 
    \mathfrak{g}^{\iota \kappa}_{\ \ , \rho} \mathfrak{g}^{\lambda \tau}_{\ \ , \sigma}
    \\
    + \frac{1}{4}
    [2 \mathfrak{g}^{\nu \sigma} \mathfrak{g}_{\lambda \iota} \mathfrak{g}_{\kappa \tau} 
    - \mathfrak{g}^{\nu \sigma} \mathfrak{g}_{\iota \kappa} \mathfrak{g}_{\lambda \tau} 
    - 4 \delta_{\lambda}^{\ \nu} \delta_{\kappa}^{\ \sigma} \mathfrak{g}_{\iota \tau}] 
    \mathfrak{g}^{\lambda \tau}_{\ \ , \sigma} \mathfrak{g}^{\iota \kappa}_{\ \ , \mu} . \label{GoldMixedNonlin}
\end{multline}

\noindent Therefore only the left raised index version $t^{\rho \lambda}_L$ of Einstein (\ref{EinsteinLeftRaise}) can be determined, as from (\ref{GoldMixedNonlin}) we can determine the linearization,

\begin{multline}
    t^{\rho \lambda} = 
    \frac{1}{4} \eta^{\rho \lambda}
    [- \partial_\sigma  h_{\alpha \beta} \partial^\sigma  h^{\alpha \beta}
    + \partial_\sigma  h^\alpha_\alpha \partial^\sigma  h^\beta_\beta
    + 2 \partial_\sigma  h_{\alpha \beta} \partial^\alpha  h^{\sigma \beta}  
    - 2 \partial_\sigma  h^\alpha_\alpha \partial_\beta  h^{\sigma \beta}] 
    \\
    + \frac{1}{2} \partial^\rho  h_{\alpha \beta} \partial^\lambda  h^{\alpha \beta} 
    - \frac{1}{2} \partial^\rho  h^\alpha_\alpha \partial^\lambda  h^\beta_\beta 
    - \partial^\rho  h_{\alpha \beta} \partial^\alpha  h^{\lambda \beta}  
    + \frac{1}{2} \partial^\rho  h^\alpha_\alpha \partial_\beta  h^{\lambda \beta} 
    + \frac{1}{2} \partial^\rho  h^{\lambda \beta} \partial_\beta  h^\alpha_\alpha ,\label{GoldbergMixedLin}
\end{multline}

\noindent which is indeed equivalent (\ref{EinsteinLeftRaise}). Therefore by the same solution as for (\ref{EinsteinLeftRaise}), the linearized Goldberg pseudotensor (\ref{GoldbergMixedLin}) is non-Noetherian.

\subsubsection{Goldberg symmetric pseudotensor}

In addition to his mixed pseudotensor, Goldberg determined a symmetric pseudotensor of arbitrary weight $n$,

\begin{equation}
    \tau_{(n)}^{\ \ \ \mu \nu} = (-g)^{\frac{n}{2}} [\tau^{\mu \nu} 
    + \frac{1}{2} n [(ln|g|)_{, \sigma \tau} + \frac{1}{2} n (ln|g|)_{, \sigma} (ln|g|)_{,\tau}] H^{[\mu \tau] [\nu \sigma]}
    + \frac{1}{2} n (ln|g|)_{,\sigma} [H^{[\mu \tau][\nu \sigma]} + H^{[\nu \tau] [\mu \sigma]}]] ,
\end{equation}

\noindent where $H^{[\lambda \sigma] [\nu \rho]}  = \mathfrak{g}^{\lambda \nu} \mathfrak{g}^{\rho \sigma} - \mathfrak{g}^{\lambda \rho} \mathfrak{g}^{\nu \sigma}$. For the weight $n = 0$ this pseudotensor reduces to $\tau^{\mu \nu} = \mathfrak{g}^{\mu \lambda}_{\ \ , \sigma} U^{\ [\nu \sigma]}_{\lambda} + \mathfrak{g}^{\mu \lambda} t^{\ \nu}_{\lambda}$, which linearizes to,

\begin{multline}
    4 \tau^{\alpha \beta} = 
    4 \partial_\sigma  h^{\alpha \lambda} \partial^\sigma  h_\lambda^{\ \beta} 
    - 6 \partial_\sigma  h^{\alpha \beta} \partial^\sigma  h^\lambda_\lambda 
    + 4 \partial_\sigma  h^{\alpha \beta} \partial_\lambda  h^{\sigma \lambda} 
    - 4 \partial^\sigma  h^{\alpha \lambda} \partial^\beta  h_{\lambda \sigma} 
    + 4 \partial_\sigma  h^{\alpha \sigma} \partial^\beta  h^\lambda_\lambda 
    \\
    - 4 \partial_\sigma  h^{\alpha \sigma} \partial_\lambda  h^{\beta \lambda}
    - 4 \partial^\alpha  h^\kappa_\kappa \partial^\beta  h^\lambda_\lambda 
    + 4 \partial_\lambda  h^{\beta \lambda} \partial^\alpha  h^\kappa_\kappa 
    + 2 \partial^\beta  h^{\alpha \sigma} \partial_\sigma  h^\lambda_\lambda 
    + 2 \partial^\beta  h_{\lambda \kappa} \partial^\alpha  h^{\lambda \kappa} 
    - 4 \partial^\lambda  h^{\beta \kappa} \partial^\alpha  h_{\lambda \kappa} 
    \\
    + 2 \partial_\kappa  h^\lambda_\lambda \partial^\alpha  h^{\beta \kappa} 
    + 3 \eta^{\alpha \beta} \partial_\sigma  h^\kappa_\kappa \partial^\sigma  h^\lambda_\lambda 
    - 4 \eta^{\alpha \beta} \partial_\sigma  h^\kappa_\kappa \partial_\lambda  h^{\sigma \lambda}
    - \eta^{\beta \alpha} \partial_\sigma  h_{\lambda \kappa} \partial^\sigma  h^{\lambda \kappa} 
    + 2 \eta^{\beta \alpha} \partial_\sigma  h_{\lambda \kappa} \partial^\kappa  h^{\sigma \lambda} . \label{GoldbergSymLin}
\end{multline}

\noindent This is equivalent to the linearized Landau-Lifshitz pseudotensor (\ref{LLLin}), therefore by the same calculations as in Section \ref{SectionLL}, the linearized symmetric Goldberg pseudotensor (\ref{GoldbergSymLin}) is non-Noetherian.

\subsubsection{Weinberg pseudotensor} 

The Weinberg pseudotensor \cite{weinberg1972} can be expressed as,

\begin{equation}
    t_{\mu \kappa} = \frac{1}{8 \pi G} 
    [R_{\mu \kappa} 
    - \frac{1}{2} g_{\mu \kappa} R^\lambda_{\ \lambda} 
    - R^{(1)}_{\ \ \ \mu \kappa} 
    + \frac{1}{2} \eta_{\mu \kappa} R^{(1) \lambda}_{\ \ \ \ \lambda}] .
\end{equation} 

In \cite{weinberg1972} the standard linearized Ricci tensor is defined as,

\begin{equation}
    R^{(1)}_{\ \ \ \mu \kappa} = \frac{1}{2} 
    (\partial_\mu \partial_\kappa \epsilon h^\lambda_\lambda 
    - \partial_\lambda \partial_\kappa \epsilon h^\lambda_{\ \mu} 
    - \partial_\lambda \partial_\mu \epsilon h^\lambda_{\ \kappa} 
    + \partial_\lambda \partial^\lambda \epsilon h_{\mu \kappa}) ,
\end{equation}

\noindent where $R_{\mu \kappa} = R^\lambda_{\ \mu \lambda \kappa}$ and $R^\lambda_{\ \mu \nu \kappa} = \partial_\kappa \Gamma^\lambda_{\mu \nu} 
    - \partial_\nu \Gamma^\lambda_{\mu \kappa}
    + \Gamma^\eta_{\mu \nu} \Gamma^\lambda_{\kappa \eta} 
    - \Gamma^\eta_{\mu \kappa} \Gamma^\lambda_{\nu \eta}$. Weinberg's book \cite{weinberg1972} distinguishes how to raise and lower indices for his expressions, with the metric on true tensors and the Minkowski metric on the rest (linearized "(1)" expressions above). Using these definitions, Weinberg's pseudotensor can be linearized as, 

\begin{multline}
    8 \pi G t_{\mu \kappa} =
    \frac{3}{8} \eta_{\mu \kappa} \partial_\lambda  h_{\gamma \nu} \partial^\lambda  h^{\gamma \nu} 
    - \frac{1}{2} \eta_{\mu \kappa} \partial_\lambda  h^{\lambda \nu} \partial^\gamma  h_{\gamma \nu} 
    + \frac{1}{2} \eta_{\mu \kappa} \partial_\gamma  h^\lambda_\lambda \partial_\nu  h^{\gamma \nu} 
    - \frac{1}{4} \eta_{\mu \kappa} \partial^\nu  h^{\gamma \lambda} \partial_\lambda  h_{\gamma \nu} 
    \\
    - \frac{1}{8} \eta_{\mu \kappa} \partial_\nu  h^\lambda_\lambda \partial^\nu  h^\gamma_\gamma 
    - \frac{1}{4} \partial_\kappa  h^{\lambda \nu} \partial_\mu  h_{\lambda \nu} 
    + \frac{1}{2} \partial_\lambda  h^{\nu \lambda} \partial_\mu  h_{\kappa \nu} 
    + \frac{1}{2} \partial_\lambda  h^{\nu \lambda} \partial_\kappa  h_{\mu \nu} 
    - \frac{1}{2} \partial_\lambda  h^{\lambda \nu} \partial_\nu  h_{\kappa \mu} 
    \\
    + \frac{1}{2} \partial^\nu  h_{\kappa \lambda} \partial^\lambda  h_{\mu \nu} 
    - \frac{1}{2} \partial_\lambda  h^\nu_{\ \kappa} \partial^\lambda  h_{\mu \nu} 
    - \frac{1}{4} \partial^\nu  h^\lambda_\lambda \partial_\mu  h_{\kappa \nu} 
    - \frac{1}{4} \partial^\nu  h^\lambda_\lambda \partial_\kappa  h_{\mu \nu} 
    + \frac{1}{4} \partial_\nu  h^\lambda_\lambda \partial^\nu  h_{\kappa \mu} 
    \\
    - \eta_{\mu \kappa}  h^{\lambda \nu} \partial_\lambda \partial^\gamma  h_{\gamma \nu} 
    + \frac{1}{2} \eta_{\mu \kappa}  h^{\lambda \nu} \partial_\gamma \partial^\gamma  h_{\lambda \nu} 
    + \frac{1}{2} \eta_{\mu \kappa}  h^{\lambda \gamma} \partial_\lambda \partial_\gamma  h^\nu_\nu 
    + \frac{1}{2}  h^{\nu \lambda} \partial_\lambda \partial_\mu  h_{\kappa \nu} 
    - \frac{1}{2}  h^{\lambda \nu} \partial_\kappa \partial_\mu  h_{\lambda \nu} 
    \\
    + \frac{1}{2}  h^{\nu \lambda} \partial_\lambda \partial_\kappa  h_{\mu \nu} 
    - \frac{1}{2}  h^{\lambda \nu} \partial_\lambda \partial_\nu  h_{\kappa \mu}  
    + \frac{1}{2}  h_{\mu \kappa} \partial_\gamma \partial_\nu  h^{\gamma \nu} 
    - \frac{1}{2}  h_{\mu \kappa} \partial_\gamma \partial^\gamma  h^\nu_\nu .  \label{LinWein}
\end{multline}

The non-zero coefficients in (\ref{FockEMT}) are: $b_1 = -\frac{1}{2}, b_2 = \frac{1}{4}, b_4 = -\frac{1}{2}, b_5 = \frac{1}{2}, b_7 = -\frac{1}{4}, b_{8_i} = -\frac{1}{4}, b_{8_{ii}} = -\frac{1}{4}, b_{10_i} = \frac{1}{2}, b_{10_{ii}} = \frac{1}{2}, c_1 = -\frac{1}{8}, c_2 = \frac{3}{8}, c_3 = -\frac{1}{2}, c_4 = -\frac{1}{4}, c_5 = \frac{1}{2}, d_1 = -\frac{1}{2}, d_2 = \frac{1}{2}, d_4 = -\frac{1}{2}, d_8 = -\frac{1}{2}, d_{12_i} = \frac{1}{2}, d_{12_{ii}} = \frac{1}{2}, a_1 = \frac{1}{2}, a_3 = -1$ and $a_5 = \frac{1}{2}$. The equations (\ref{eqn:d_4 Coefficient}) and (\ref{eqn:d_5_i Coefficient}) conflict as $ -\frac{1}{2} = d_4 = \frac{1}{4} B_4 D_2 + \frac{1}{4} B_5 D_2 $ and $0 = d_{5_i} = \frac{1}{4} B_4 D_2 + \frac{1}{4} B_5 D_2 $. Therefore the linearized Weinberg pseudotensor (\ref{LinWein}) is non-Noetherian.

\subsection{Complexes, linearization and comparison}\label{Complexes, linearization and comparison}

We will now continue with our linearization of various common pseudotensors, however we sort the expressions in this Section \ref{Complexes, linearization and comparison} based on the complex derivations as defined in \cite{moller1961}. Commonly in the literature these are just referred to interchangeably as gravitational energy-momentum pseudotensors, and for our purposes, either terminology is suitable.

\subsubsection{Papapetrou pseudotensor} 

Papapetrou developed a symmetric energy-momentum pseudotensor \cite{Papapetrou1948} by developing a superpotential in the same form as the Belinfante symmetrization procedure \cite{belinfante1940},

\begin{equation}
    16 \pi \theta^{\rho \lambda} = \partial_\alpha \partial_\beta 
    [\sqrt{-g}(\eta^{\alpha \beta} g^{\rho \lambda}  
    - \eta^{\lambda \beta} g^{\rho \alpha}
    + \eta^{\rho \lambda} g^{\alpha \beta} 
    - \eta^{\rho \beta} g^{\lambda \alpha})] .
\end{equation}

\noindent Linearizing this pseudotensor we obtain,

\begin{multline}
    32 \pi \theta^{\rho \lambda} = 
    \partial^\lambda  h^{\rho \alpha} \partial_\alpha  h^\gamma_\gamma 
    + \partial^\rho  h^{\lambda \alpha} \partial_\alpha  h^\gamma_\gamma
    + \partial_\alpha  h^{\rho \alpha} \partial^\lambda  h^\gamma_\gamma 
    + \partial_\alpha  h^{\lambda \alpha} \partial^\rho  h^\gamma_\gamma
    - 2 \partial_\alpha  h^{\rho \lambda} \partial^\alpha  h^\gamma_\gamma 
    \\
    - 2 \eta^{\rho \lambda} \partial_\alpha  h^{\alpha \beta} \partial_\beta  h^\gamma_\gamma  
    -  h^\gamma_\gamma \partial_\alpha \partial^\alpha  h^{\rho \lambda} 
    -  h^{\rho \lambda} \partial_\alpha \partial^\alpha  h^\gamma_\gamma 
    +  h^\gamma_\gamma \partial_\alpha \partial^\lambda  h^{\rho \alpha} 
    +  h^\gamma_\gamma \partial_\alpha \partial^\rho  h^{\lambda \alpha} 
    \\
    +  h^{\rho \alpha} \partial_\alpha \partial^\lambda  h^\gamma_\gamma 
    +  h^{\lambda \alpha} \partial_\alpha \partial^\rho  h^\gamma_\gamma 
    - \eta^{\rho \lambda}  h^\gamma_\gamma \partial_\alpha \partial_\beta  h^{\alpha \beta} 
    - \eta^{\rho \lambda}  h^{\alpha \beta} \partial_\alpha \partial_\beta  h^\gamma_\gamma .  \label{PapaLin}
\end{multline}

Comparing to (\ref{FockEMT}) the non-zero coefficients are: $b_2 = -2, b_{8_i} = 1, b_{8_{ii}} = 1, b_{9_i} = 1, b_{9_{ii}} = 1, c_5 = -2, d_1 = -1, d_3 = -1, d_{9_i} = 1, d_{9_{ii}} = 1, d_{11_i} = 1, d_{11_{ii}} = 1, a_1 = -1$ and $a_2 = -1$. We know that $C_1 = 0, C_5 = 0, D_3 = 0$ from (\ref{eqn:c_2 Coefficient}), (\ref{eqn:c_4 Coefficient}), and (\ref{eqn:a_5 Coefficient}). These cause equations (\ref{eqn:d_1 Coefficient}) and (\ref{eqn:d_9_ii Coefficient}) to conflict, namely $-1 = d_1 = \frac{1}{4} B_6 D_2 + \frac{1}{4} B_7 D_2 $ and $1 = d_{9_{ii}} = \frac{1}{4} B_6 D_2 + \frac{1}{4} B_7 D_2 + B_6 D_3 + B_7 D_3 $. Therefore, the linearized Papapetrou pseudotensor (\ref{PapaLin}) is non-Noetherian. 

\subsubsection{Bergmann-Thomson pseudotensor} 

The Bergmann-Thomson pseudotensor was introduced as an expression which they state can be interpreted as the total energy-momentum of all fields present \cite{Bergmann1953}; it can be expressed in compact form as,

\begin{equation}
    T_\rho^{\ \lambda} = - U^{[\gamma \lambda]}_{\ \ \ \ \rho, \gamma} ,
\end{equation}

\noindent where $U^{[\gamma \lambda]}_{\ \ \ \rho}$ is attributed to Freud \cite{Freud1939}, which is defined in terms of the following determinant,

\begin{equation}
    U^{[\gamma \lambda]}_{\ \ \ \rho} = 
    \frac{1}{2} det
    \begin{vmatrix}
     \delta^\gamma_\rho & \delta^\lambda_\rho & \delta^\mu_\rho
     \\ 
     \mathfrak{g}^{\gamma \nu} & \mathfrak{g}^{\lambda \nu} & \mathfrak{g}^{\mu \nu}
     \\
     \Gamma^\gamma_{\nu \mu} & \Gamma^\lambda_{\nu \mu} & \Gamma^\mu_{\nu \mu} 
\end{vmatrix}
,
\end{equation}

\noindent which expands to $U^{[\gamma \lambda]}_{\ \ \ \rho} = \frac{1}{2} [\delta^\gamma_\rho (\mathfrak{g}^{\lambda \nu} \Gamma^\mu_{\nu \mu} - \mathfrak{g}^{\mu \nu} \Gamma^\lambda_{\nu \mu})
     - \delta^\lambda_\rho (\mathfrak{g}^{\gamma \nu} \Gamma^\mu_{\nu \mu} - \mathfrak{g}^{\mu \nu} \Gamma^\gamma_{\nu \mu} )
     + \delta^\mu_\rho (\mathfrak{g}^{\gamma \nu} \Gamma^\lambda_{\nu \mu} - \mathfrak{g}^{\lambda \nu} \Gamma^\gamma_{\nu \mu})] $. It is important to note that $\mathfrak{g}^{\alpha \beta}$ is the gothic metric $\mathfrak{g}^{\alpha \beta} = \sqrt{-g} g^{\alpha\beta}$. Linearizing the Bergman-Thomson pseudotensor we obtain,

\begin{multline}
    4 T_{G}^{\rho \lambda} = 
    2 \partial_\iota  h^\mu_\mu \partial^\rho  h^{\iota \lambda} 
    - \partial^\rho  h^\iota_\iota \partial^\lambda  h^\mu_\mu 
    + 2 \partial^\rho  h_{\mu \iota} \partial^\lambda  h^{\mu \iota} 
    - 2 \partial^\rho  h_{\mu \iota} \partial^\mu  h^{\iota \lambda} 
    + \partial^\rho  h^\mu_\mu \partial_\iota  h^{\lambda \iota} 
    - 2 \partial^\rho  h^{\iota \lambda} \partial^\mu  h_{\mu \iota} 
    \\
    - 3 \eta^{\lambda \rho} \partial_\iota  h^{\iota \sigma} \partial_\sigma  h^\mu_\mu 
    + \eta^{\lambda \rho} \partial_\iota  h^\sigma_\sigma \partial^\iota  h^\mu_\mu 
    - 2 \eta^{\lambda \rho} \partial_\iota  h^{\mu \sigma} \partial^\iota  h_{\mu \sigma} 
    + 2 \eta^{\lambda \rho} \partial_\iota  h_{\mu \sigma} \partial^\sigma  h^{\mu \iota} 
    + 2 \eta^{\lambda \rho} \partial_\iota  h^{\iota \sigma} \partial^\mu  h_{\mu \sigma} 
    \\
    + 2 \partial_\iota  h^{\iota \mu} \partial_\mu  h^{\lambda \rho} 
    - \partial_\iota  h^\mu_\mu \partial^\iota  h^{\lambda \rho} 
    - 2 \partial^\iota  h_{\iota \mu} \partial^\lambda  h^{\rho \mu} 
    + \partial_\iota  h^\mu_\mu \partial^\lambda  h^{\iota \rho} 
    + 2 \partial_\iota  h^{\lambda \mu} \partial^\iota  h^\rho_{\ \mu} 
    - 2 \partial_\iota  h^{\lambda \mu} \partial_\mu  h^{\iota \rho}
    \\
    + 2  h^{\iota \lambda} \partial_\iota \partial^\rho  h^\mu_\mu 
    -  h^\iota_\iota \partial^\lambda \partial^\rho  h^\mu_\mu 
    + 2  h_{\mu \iota} \partial^\lambda \partial^\rho  h^{\mu \iota} 
    - 2  h_{\mu \iota} \partial^\mu \partial^\rho  h^{\iota \lambda} 
    +  h^\mu_\mu \partial_\iota \partial^\rho  h^{\lambda \iota} 
    - 2  h^{\iota \lambda} \partial^\mu \partial^\rho  h_{\mu \iota} 
    \\
    - 2 \eta^{\lambda \rho}  h^{\iota \sigma} \partial_\sigma \partial_\iota  h^\mu_\mu 
    + \eta^{\lambda \rho}  h^\sigma_\sigma \partial^\iota \partial_\iota  h^\mu_\mu 
    - 2 \eta^{\lambda \rho}  h^{\mu \sigma} \partial^\iota \partial_\iota  h_{\mu \sigma} 
    + 4 \eta^{\lambda \rho}  h^{\sigma \mu} \partial^\iota \partial_\sigma  h_{\mu \iota} 
    - \eta^{\lambda \rho}  h^\sigma_\sigma \partial_\mu \partial_\iota  h^{\mu \iota} 
    + 2  h^{\iota \mu} \partial_\mu \partial_\iota  h^{\lambda \rho}
    \\
    -  h^\mu_\mu \partial_\iota \partial^\iota  h^{\lambda \rho} 
    - 2  h_{\iota \mu} \partial^\lambda \partial^\iota  h^{\rho \mu} 
    +  h^\mu_\mu \partial^\lambda \partial_\iota  h^{\iota \rho} 
    + 2  h^{\lambda \mu} \partial_\iota \partial^\iota  h^\rho_{\ \mu} 
    - 2  h^{\lambda \mu} \partial_\mu \partial_\iota  h^{\iota \rho} . \label{BTPseudo}
\end{multline}

The non-zero coefficients in (\ref{FockEMT}) are: $b_1 = 2, b_2 = -1, b_4 = 2, b_5 = -2, b_6 = -1, b_7 = 2, b_{8_i} = 2, b_{8_{ii}} = 1, b_{9_i} = 1, b_{10_i} = -2, b_{10_{ii}} = -2, b_{11_i} = -2, c_1 = 1, c_2 = -2, c_3 = 2, c_4 = 2, c_5 = -3, d_3 = -1, d_4 = 2, d_{5_{ii}} = 2, d_{6_{ii}} = -2, d_7 = -1, d_8 = 2, d_{9_{ii}} = 2, d_{10_{ii}} = -2, d_{11_i} = 1, d_{11_{ii}} = 1, d_{12_i} = -2, d_{12_{ii}} = -2, a_1 = -2, a_2 = -1, a_3 = 4, a_4 = 1$ and $a_5 = -2$. The coefficient $d_{5_{ii}} = 0$ in Appendix \ref{Fock to Noether Coefficient Comparison}, thus the coefficient equation (\ref{eqn:d_5_ii Coefficient}) is not consistent with Bergmann-Thomson's result, as it yields $2 = d_{5_{ii}} = 0$. Therefore, the linearized Bergmann-Thomson pseudotensor (\ref{BTPseudo}) is non-Noetherian.

\subsubsection{Imitation Einstein pseudotensor}

We now turn to an expression that is sometimes mistaken as the Einstein pseudotensor in the literature, which we will refer to as the Imitation Einstein pseudotensor. The Einstein pseudotensor (\ref{Einsteinpseudo}) was clearly defined by Einstein \cite{Einstein1916}. Another "Einstein" pseudotensor can be found in the literature, for example in \cite{matyjasek2008,padmanabhan2010}, defined compactly as,

\begin{equation}
    \theta_\rho^{\ \lambda} = \partial_\sigma h^{\ \lambda \sigma}_\rho ,\label{ImitationEinsteinPseudo}
\end{equation}

\noindent where $h_\rho^{\ \lambda \sigma} = \frac{g_{\rho \nu}}{2 k \sqrt{-g}} \partial_\mu [(-g) (g^{\lambda \nu} g^{\sigma \mu} - g^{\sigma \nu} g^{\lambda \mu})]$. As we will see, the result of linearizing (\ref{ImitationEinsteinPseudo}) is not equivalent to the linerization (left or right) of the Einstein pseudotensor (\ref{Einsteinpseudo}). This can cause some confusion as this expression is incorrectly attributed to Einstein; it was first introduced by M{\o}ller in \cite{Moller1958}, who modified the original Einstein expression using his complex definition. However, this confusion cannot be attributed to M{\o}ller himself, who merely introduced this expression, without claiming the modified version was the "Einstein" pseudotensor. The relationship between the Einstein and Imitation Einstein pseudotensors was detailed by Rosen in \cite{Rosen1994}. We have included in tables in Appendix \ref{Table of Fock Coefficients vs. Pseudotensors} the nonzero terms in (\ref{FockEMT}) for all of our comparisons, which can be used to check the distinction between the linearization of the Einstein (\ref{Einsteinpseudo}) and Imitation Einstein (\ref{ImitationEinsteinPseudo}) pseudotensors.\\

Linearizing the Imitation Einstein pseudotensor (\ref{ImitationEinsteinPseudo}) we obtain,

\begin{multline}
    2 k \theta^{\rho \lambda} =  
    \partial_\mu  h^{\mu \nu} \partial_\nu  h^{\rho \lambda} 
    - \frac{1}{2} \partial_\mu  h^\nu_\nu \partial^\mu  h^{\rho \lambda} 
    + \frac{1}{2} \partial^\rho  h^\mu_\mu \partial^\lambda  h^\nu_\nu 
    - \partial_\mu  h^{\lambda \nu} \partial_\nu  h^{\rho \mu} 
    - \partial_\mu  h^\rho_\nu \partial^\mu  h^{\lambda \nu} 
    + \frac{1}{2} \partial^\rho  h^\mu_\mu \partial_\nu  h^{\lambda \nu} 
    \\
    + \partial^\lambda  h_{\mu \nu} \partial^\mu  h^{\rho \nu} 
    + \partial^\rho  h^{\lambda \mu} \partial_\mu  h^\nu_\nu 
    + \frac{1}{2} \partial^\lambda  h^{\mu \rho} \partial_\mu  h^\nu_\nu 
    - \frac{3}{2} \eta^{\rho \lambda} \partial_\mu  h^{\mu \nu} \partial_\nu  h^\alpha_\alpha 
    - \frac{1}{2} \eta^{\rho \lambda} \partial_\mu  h^\nu_\nu \partial^\mu  h^\alpha_\alpha 
    + \frac{1}{2}  h^\mu_\mu \partial^\rho \partial^\lambda  h^\nu_\nu 
    \\
    +  h_{\mu \nu} \partial^\mu \partial^\nu  h^{\rho \lambda} 
    - \frac{1}{2}  h^\mu_\mu \partial_\nu \partial^\nu  h^{\rho \lambda} 
    -  h^{\lambda \mu} \partial_\mu \partial_\nu  h^{\nu \rho} 
    +  h^{\lambda \mu} \partial^\rho \partial_\mu  h^\nu_\nu 
    -  h^\rho_\mu \partial_\nu \partial^\nu  h^{\lambda \mu} 
    +  h^{\rho \mu} \partial^\lambda \partial^\nu  h_{\mu \nu} 
    \\
    + \frac{1}{2}  h^\mu_\mu \partial_\nu \partial^\rho  h^{\lambda \nu} 
    + \frac{1}{2}  h^\mu_\mu \partial_\nu \partial^\lambda  h^{\rho \nu} 
    - \eta^{\rho \lambda}  h_{\mu \nu} \partial^\mu \partial^\nu  h^\alpha_\alpha  
    - \frac{1}{2} \eta^{\rho \lambda}  h^\alpha_\alpha \partial_\mu \partial_\nu  h^{\mu \nu} 
    - \frac{1}{2} \eta^{\rho \lambda}  h^\mu_\mu \partial_\nu \partial^\nu  h^\alpha_\alpha . \label{LinImEin}
\end{multline}

The non-zero coefficients in (\ref{FockEMT}) are: $b_1 = 1, b_2 = -\frac{1}{2}, b_4 = -1, b_5 = -1, b_6 = \frac{1}{2}, b_{8_i} = 1, b_{8_{ii}} = \frac{1}{2}, b_{9_i} = \frac{1}{2}, b_{11_{ii}} = 1, c_1 = -\frac{1}{2}, c_5 = -\frac{3}{2}, d_3 = -\frac{1}{2}, d_4 = 1, d_{5_i} = -1, d_{6_{ii}} = -1, d_7 = \frac{1}{2}, d_{9_{ii}} = 1, d_{10_i} = 1, d_{11_i} = \frac{1}{2}, d_{11_{ii}} = \frac{1}{2}, a_1 = -1, a_2 = -\frac{1}{2}$ and $a_4 = -\frac{1}{2}$. Two conflicting equations are (\ref{eqn:d_4 Coefficient}) and (\ref{eqn:d_5_i Coefficient}), which are $d_4 = 1 = \frac{1}{4} B_4 D_2 + \frac{1}{4} B_5 D_2 $ and $d_{5_{i}} = -1 = \frac{1}{4} B_4 D_2 + \frac{1}{4} B_5 D_2$. Therefore, the linearized Imitation Einstein pseudotensor (\ref{LinImEin}) is non-Noetherian. 

\subsubsection{M{\o}ller pseudotensor} 

M{\o}ller also produced a pseudotensor which is correctly attributed to his name as the M{\o}ller pseudotensor \cite{moller1961}, which can be defined compactly defined as,

\begin{equation}
    T_\rho^{\ \lambda} = \mathcal{X}^{\ \lambda \sigma}_{\rho \ \ , \sigma} . \label{MollerPseudo}
\end{equation}

\noindent where $\mathcal{X}_\rho^{\ \lambda \sigma} = \frac{\sqrt{-g}}{k} (g_{\rho \nu, \mu} - g_{\rho \mu, \nu}) g^{\lambda \mu} g^{\sigma \nu}$. Linearizing the M{\o}ller pseudotensor (\ref{MollerPseudo}) we obtain,

\begin{multline}
    k T^{\rho \lambda} = 
    \frac{1}{2} \partial^\lambda  h^{\rho \nu} \partial_\nu  h^\alpha_\alpha  
    - \frac{1}{2} \partial_\nu  h^{\lambda \rho} \partial^\nu  h^\alpha_\alpha 
    - \partial_\nu  h^{\lambda \alpha} \partial_\alpha  h^{\rho \nu} 
    - \partial^\alpha  h_{\alpha \nu} \partial^\lambda  h^{\rho \nu}
    + \partial^\nu  h^{\lambda \alpha} \partial_\nu  h^\rho_{\ \alpha}
    + \partial_\alpha  h^{\alpha \nu} \partial_\nu  h^{\lambda \rho}
    \\
    -  h^{\lambda \alpha} \partial_\alpha \partial_\nu  h^{\rho \nu} 
    -  h^{\alpha \nu} \partial_\alpha \partial^\lambda  h^\rho_{\ \nu} 
    +  h^{\lambda \alpha} \partial_\nu \partial^\nu  h^\rho_{\ \alpha} 
    +  h^{\alpha \nu} \partial_\alpha \partial_\nu  h^{\lambda \rho} 
    + \frac{1}{2}  h^\alpha_\alpha \partial_\nu \partial^\lambda  h^{\rho \nu} 
    - \frac{1}{2}  h^\alpha_\alpha \partial_\nu \partial^\nu  h^{\lambda \rho} . \label{LinMoller}
\end{multline}

The non-zero coefficients in (\ref{FockEMT}) are: $b_1 = 1, b_2 = -\frac{1}{2}, b_4 = 1, b_5 = -1, b_{8_{ii}} = \frac{1}{2}, b_{10_{ii}} = -1, d_3 = -\frac{1}{2}, d_4 = 1, d_{5_{ii}} = 1, d_{6_{ii}} = -1, d_{11_{ii}} = \frac{1}{2}$ and $d_{12_{ii}} = -1$. The term $d_{5_{ii}} = 0$ in Appendix \ref{Fock to Noether Coefficient Comparison}, thus the coefficient equation (\ref{eqn:d_5_ii Coefficient}) conflicts as it yields $1 = d_{5_{ii}} = 0$. Therefore the linearized M{\o}ller pseudotensor (\ref{LinMoller}) is non-Noetherian.

\subsection{Linearized gravity energy-momentum tensor comparison} \label{SectionLinGravCompare}

In our final comparison Section \ref{SectionLinGravCompare}, we will now compare our general set of energy-momentum tensors in (\ref{FockEMT}) and Appendix \ref{Fock to Noether Coefficient Comparison} to various linearized gravity energy-momentum tensors. These expressions have been developed within the theory of linearized gravity, therefore require no linearization themselves. 

\subsubsection{Hilbert energy-momentum tensor}

The Hilbert energy-momentum tensor is commonly derived from the Fierz-Pauli Lagrangian density using the Hilbert method for deriving an energy-momentum tensor in Minkowski spacetime \cite{blaschke2016}. This Lagrangian and definition can be found in \cite{Baker2021b}, resulting in,

\begin{multline}
    T^{\rho \lambda} = \frac{1}{4} \eta^{\rho \lambda} [\partial_\alpha h^\beta_\beta \partial^\alpha h^\gamma_\gamma 
    - \partial_\alpha h_{\beta \gamma} \partial^\alpha h^{\beta \gamma} 
    + 2 \partial_\alpha h_{\beta \gamma} \partial^\gamma h^{\beta \alpha}
    + 2 h^{\alpha \beta} \partial_\alpha \partial_\beta h^\gamma_\gamma]
    - \partial^\rho h_{\beta \alpha} \partial^\alpha h^{\beta \lambda}
    - \partial_\alpha h^\rho_\beta \partial^\lambda h^{\beta \alpha}
    \\
    + \partial_\alpha h^\rho_\beta \partial^\alpha h^{\lambda \beta}
    + \partial_\alpha h^{\rho \beta} \partial_\beta h^{\lambda \alpha}
    - \partial_\alpha h^{\rho \lambda} \partial_\beta h^{\alpha \beta}
    - \frac{1}{2} \partial_\alpha h^{\rho \lambda} \partial^\alpha h^\beta_\beta 
    - \frac{1}{2} \partial^\rho h^\beta_\beta \partial^\lambda h^\alpha_\alpha 
    + \frac{1}{2} \partial^\rho h_{\alpha \beta} \partial^\lambda h^{\alpha \beta} 
    \\
    + \frac{1}{2} \partial^\lambda h^\beta_\beta \partial^\alpha h^\rho_\alpha 
    + \frac{1}{2} \partial^\rho h^\beta_\beta \partial^\alpha h^\lambda_\alpha
    + h^{\rho \alpha} \partial_\alpha \partial_\beta h^{\lambda \beta}
    + h^{\lambda \alpha} \partial_\alpha \partial_\beta h^{\rho \beta}
    - h^{\rho \lambda} \partial_\alpha \partial_\beta h^{\alpha \beta} 
    - h^{\alpha \beta} \partial_\alpha \partial_\beta h^{\rho \lambda}
    \\
    + \frac{1}{2} h^{\rho \lambda} \partial_\alpha \partial^\alpha h^\beta_\beta
    - \frac{1}{2} h^{\rho \alpha} \partial^\lambda \partial_\alpha h^\beta_\beta
    - \frac{1}{2} h^{\lambda \alpha} \partial^\rho \partial_\alpha h^\beta_\beta . \label{HilbertEMT}
\end{multline}

The non-zero coefficients in (\ref{FockEMT}) are: $b_1 = -1, b_2 = -\frac{1}{2}, b_4 = 1, b_5 = 1, b_6 = -\frac{1}{2}, b_7 = \frac{1}{2}, b_{9_i} = \frac{1}{2}, b_{9_{ii}} = \frac{1}{2}, b_{11_i} = -1, b_{11_{ii}} = -1, c_1 = \frac{1}{4}, c_2 = -\frac{1}{4}, c_4 = \frac{1}{2}, d_1 = \frac{1}{2}, d_2 = -1, d_4 = -1, d_{6_i} = 1, d_{6_{ii}} = 1, d_{9_i} = -\frac{1}{2}, d_{9_{ii}} = -\frac{1}{2}$ and $a_1 = \frac{1}{2}$. Two conflicting equations are (\ref{eqn:d_4 Coefficient}) and (\ref{eqn:d_5_i Coefficient}), which are $-1 = d_4 = \frac{1}{4} B_4 D_2 + \frac{1}{4} B_5 D_2 $ and $0 = d_{5_{i}} = \frac{1}{4} B_4 D_2 + \frac{1}{4} B_5 D_2$. Therefore the Hilbert energy-momentum tensor (\ref{HilbertEMT}) is non-Noetherian; a result which is expected based on the results of \cite{baker2021}.

\subsubsection{Fierz energy-momentum tensor}

The Fierz energy-momentum tensor was recently developed in \cite{Toth2022} by using the Fierz formalism of linearized gravity to define an energy-momentum tensor in analogous form to that of classical electrodynamics. The basic idea is to define analogous field strength tensors from which the Lagrangian and equation of motion can be determined,

\begin{gather}
        F_{\alpha \beta \gamma} = \frac{1}{2} 
    (\partial_\alpha h_{\beta \gamma} 
    - \partial_\beta h_{\alpha \gamma} 
    + \partial_\kappa h^\kappa_{\ \alpha} \eta_{\beta \gamma} 
    - \partial_\kappa h^\kappa_{\ \beta} \eta_{\alpha \gamma} 
    - \partial_\alpha h^\kappa_\kappa \eta_{\beta \gamma} 
    + \partial_\beta h^\kappa_\kappa \eta_{\alpha \gamma}) ,
    \\
        \mathring{F}_{\alpha \beta \gamma} = \frac{1}{2} 
    (\partial_\alpha h_{\beta \gamma} 
    - \partial_\beta h_{\alpha \gamma}) ,
\end{gather}

\noindent and to use these to derive a corresponding energy-momentum tensor,

\begin{equation}
    T^{\rho \lambda} = 2 
    (F^{\rho \alpha \beta} \mathring{F}^\lambda_{\ \alpha \beta} 
    - \frac{1}{4} \eta^{\rho \lambda} F^{\gamma \alpha \beta} \mathring{F}_{\gamma \alpha \beta}) .  \label{FierzEMT}
\end{equation}

\noindent Using these definitions we can expand (\ref{FierzEMT}) to the form of our (\ref{FockEMT}) for the purpose of comparison,

\begin{multline}
    2 T^{\rho \lambda} =
    \partial^\rho h_{\mu \nu} \partial^\lambda h^{\mu \nu} 
    - \partial^\rho h^\mu_\mu \partial^\lambda h^\nu_\nu 
    + \partial_\mu h^{\rho \lambda} \partial_\nu h^{\mu \nu} 
    - \partial_\mu h^{\rho \lambda} \partial^\mu h^\nu_\nu 
    - \partial_\mu h^{\rho \mu} \partial_\nu h^{\lambda \nu} 
    + \partial_\mu h^{\rho \nu} \partial^\mu h^\lambda_{\ \nu}
    \\
    - \partial^\lambda h^{\mu \rho} \partial^\nu h_{\mu \nu} 
    + \partial^\lambda h^{\mu \rho} \partial_\mu h^\nu_\nu 
    - \partial^\lambda h_{\mu \nu} \partial^\mu h^{\rho \nu} 
    - \partial^\rho h_{\mu \nu} \partial^\mu h^{\lambda \nu} 
    + \partial^\lambda h^\mu_\mu \partial_\nu h^{\rho \nu} 
    + \partial^\rho h^\mu_\mu \partial_\nu h^{\lambda \nu} 
    \\
    + \frac{1}{2} \eta^{\rho \lambda} \partial_\mu h^\nu_\nu \partial^\mu h^\sigma_\sigma 
    - \frac{1}{2} \eta^{\rho \lambda} \partial_\mu h_{\nu \sigma} \partial^\mu h^{\nu \sigma} 
    - \eta^{\rho \lambda} \partial_\mu h^{\mu \nu} \partial_\nu h^\sigma_\sigma 
    + \frac{1}{2} \eta^{\rho \lambda} \partial_\mu h^{\mu \nu} \partial^\sigma h_{\nu \sigma} 
    + \frac{1}{2} \eta^{\rho \lambda} \partial_\mu h_{\nu \sigma} \partial^\nu h^{\mu \sigma} . \label{FierzEMTExpand}
\end{multline}

The non-zero coefficients in (\ref{FockEMT}) are: $b_1 = 1, b_2 = -1, b_3 = -1, b_4 = 1, b_6 = -1, b_7 = 1, b_{8_{ii}} = 1, b_{9_i} = 1, b_{9_{ii}} = 1, b_{10_{ii}} = -1, b_{11_i} = -1, b_{11_{ii}} = -1, c_1 = \frac{1}{2}, c_2 = -\frac{1}{2}, c_3 = \frac{1}{2}, c_4 = \frac{1}{2}$ and $c_5 = -1$. Immediately we have two conflicting equations (\ref{eqn:b_4 Coefficient}) and (\ref{eqn:b_5 Coefficient}), which are $1 = b_4 = B_4 C_5 + B_5 C_5 - \frac{1}{4} B_4 D_2 - \frac{1}{4} B_5 D_2 $ and $ 0 = b_5 = B_4 C_5 + B_5 C_5 - \frac{1}{4} B_4 D_2 - \frac{1}{4} B_5 D_2 $, therefore the Fierz energy-momentum tensor (\ref{FierzEMTExpand}) is non-Noetherian.

\subsubsection{Butcher energy-momentum tensor}

Butcher developed two linearized gravity energy-momentum tensors \cite{Butcher2012PHD} which we will compare, as part of a response to the Padmanabhan-Deser debate \cite{padmanabhan2008,deser2010} regarding the requirement of a unique energy-momentum tensor for calculation purpose in linearized gravity. The first one we will consider is defined as, in terms of the notation used in our paper (Butcher uses notation $\breve{\nabla}_\rho = \partial_\rho$ and $\breve{g}_{\mu \nu} = \eta_{\mu \nu}$),

\begin{equation}
   k \tau_{\rho \lambda} 
    = \frac{1}{4} \partial_\rho h_{\alpha \beta} \partial_\lambda h^{\alpha \beta} 
    - \frac{1}{8} \partial_\rho h^\alpha_\alpha \partial_\lambda h^\beta_\beta
    - \frac{1}{8} \eta_{\rho \lambda} \partial_\gamma h_{\alpha \beta} \partial^\gamma h^{\alpha \beta} 
    + \frac{1}{16} \eta_{\rho \lambda} \partial_\gamma h^\alpha_\alpha \partial^\gamma h^\beta_\beta . \label{Butcher1}
\end{equation}

The non-zero coefficients in (\ref{FockEMT}) are: $b_6 = -\frac{1}{8}, b_7 = \frac{1}{4}, c_1 = \frac{1}{16}$ and $c_2 = -\frac{1}{8}$. When solving the coefficient equations we know that $C_1 = -\frac{1}{8}, C_5 = 0, D_3 = 0$ from (\ref{eqn:c_2 Coefficient}), (\ref{eqn:c_4 Coefficient}) and (\ref{eqn:a_5 Coefficient}). From the equations (\ref{eomcondition2}) and (\ref{eqn:b_7 Coefficient}) we get that $n = \frac{1}{2}, B_1 = -1$, which yield equations $D_3 - C_1 = \frac{n}{4}$ and $\frac{1}{4} = b_7 = 2B_1 C_1 - B_1 D_3 $. From the equation (\ref{eqn:b_11_ii Coefficient}) we get that $D_2 = 0$, thus $0 = b_{11_{ii}} = 2 B_1 C_5 - \frac{1}{2} B_1 D_2 $. Then we have two conflicting equations (\ref{eomcondition4}) and (\ref{eqn:b_10_ii Coefficient}) which are $D_2 - C_4 - C_5 = - \frac{n}{2}$ and $ 0 = b_{10_{ii}} = 2 B_1 C_4 + B_2 C_5 + B_3 C_5 - \frac{1}{2} B_1 D_2 - \frac{1}{4} B_2 D_2 - \frac{1}{4} B_3 D_2 $; these equations give $C_4 = \frac{1}{4}$ and $C_4 = 0$. Therefore, the Butcher energy-momentum tensor (\ref{Butcher1}) is non-Noetherian.

\subsubsection{Butcher modified energy-momentum tensor}

Butcher introduced a modified linearized gravity energy-momentum tensor in \cite{Butcher2012PHD} as, 

\begin{equation}
    t^{\mu\nu} = \tau^{\mu\nu} + \partial_\alpha \Psi^{[\nu\alpha]\mu} \label{Butcher2}
\end{equation}

\noindent where the $\tau^{\mu\nu}$ is from (\ref{Butcher1}), $\Psi^{[\nu\alpha]\mu} = \frac{1}{2} (s^{\mu[\nu\alpha]} + s^{\nu[\mu\alpha]} - s^{\alpha [\mu\nu]})$ and $\kappa s^\alpha_{\ \mu\nu} =  2 \bar{h}_{\beta [ \nu}   \partial^{[ \alpha } \bar{h}_{\mu ]}^{\ \beta]}$. We interpret $2 \bar{h}_{\beta [ \nu}   \partial^{[ \alpha } \bar{h}_{\mu ]}^{\ \beta]}$ to imply antisymmetric pairs in $\mu\nu$ and $\alpha\beta$. We also note that Butcher works in terms of the trace-reversed potential $\bar{h}_{\mu\nu} = h_{\mu\nu} - \frac{1}{2} \eta_{\mu\nu} h^\alpha_\alpha$. Inserting all of these into (\ref{Butcher2}) to write it in the form of (\ref{FockEMT}) we obtain,

\begin{multline}
k t^{\rho \lambda} 
= \frac{1}{4} \partial^\rho h_{\alpha \beta} \partial^\lambda h^{\alpha \beta} 
+ \frac{3}{8} \partial^\rho h^\alpha_\alpha \partial^\lambda h^\beta_\beta 
- 2 \partial_\alpha h^{\alpha \beta} \partial_\beta h^{\rho \lambda} 
+ \partial_\alpha h^\beta_\beta \partial^\alpha h^{\rho \lambda} 
+ \partial^\alpha h_{\alpha \beta} \partial^\rho h^{\beta \lambda} 
+ \partial^\alpha h_{\alpha \beta} \partial^\lambda h^{\beta \rho} 
\\
- \partial^\alpha h^{\beta \lambda} \partial^\rho h_{\alpha \beta} 
- \partial^\alpha h^{\beta \rho} \partial^\lambda h_{\alpha \beta} 
+ 2 \partial_\alpha h^{\beta \lambda} \partial_\beta h^{\alpha \rho} 
- \partial_\alpha h^\beta_\beta \partial^\lambda h^{\alpha \rho} 
- \partial_\alpha h^\beta_\beta \partial^\rho h^{\alpha \lambda} 
- \frac{1}{8} \eta^{\rho \lambda} \partial_\gamma h_{\alpha \beta} \partial^\gamma h^{\alpha \beta} 
\\
- \frac{7}{16} \eta^{\rho \lambda} \partial_\alpha h^\beta_\beta \partial^\alpha h^\gamma_\gamma 
+ \eta^{\rho \lambda} \partial_\alpha h^{\alpha \beta} \partial_\beta h^\gamma_\gamma 
+ \frac{1}{2} h^\alpha_\alpha \partial^\rho \partial^\lambda h^\beta_\beta 
- 2 h_{\alpha \beta} \partial^\alpha \partial^\beta h^{\rho \lambda} 
+ h^\alpha_\alpha \partial_\beta \partial^\beta h^{\rho \lambda} 
+ 2 h^{\alpha \lambda} \partial_\alpha \partial_\beta h^{\beta \rho} 
\\
+ h_{\alpha \beta} \partial^\alpha \partial^\rho h^{\beta \lambda} 
+ h_{\alpha \beta} \partial^\alpha \partial^\lambda h^{\beta \rho} 
- h^{\alpha \lambda} \partial^\beta \partial^\rho h_{\alpha \beta} 
- h^{\alpha \rho} \partial^\beta \partial^\lambda h_{\alpha \beta} 
- h^\lambda_{\ \beta} \partial_\alpha \partial^\alpha h^{\beta \rho} 
+ h^\rho_{\ \beta} \partial_\alpha \partial^\alpha h^{\beta \lambda} 
\\
- h^\alpha_\alpha \partial_\beta \partial^\lambda h^{\beta \rho} 
- h^{\alpha \lambda} \partial_\alpha \partial^\rho h^\beta_\beta 
+ \eta^{\rho \lambda} h_{\alpha \beta} \partial^\alpha \partial^\beta h^\gamma_\gamma 
- \frac{1}{2} \eta^{\rho \lambda} h^\alpha_\alpha \partial_\beta \partial^\beta h^\gamma_\gamma . \label{Butcher3}
\end{multline}

The non-zero coefficients in (\ref{FockEMT}) are: $b_1 = -2, b_2 = 1, b_5 = 2, b_6 = \frac{3}{8}, b_7 = \frac{1}{4}, b_{8_i} = -1, b_{8_{ii}} = -1, b_{10_i} = 1, b_{10_{ii}} = 1, b_{11_i} = -1, b_{11_{ii}} = -1, c_1 = -\frac{7}{16}, c_2 = -\frac{1}{8}, c_5 = 1, d_3 = 1, d_4 = -2, d_{5_i} = 1, d_{5_{ii}} = -1 d_{6_{ii}} = 2, d_7 = \frac{1}{2}, d_{9_{ii}} = -1, d_{10_i} = -1, d_{10_{ii}} = -1, d_{11_{ii}} = -1, d_{12_i} = 1, d_{12_{ii}} = 1, a_1 = 1$ and $a_4 = -\frac{1}{2}$. There is no solution as $d_{5_{ii}} = -1$ which conflicts with the requirement in equation (\ref{eqn:d_5_ii Coefficient}), therefore the Butcher energy-momentum tensor (\ref{Butcher3}) is non-Noetherian. We note that the unusual $-\frac{7}{16}$ coefficient comes from the $\frac{1}{16}$ in (\ref{Butcher1}) combined with the same term in the superpotential with coefficient $- \frac{1}{2}$.

\subsubsection{Padmanabhan energy-momentum tensor}

In \cite{padmanabhan2008}, Padmanabhan introduces a new energy-momentum tensor $S^{\rho\sigma}$ which is a modification of the standard Hilbert method \cite{blaschke2016}. Recall that the Hilbert energy-momentum tensor in Minkowski spacetime is defined as $T^{\gamma\rho}_H = \frac{2}{\sqrt{-g}} \frac{\delta \mathcal{L}}{\delta  g_{\gamma\rho}} \Big|_{g = \eta} $, derived from a Lagrangian density $\mathcal{L}$ by replacing all partial derivatives with covariant derivatives $\partial \to \nabla$, replacing the Minkowski metric with the general metric tensor $\eta \to g$, and inserting the Jacobian contribution $\sqrt{-g}$. Padmanabhan, for purposes of his debate with Deser \cite{padmanabhan2008,deser2010}, defined a modification of this definition, which has an identical prescription, except for not modifying the partial derivatives to covariant derivatives. This Padmanabhan energy-momentum tensor is defined as $S^{\gamma\rho} = \frac{2}{\sqrt{-g}} \frac{\delta \mathcal{L}}{\delta  g_{\gamma\rho}} \Big|_{g = \eta} $, replacing the Minkowski metric with the general metric tensor $\eta \to g$, and inserting the Jacobian contribution $\sqrt{-g}$. The Fierz-Pauli Lagrangian $\mathcal{L}_{FP} = \frac{1}{4} [\partial_\alpha h_\beta^\beta \partial^\alpha h_\gamma^\gamma 
- \partial_\alpha h_{\beta\gamma} \partial^\alpha h^{\beta\gamma} 
+ 2 \partial_\alpha h_{\beta\gamma} \partial^\gamma h^{\beta\alpha}
- 2 \partial^\alpha h_{\beta}^\beta \partial^\gamma h_{\gamma\alpha}] $ (the starting point for his calculation) can be expressed under this prescription as,

\begin{equation}
\mathcal{L} = \frac{1}{4} \sqrt{-g} g^{\mu\alpha} g^{\nu\beta} g^{\lambda\gamma} [\partial_\alpha h_{\nu\beta} \partial_\mu h_{\lambda\gamma} 
- \partial_\alpha h_{\beta\gamma} \partial_\mu h_{\nu\lambda} 
+ 2 \partial_\alpha h_{\beta\gamma} \partial_\lambda h_{\nu\mu}
- 2 \partial_\mu h_{\nu\beta} \partial_\lambda h_{\gamma\alpha}] . \label{PaddyStart}
\end{equation}

\noindent This is identical to the starting point of the Hilbert energy-momentum tensor (\ref{HilbertEMT}) derivation, except instead of covariant derivatives $\nabla_\alpha$ in (\ref{PaddyStart}), we have partial derivatives $\partial_\alpha$. From here, following the standard procedure for the Hilbert energy-momentum tensor for the Padmanabhan modified Lagrangian, one can calculate $S^{\gamma\rho}$ as,

\begin{multline}
S^{\rho\sigma} = \frac{1}{4}\eta^{\rho\sigma} \partial_\alpha h^\beta_{\beta} \partial^\alpha h^\gamma_{\gamma} 
- \frac{1}{4}\eta^{\rho\sigma} \partial_\alpha h_{\beta\gamma} \partial^\alpha h^{\beta\gamma} 
+ \frac{1}{2}\eta^{\rho\sigma}  \partial_\alpha h_{\beta\gamma} \partial^\gamma h^{\alpha \beta} 
- \frac{1}{2}\eta^{\rho\sigma}  \partial^\alpha h^\beta_{\beta} \partial^\gamma h_{\gamma\alpha}
\\
- \frac{1}{2} \partial^\sigma h^\beta_{\beta} \partial^\rho h^\gamma_{\gamma} 
+ \frac{1}{2} \partial^\sigma h_{\beta\gamma} \partial^\rho h^{\beta\gamma} 
-  \partial_\alpha h^{\rho\sigma} \partial^\alpha h^\gamma_{\gamma}
+  \partial^\alpha h^{\rho\sigma} \partial^\gamma h_{\gamma\alpha}
+  \partial_\alpha h^\sigma_{\gamma} \partial^\alpha h^{\rho\gamma}
-  \partial^\alpha h^\sigma_{\gamma} \partial^\gamma h^\rho_{\alpha}
\\
-  \partial^\sigma h^{\nu\lambda} \partial_\lambda h^\rho_{\nu} 
+ \frac{1}{2} \partial^\sigma h^\beta_{\beta} \partial^\gamma h^\rho_{\gamma}
+ \frac{1}{2} \partial^\alpha h^\beta_{\beta} \partial^\sigma h^\rho_{\alpha}
-  \partial^\rho h^{\nu\lambda} \partial_\lambda h^\sigma_{\nu} 
+ \frac{1}{2} \partial^\rho h^\beta_{\beta} \partial^\gamma h^\sigma_{\gamma}
+ \frac{1}{2} \partial^\alpha h^\beta_{\beta} \partial^\rho h^\sigma_{\alpha} . \label{PaddyEMT}
\end{multline}

\noindent We note the required $\frac{\delta \sqrt{-g}}{\delta g_{\rho\sigma}} = \frac{1}{2} g^{\rho\sigma} \sqrt{-g}$ and $\frac{\delta g^{\mu\alpha}}{\delta g_{\rho\sigma}} = - \frac{1}{2} (g^{\mu\rho} g^{\alpha\sigma} + g^{\mu\sigma} g^{\alpha\rho})$ which are essential for this calculation.\\

The non-zero coefficients in (\ref{FockEMT}) are: $b_1 = 1, b_2 = -1, b_4 = 1, b_5 = -1, b_6 = -\frac{1}{2}, b_7 = \frac{1}{2}, b_{8_i} = \frac{1}{2}, b_{8_{ii}} = \frac{1}{2}, b_{9_i} = \frac{1}{2}, b_{9_{ii}} = \frac{1}{2}, b_{11_i} = -1, b_{11_{ii}} = -1, c_1 = \frac{1}{4}, c_2 = -\frac{1}{4}, c_4 = \frac{1}{2}$ and $c_5 = -\frac{1}{2}$. We immediately have two conflicting equations (\ref{eqn:b_4 Coefficient}) and (\ref{eqn:b_5 Coefficient}), explicitly $1 = b_4 = B_4 C_5 + B_5 C_5 - \frac{1}{4} B_4 D_2 - \frac{1}{4} B_5 D_2 $ and $-1 = b_5 = B_4 C_5 + B_5 C_5 - \frac{1}{4} B_4 D_2 - \frac{1}{4} B_5 D_2 $, therefore the Padmanabhan energy-momentum tensor (\ref{PaddyEMT}) is non-Noetherian.

\section{Restricting the general set of energy-momentum tensors derived from the Noether current on physical grounds}   \label{ResultsRestriction}

Due to the fact that a gauge invariant expression under the spin-2 gauge transformation cannot be obtained \cite{magnano2002}, it is not clear what uniqueness criteria is possible to severely restrict the general set of energy-momentum tensors derived from the Noether current. There are three reasonable criteria common amongst numerous physical energy-momentum tensors: symmetry, tracelessness, and particular symmetry transformations $\delta h_{\mu\nu}$. We will consider each of these cases in this section.\\

Note that the conditions in this section are restricted by the complete set of conditions in Appendix \ref{Fock to Noether Coefficient Comparison}, however results in Section \ref{simulchrissymtrace} are valid independent of the conditions in Appendix \ref{Restriction1} and \ref{Restriction2}. In Section \ref{symmtracecombo} the symmetry and tracelessness conditions are simultaneously applied in order to determine a symmetric and traceless energy-momentum tensor. Finally, in Section \ref{SectionConservedIdentity} we discuss general symmetry transformation restrictions required for the general set of Noetherian energy-momentum tensors in linearized gravity.

\subsection{Symmetric energy-momentum tensor solutions} \label{SectionSymmetric}

There are ten symmetric terms in (\ref{FockEMT}). For $p = 8,9,10,11$ and $m = 5,6,9,10,11,12$ we have the symmetric conditions $b_p = b_{p_{i}} = b_{p_{ii}}  $ and $d_m = d_{m_{i}} = d_{m_{ii}} $. Their coefficient equations can be found in Appendix \ref{Symmetry conditions}. There are infinitely many solutions to this system of equations, therefore we have shown that there are infinitely many symmetric energy-momentum tensors in linearized gravity which can be derived directly from the Noether current.

\subsection{Traceless energy-momentum tensor solutions} \label{SectionTraceless}

We can determine conditions such that the energy-momentum tensor (\ref{FockEMT}) is traceless ($\eta_{\mu\nu} T^{\mu\nu} = T^\rho_{\rho} = 0$). Performing this contraction on (\ref{Fock to Noether Coefficient Comparison}) we have,

\begin{multline}
    T^\rho_{\rho} = 
    (b_1 + b_{8_i} + b_{8_{ii}} + b_{9_i} + b_{9_{ii}} + 4 c_5 ) \partial_\alpha h \partial_\beta h^{\alpha \beta} 
    + (b_2 + b_6 + 4 c_1 ) \partial_\alpha h \partial^\alpha h
    \\
    + (b_3 + b_{10_i} + b_{10_{ii}} + 4 c_3 ) \partial_\rho h^{\rho \alpha} \partial^\beta h_{\alpha \beta} 
    + (b_4 + b_7 + 4 c_2 ) \partial_\rho h_{\alpha \beta} \partial^\rho h^{\alpha \beta} 
    + (b_5 + b_{11_i} + b_{11_{ii}} + 4 c_4 ) \partial^\alpha h^{\rho \beta} \partial_\beta h_{\rho \alpha} 
    \\
    + (d_1 + d_3 + d_7 + 4 a_4 ) h \partial_\alpha \partial^\alpha h 
    + (d_2 + d_{11_{i}} + d_{11_{ii}} + 4 a_2 ) h \partial_\alpha \partial_\beta h^{\alpha \beta}
    + (d_4 + d_{9_i} + d_{9_{ii}} + 4 a_1 ) h_{\alpha \beta} \partial^\alpha \partial^\beta h 
    \\
    + (d_{5_i} + d_{5_{ii}} + d_8 + 4 a_5 ) h_{\alpha \beta} \partial_\rho \partial^\rho h^{\alpha \beta} 
    + (d_{6_i} + d_{6_{ii}} + d_{10_{i}} + d_{10_{ii}} + d_{12_{i}} + d_{12_{ii}} + 4 a_3 ) h_{\alpha \beta} \partial^\alpha \partial_\rho h^{\rho \beta} = 0 .
\end{multline}

\noindent Therefore the coefficients above must each be identically zero, yielding a system of equations that must be satisfied for a traceless energy-momentum tensor which we give in Appendix \ref{Traceless conditions}; these also need to be solved simultaneously with the conditions in Appendix \ref{Fock to Noether Coefficient Comparison}. Since there are infinitely many solutions here, we have shown that there are infinitely many traceless energy-momentum tensors in linearized gravity which can be derived directly from the Noether current. We note here that tracelessness is often considered an essential property of a physical energy-momentum tensor is related to scale invariance and dilatation transformations, which is a 1-parameter transformation that is a subset of the 15-parameter conformal group in 4D. In order for the corresponding object derived from Noether's first theorem (the dilatation tensor) to be conserved, the energy-momentum tensor associated to the 4-parameter Poincar{\'e} translation must be traceless.

\subsection{Linearized Christoffel symmetry transformations} \label{LinChrisTransSec}

Symmetry transformations proportional to the linearized Christoffel symbol, similar to numerous standard field theories where the symmetry transformations are proportional to the field strength tensor \cite{Baker2021PHD}, are found in the case of the symmetry transformations linearized Gauss-Bonnet gravity model \cite{baker2019} using the Bessel-Hagen method as $\delta h_{\rho\sigma} = - 2 \bar{\Gamma}^\nu_{\rho\sigma} \delta x_\nu$; therefore it is of interest which energy-momentum tensors this transformation yields in the case of linearized gravity. The linearized Christoffel symbol is given by $\bar{\Gamma}^\nu_{\rho\sigma} = \frac{1}{2} (\partial^\nu h_{\rho\sigma} - \partial_\rho h^\nu_\sigma - \partial_\sigma h^\nu_\rho)$. From the general field transformations (\ref{GenSymTransform}) this yields the particular non-zero coefficients $B_m$ as $B_1 = -1$ $B_4 = 1$ and $B_5 = 1$. This yields an infinite set of possible energy-momentum tensors whose modified coefficient conditions from Appendix \ref{Restriction3} and given in Appendix \ref{Christoffel symmetry solution}.

\subsection{Canonical transformations}  \label{SectionCanonical}

For completeness we will also give the general set of canonical Noether energy-momentum tensors possible in linearized gravity, due to the common use of this symmetry transformation ($\delta h_{\rho\sigma} = - \partial^\alpha h_{\rho\sigma} \delta x_\alpha$) in energy-momentum tensor derivations. From the general field transformations (\ref{GenSymTransform}) this yields $B_1 = -1$ and all other $B_m = 0$. This produces an infinite number of possible energy-momentum tensors (all of which are canonical Noether energy-momentum tensors) whose modified coefficient conditions from Appendix \ref{Restriction3} are given in Appendix \ref{Canonical symmetry solution}. The linearized Einstein pseudotensor (\ref{EinsteinRightRaise}) has particular coefficients $C_1 = -\frac{1}{4}, C_2 = \frac{1}{4}, C_3 = -\frac{1}{2}, C_4 = 0, C_5 = \frac{1}{2}$ and $D_m = 0$, which is indeed a solution to this system of equations.

\subsection{Simultaneously symmetric, traceless and with the Christoffel or canonical symmetry} \label{simulchrissymtrace}

We will now raise questions about the simultaneity of some of these standard properties of physical energy-momentum tensors, starting with the possibility of having a linearized gravity energy-momentum tensor which is simultaneously symmetric, traceless and uses the Christoffel transformations in Section \ref{LinChrisTransSec}. Therefore we need to simultaneously satisfy the system of equations for each in Appendix \ref{Symmetry conditions}, \ref{Traceless conditions} and \ref{Christoffel symmetry solution}. After solving, there are no non-trivial solutions ($T^{\mu\nu} \neq 0$), therefore it is not possible to have a Noetherian energy-momentum tensor in linearized gravity with these properties. We note that this is also not possible if we were to consider the canonical symmetry conditions instead in Appendix \ref{Canonical symmetry solution} due to the lack of symmetry of this expression; a result that can also be verified using Appendix \ref{Symmetry conditions} and \ref{Traceless conditions}.

\subsection{Simultaneously symmetric and traceless} \label{symmtracecombo}

We will now consider the possibility of deriving an energy-momentum tensor from the Noether current which is simultaneously symmetric and traceless; therefore we need to apply conditions from Appendix \ref{Symmetry conditions} and \ref{Traceless conditions} (discussed in Section \ref{SectionSymmetric} and Section \ref{SectionTraceless}) as well as the system of equations from Appendix \ref{Fock to Noether Coefficient Comparison}. From this we obtain the system of equations which we present in Appendix \ref{Symmetric Traceless Conditions}. There are infinitely many solutions, therefore there are infinitely many possible linearized gravity energy-momentum tensors (\ref{FockEMT}) derivable from the Noether current which are simultaneously symmetric and traceless. \\

We will give one particular solution as an illustrative example; we will select one using only the Lagrangian terms of the form $\partial h \partial h$. If we start from the Lagrangian $    \mathcal{L} = 
    - \frac{1}{4} \partial_\beta h_{\mu \nu} \partial^\beta h^{\mu \nu} 
    + \frac{1}{4} \partial_\beta h^\mu_\mu \partial^\beta h^\nu_\nu 
    - \frac{1}{2} \partial_\beta h^{\beta \mu} \partial_\mu h^\nu_\nu 
    + \frac{1}{4} \partial_\beta h^{\beta \mu} \partial^\nu h_{\mu \nu} 
    + \frac{1}{4} \partial_\mu h_{\nu \beta} \partial^\nu h^{\mu \beta} $ for $n=1$ in Appendix \ref{Fock to Noether Coefficient Comparison} we have coefficients     $C_1 = - \frac{1}{4},
     C_2 = \frac{1}{4},
     C_3 = - \frac{1}{2},
     C_4 = \frac{1}{4},
     C_5 = \frac{1}{4}$ and $D_m = 0$. Selecting symmetry coefficients which satisfy Appendix \ref{Fock to Noether Coefficient Comparison} and \ref{Symmetric Traceless Conditions} we can select $B_1 = -\frac{4}{3}, 
     B_2 = B_3 = \frac{2}{3},
     B_4 = B_5 = \frac{2}{3},
     B_6 = B_7 = -\frac{2}{3},
     B_8 = \frac{2}{3},
     B_9 = -\frac{2}{3}$. This yields the nonzero coefficients in (\ref{FockEMT}) as $b_1 = \frac{2}{3}, b_2 = -\frac{2}{3}, b_3 = - \frac{5}{3}, b_4 = \frac{1}{3}, b_5 = \frac{1}{3}, b_6 = - \frac{5}{3}, b_7 = \frac{2}{3}, b_{8_i} = b_{8_{ii}} = \frac{1}{3}, b_{9_i} = b_{9_{ii}} = \frac{5}{3}, b_{10_i} = b_{10_{ii}} = - \frac{1}{3}, b_{11_i} = b_{11_{ii}} = - \frac{2}{3}, c_1 = \frac{7}{12}, c_2 = -\frac{1}{4}, c_3 = \frac{7}{12}, c_4 = \frac{1}{4}$ and $c_5 = -\frac{7}{6}$, therefore we are left with the energy-momentum tensor,

\begin{multline}
    t^{\rho \lambda} = 
    \frac{2}{3} \partial_\alpha h^{\rho \lambda} \partial_\beta h^{\alpha \beta} 
    - \frac{2}{3} \partial_\alpha h^{\rho \lambda} \partial^\alpha h
    - \frac{5}{3} \partial_\alpha h^{\rho \alpha} \partial_\beta h^{\lambda \beta} 
    + \frac{1}{3} \partial_\alpha h^\rho_{\ \beta} \partial^\alpha h^{\lambda \beta} 
    + \frac{1}{3} \partial_\alpha h^{\rho \beta} \partial_\beta h^{\lambda \alpha} 
    - \frac{5}{3} \partial^\rho h \partial^\lambda h 
    \\
    + \frac{2}{3} \partial^\rho h_{\alpha \beta} \partial^\lambda h^{\alpha \beta}
    + \frac{1}{3} \partial^\rho h^{\lambda \alpha} \partial_\alpha h
    + \frac{1}{3} \partial^\lambda h^{\rho \alpha} \partial_\alpha h
    + \frac{5}{3} \partial^\rho h \partial_\alpha h^{\lambda \alpha} 
    + \frac{5}{3} \partial^\lambda h \partial_\alpha h^{\rho \alpha} 
    - \frac{1}{3} \partial^\rho h^{\lambda \alpha} \partial^\beta h_{\alpha \beta} 
    \\
    - \frac{1}{3} \partial^\lambda h^{\rho \alpha} \partial^\beta h_{\alpha \beta}
    - \frac{2}{3} \partial^\rho h_{\alpha \beta} \partial^\alpha h^{\lambda \beta} 
    - \frac{2}{3} \partial^\lambda h_{\alpha \beta} \partial^\alpha h^{\rho \beta}
    + \frac{7}{12} \eta^{\rho \lambda} \partial_\alpha h \partial^\alpha h
    - \frac{1}{4} \eta^{\rho \lambda} \partial_\alpha h_{\beta \sigma} \partial^\alpha h^{\beta \sigma} 
    + \frac{7}{12} \eta^{\rho \lambda} \partial_\alpha h^{\alpha \beta} \partial_\sigma h^\sigma_{\ \beta}
    \\
    + \frac{1}{4} \eta^{\rho \lambda} \partial_\alpha h_{\sigma \beta} \partial^\sigma h^{\alpha \beta}
    - \frac{7}{6} \eta^{\rho \lambda} \partial_\alpha h^{\alpha \beta} \partial_\beta h . \label{RoseXavier}
\end{multline}

\noindent We will refer to this linearized gravity energy-momentum tensor as the Rose-Xavier energy-momentum tensor, which is symmetric, traceless and derivable directly from the Noether current.

\subsection{The general set of possible symmetry transformations} \label{SectionConservedIdentity}

We will discuss one final restriction in our article: the set of restrictions on the general field transformations in (\ref{GenSymTransform}) which limit possible transformations to those symmetry transformations which have two-way correspondence with conserved Noetherian energy-momentum tensors (due to Noether's first theorem, and its converse \cite{noether1918,kosmann2011}). The basic idea for the required calculation is that the Noether identity (\ref{CompactNoetherIdentityT}) is satisfied for a set of symmetry transformations $\delta h_{\mu\nu}$, which correspond to a set of conserved energy-momentum tensors. Therefore values of $B_m$ in $E^{\omega\sigma} \delta h_{\omega \sigma} + a_\lambda \partial_\rho T^{\rho\lambda} = 0 $ must be restricted based on the results in (\ref{FockEMT}) and Appendix \ref{Fock to Noether Coefficient Comparison} coefficients simultaneously; this will yield the most general set of possible symmetry transformations. Doing so will provide an additional set of restrictions which can be used in conjunction with the results of Appendix \ref{Fock to Noether Coefficient Comparison} to yield the general set of Noetherian energy-momentum tensors in linearized gravity from (\ref{FockEMT}). However, this is a highly lengthy and nontrivial calculation, because each of the many terms in (\ref{FockEMT}) can be present in multiple terms proportional to the equations of motion in the Noether identity. Therefore to do this calculation properly, each of the 86 possible terms arising from the divergence of (\ref{FockEMT}) need to be collected and then separated throughout each possible equation of motion contribution with an additional free parameter for each common term. This will yield many additional restrictions to be added to Appendix \ref{Fock to Noether Coefficient Comparison}, that do not impact any of the results to this point in the article. What we have developed to this point is the mathematical framework for the general set of Noetherian energy-momentum tensors in linearized gravity, with only the restrictions described in this section remaining for the most general possible set of Noetherian results.\\

The relationship between symmetry transformations and conservation laws for an arbitrary system of Lagrangians and field transformations, to our knowledge, has not been worked out in the literature for any model, let alone one with the complexities of linearized gravity. The intricate nature of these relationships is of interest not just to linearized gravity, but the application of Noether's first theory to field theories as a whole. In addition there remains the question of non-exact gauge symmetries in the action despite the lack of gauge invariant energy-momentum tensor, using methods also introduced by Bessel-Hagen \cite{besselhagen1921,besseltranslation2006}, not to be confused with his distinct transformation methodology discussed early in the article. These topics are on their own a lengthy study that would dramatically increase the length of the current article, as we discussed in Section \ref{SectionImportantNote}. Therefore we have decided to leave the calculations described in this section to the subject of future work.

\section{Conclusion} \label{SectionSummaryDiscussion}

Noether's first theorem is often praised as one of the most significant theorems in physics; the ability for one to simultaneously obtain the equations of motion, conservation laws (and force densities) directly from a Lagrangian in one compact identity is an idealistic presentation of the most foundational equations in a given theory. In field theory, the Bessel-Hagen method \cite{besselhagen1921,besseltranslation2006} allows one to derive standard physical conserved tensors, such as the energy-momentum tensor, directly from the Noether identity \cite{baker2021noether,Baker2021PHD}. Our current accepted theory of gravitation, general relativity, to some degree conflicts with Noether's first theorem, primarily due to the lack of a finite global group of symmetry transformations such as the Poincar{\'e} group in Minkowski spacetime; the 4-parameter Poincar{\'e} translation of this group is the coordinate transformation from which an energy-momentum tensor derived from Noether's first theorem is defined. Therefore to assess the gravitational energy-momentum tensors derived from Noether's first theorem, one must turn to linearized gravity (the conventional name for linearized general relativity, despite the existence of other linearized gravity theories), which is a theory in Minkowski spacetime and therefore has a well defined energy-momentum tensor which can be derived from the Noether current. The problem for linearized gravity is the lack of gauge invariance of the energy-momentum tensor \cite{magnano2002}, which is one of the primary uniqueness criteria when deriving the unique physical energy-momentum tensor is theories such as electrodynamics and Yang-Mills theory \cite{Baker2021PHD}. This lack of uniqueness criteria has contributed to the numerous energy-momentum expressions that exist for gravitation in the literature, in both linearized and non-linearized gravity. Without concrete uniqueness criteria (i.e. invariance under the spin-2 gauge transformation, linearized diffeomorphisms) to guide calculations, we are left with two freedoms in the Noether identity to derive an energy-momentum tensor for linearized gravity from the Noether current: i) all possible Lagrangian densities (\ref{genlag}) which yield the linearized Einstein field equations, and ii) all possible field transformations which can be inserted into the Noether identity (\ref{GenSymTransform}). Using these most general possible freedoms we developed the general set of energy-momentum tensors in linearized gravity (\ref{FockEMT}) that can be derived from the Noether current, which are restricted by the Lagrangian and symmetry coefficients given in Appendix \ref{Fock to Noether Coefficient Comparison}; all solutions to this system of equations are possible energy-momentum tensors for linearized gravity which can be derived directly from the Noether current.\\

A natural question arises from the general set of energy-momentum tensors in linearized gravity which can be derived from the Noether current: which of the many published gravitational energy-momentum expressions in the literature, if any, can be classified as Noetherian. In Section \ref{LiteratureComparison} we compared our results to numerous common linearized pseudotensor expressions of Einstein, Landau-Lifshitz, Goldberg, Weinberg, Papapetrou, Bergmann-Thomson and M{\o}ller. From this comparison we have shown that only Einstein's linearized pseudotensor can be classified as Noetherian; a result which is expected as it is connected to the canonical Noether tensor which is derived from only a restricted set of (canonical) action symmetries \cite{baker2021noether,Baker2021PHD}, therefore can be classfied as Noetherian without requiring results from Section \ref{SectionConservedIdentity}. We then compared our results to several energy-momentum tensors of linearized gravity including Hilbert, Fierz, Butcher and Padmanabhan, and showed that none of these expressions can be classified as Noetherian. In cases where resulting energy-momentum tensors were non-Noetherian, for ease of reading we presented a simple example of conflicting coefficient restrictions in Appendix \ref{Fock to Noether Coefficient Comparison}, despite numerous possible conflicting restriction equations in each case. In our view the lack of uniqueness of an energy-momentum tensor in linearized gravity may ultimately be attributed to the lack of consideration of Noether's first theorem when producing the various results that exist in the literature. We note that a complete list of our comparisons, and which nonzero coefficients of (\ref{FockEMT}) are in each, is included in Appendix \ref{Table of Fock Coefficients vs. Pseudotensors}. In Section \ref{ResultsRestriction} we have imposed various restrictions on our general set of energy-momentum tensors derived from the Noether current based on common physical criteria of an energy-momentum tensor such as symmetry, tracelessness and particular transformations in the Noether identity. What we have shown is that it is possible to derive from the Noether current linearized gravity energy-momentum tensors which are simultaneously symmetric and traceless; we gave a particular example in (\ref{RoseXavier}). Therefore it is possible to yield symmetric and traceless energy-momentum tensors from the Noether current without resorting to any ad-hoc \say{improvements} in the case of linearized gravity.\\

Our study was largely motivated by \cite{Baker2021b} where we demonstrated several issues with the ad-hoc \say{improvement} of energy-momentum tensors in the Noether identity; the \say{improvement} freedom leads to an absurd multiplicity of results, with infinitely many possible improved expressions, even for particular superpotentials in the off-shell case. The origins of the \say{improvement} method are often justified as a requirement of the canonical Noether energy-momentum tensor not coinciding with known physical results, thereby requiring terms to be ad-hoc added to the Noether identity to obtain the desired result. For example, since the canonical Noether energy-momentum tensor is not symmetric, it is claimed that an external symmetrization \say{improvement} procedure is required to make it symmetric. This can give the impression that Noether's first theorem is in some sense not working to derive physical energy-momentum tensors; this impression however is not true, as the Bessel-Hagen method \cite{besselhagen1921,besseltranslation2006} uses the complete set of field transformations to derive the physical energy-momentum tensor directly from the Noether identity in foundational theories such as electrodynamics and Yang-Mills theory \cite{baker2021noether,Baker2021PHD}. In this article, we have demonstrated that even when gauge invariance of the energy-momentum tensor cannot be used as a guiding principle, the Bessel-Hagen approach can be used to directly obtain desired physical properties such as symmetry and tracelessness of the energy-momentum tensor; see (\ref{RoseXavier}) for a particular example of this for linearized gravity. There are additional features of our results which are superior to the \say{improvement} method, in particular, that we have shown it is possible to impact the terms proportional to the Minkowski metric in (\ref{FockEMT}) directly by the Lagrangian and field transformation coefficients in Appendix \ref{Fock to Noether Coefficient Comparison}. In the \say{improvement} method, no superpotential terms can impact the Minkowski proportional piece; this piece is always exactly proportional to the Lagrangian, which is one of the reasons why the Noether and Hilbert energy-momentum tensors cannot be reconciled by \say{improvement} terms \cite{baker2021}. From the general set of energy-momentum tensors (\ref{FockEMT}), this Minkowski proportional contribution can be directly impacted, leading to the possibility that an energy-momentum tensor can be derived with this contribution being proportional to terms which are not the Lagrangian density; we note that the exception are terms $c_2$, $c_4$ and $a_5$ in (\ref{FockEMT}) which cannot be impacted.\\

There is an intricate relationship between i) the arbitrary possible energy-momentum tensors (\ref{FockEMT}) without restrictions (all terms that one could write down in linearized gravity), ii) ones which are conserved on-shell up to the linearized Einstein field equations, and iii) ones which can be derived explicitly from the Noether current based on restrictions in Appendix \ref{Fock to Noether Coefficient Comparison}. What we have shown is that determination of an arbitrary expression that satisfies criteria i) and ii) by no means implies the satisfaction of criteria iii). To have a result which is Noetherian is special and may be an important starting point in any discussion of uniqueness. One of the goals of this project was to further investigate the topic of uniqueness of energy-momentum tensors in linearized gravity \cite{babak1999,magnano2002,Butcher2012PHD,bicak2016,Baker2021b}. We have shown that many of the standard gravitational energy-momentum expressions that exist in the literature are non-Noetherian even when considering the most charitable domain of possible Lagrangians and field transformations, thus from the perspective of Noether's first theorem it does not appear possible to determine strict uniqueness criteria when it comes to conventional published expressions. What we have shown, however, is that there is a set of solutions which simultaneously contain crucial properties of an energy-momentum in standard physical theories: symmetry and tracelessness. This is notable because this result is obtained directly from the Noether current, without any ad-hoc \say{improvement} terms required. Since a gauge invariant energy-momentum tensor for linearized gravity is not possible (another common feature of the unique energy-momentum tensor in standard physical gauge theories such as electrodynamics and Yang-Mills theory \cite{Baker2021PHD}), further restriction towards a unique result is problematic, and perhaps not possible. This may ultimately be an unavoidable flaw of linearized general relativity as a whole: despite the presence of global Poincar{\'e} coordinate symmetries and a clearly defined procedure for deriving an energy-momentum tensor from Noether's first theorem, the lack of gauge invariance hinders a unique expression from being determined from a larger set of possible energy-momentum tensors. Many authors in the past have criticized the possibility of a true energy expression in general relativity, which is in part due to the fact that the global 4-parameter Poincar{\'e} translation cannot be applied to Noether's first theorem for a general metric spacetime, thus rendering the Noetherian view of an energy-momentum tensor ill-defined. What we have shown is that even in the case of linearized gravity, under the most charitable circumstances, the ill-defined nature of the gravitational energy-momentum tensor persists; it is possible that these two problems are intertwined, and regardless of any prescribed simplification of general relativity, that a standard unique notation of energy-momentum tensor for linearized gravity as in the case of standard physical gauge theories is simply not possible.

\section*{Acknowledgement}

We would like to thank the Department of Physics at St. Francis Xavier University and the Department of Physics and Optical Engineering at Rose-Hulman Institute of Technology for their support which was essential for the completion of this project.

\bibliographystyle{unsrtnat}
\bibliography{arXivBib}

\begin{thebibliography}{59}
\providecommand{\natexlab}[1]{#1}
\providecommand{\url}[1]{\texttt{#1}}
\expandafter\ifx\csname urlstyle\endcsname\relax
  \providecommand{\doi}[1]{doi: #1}\else
  \providecommand{\doi}{doi: \begingroup \urlstyle{rm}\Url}\fi

\bibitem[Baker(2021{\natexlab{a}})]{Baker2021b}
M.R. Baker.
\newblock Canonical {N}oether and the energy--momentum non-uniqueness problem
  in linearized gravity.
\newblock \emph{Classical and Quantum Gravity}, 38\penalty0 (9):\penalty0
  095007, 2021{\natexlab{a}}.

\bibitem[Belinfante(1940)]{belinfante1940}
F.J. Belinfante.
\newblock On the current and the density of the electric charge, the energy,
  the linear momentum and the angular momentum of arbitrary fields.
\newblock \emph{Physica}, 7\penalty0 (5):\penalty0 449 -- 474, 1940.

\bibitem[Padmanabhan(2008)]{padmanabhan2008}
T.~Padmanabhan.
\newblock From gravitons to gravity: Myths and reality.
\newblock \emph{International Journal of Modern Physics D}, 17\penalty0
  (03n04):\penalty0 367--398, 2008.

\bibitem[Deser(2010)]{deser2010}
S.~Deser.
\newblock Gravity from self-interaction redux.
\newblock \emph{General Relativity and Gravitation}, 42\penalty0 (3):\penalty0
  641--646, 2010.

\bibitem[Butcher et~al.(2009)Butcher, Hobson, and Lasenby]{butcher2009}
L.M. Butcher, M.~Hobson, and A.~Lasenby.
\newblock Bootstrapping gravity: A consistent approach to energy-momentum
  self-coupling.
\newblock \emph{Physical Review D}, 80\penalty0 (8):\penalty0 084014, 2009.

\bibitem[Barcel{\'o} et~al.(2014)Barcel{\'o}, Carballo-Rubio, and
  Garay]{barcelo2014}
C.~Barcel{\'o}, R.~Carballo-Rubio, and L.J. Garay.
\newblock Unimodular gravity and general relativity from graviton
  self-interactions.
\newblock \emph{Physical Review D}, 89\penalty0 (12):\penalty0 124019, 2014.

\bibitem[Baker(2021{\natexlab{b}})]{Baker2021PHD}
M.R. Baker.
\newblock \emph{Field theories from physical requirements: Noether's first
  theorem, energy-momentum tensors and the question of uniqueness}.
\newblock PhD thesis, The University of Western Ontario (Canada),
  2021{\natexlab{b}}.

\bibitem[Szabados(1992)]{szabados1992}
L.B. Szabados.
\newblock On canonical pseudotensors, {S}parling's form and {N}oether currents.
\newblock \emph{Classical and Quantum Gravity}, 9\penalty0 (11):\penalty0 2521,
  1992.

\bibitem[Babak and Grishchuk(1999)]{babak1999}
S.V. Babak and L.P. Grishchuk.
\newblock Energy-momentum tensor for the gravitational field.
\newblock \emph{Physical Review D}, 61\penalty0 (2):\penalty0 024038, 1999.

\bibitem[Magnano and Sokolowski(2002)]{magnano2002}
G.~Magnano and L.M. Sokolowski.
\newblock Symmetry properties under arbitrary field redefinitions of the metric
  energy--momentum tensor in classical field theories and gravity.
\newblock \emph{Classical and Quantum Gravity}, 19\penalty0 (2):\penalty0 223,
  2002.

\bibitem[Bi{\v{c}}{\'a}k and Schmidt(2016)]{bicak2016}
J.~Bi{\v{c}}{\'a}k and J.~Schmidt.
\newblock Energy-momentum tensors in linearized {E}instein’s theory and
  massive gravity: The question of uniqueness.
\newblock \emph{Physical Review D}, 93\penalty0 (2):\penalty0 024009, 2016.

\bibitem[T{\'o}th(2022)]{Toth2022}
G.Z. T{\'o}th.
\newblock Energy--momentum tensor and duality symmetry of linearized gravity in
  the {F}ierz formalism.
\newblock \emph{Classical and Quantum Gravity}, 39\penalty0 (7):\penalty0
  075003, 2022.

\bibitem[Forger and R{\"o}mer(2004)]{forger2004}
M.~Forger and H.~R{\"o}mer.
\newblock Currents and the energy-momentum tensor in classical field theory: a
  fresh look at an old problem.
\newblock \emph{Annals of Physics}, 309\penalty0 (2):\penalty0 306--389, 2004.

\bibitem[Blaschke et~al.(2016)Blaschke, Gieres, Reboud, and
  Schweda]{blaschke2016}
D.N. Blaschke, F.~Gieres, M.~Reboud, and M.~Schweda.
\newblock The energy--momentum tensor(s) in classical gauge theories.
\newblock \emph{Nuclear Physics B}, 912:\penalty0 192--223, 2016.

\bibitem[Noether(1918)]{noether1918}
E.~Noether.
\newblock Invariante variationsprobleme.
\newblock \emph{König. Gesellsch. d. Wiss. zu Göttingen, Math.-Phys. Klasse},
  pages 235--257, 1918.

\bibitem[Kosmann-Schwarzbach(2010)]{kosmann2011}
Y.~Kosmann-Schwarzbach.
\newblock \emph{The {N}oether Theorems: Invariance and Conservation Laws in the
  Twentieth Century}.
\newblock Springer New York, 2010.

\bibitem[Baker et~al.(2022)Baker, Linnemann, and Smeenk]{baker2021noether}
M.R. Baker, N.~Linnemann, and C.~Smeenk.
\newblock Noether's first theorem and the energy-momentum tensor ambiguity
  problem.
\newblock \emph{The Physics and Philosophy of Noether's Theorems}, 2022.

\bibitem[Bessel-Hagen(1921)]{besselhagen1921}
E.~Bessel-Hagen.
\newblock {\"U}ber die erhaltungss{\"a}tze der elektrodynamik.
\newblock \emph{Mathematische Annalen}, 84\penalty0 (3-4):\penalty0 258--276,
  1921.

\bibitem[Ibragimov(2006)]{besseltranslation2006}
N.~Ibragimov.
\newblock \emph{English translation of {\it{{\"U}ber die erhaltungss{\"a}tze
  der elektrodynamik}} by E. Bessel-Hagen}.
\newblock Archives of ALGA, vol. 3, 2006.

\bibitem[Baker et~al.(2021)Baker, Kiriushcheva, and Kuzmin]{baker2021}
M.R. Baker, N.~Kiriushcheva, and S.~Kuzmin.
\newblock Noether and {H}ilbert (metric) energy-momentum tensors are not, in
  general, equivalent.
\newblock \emph{Nuclear Physics B}, page 115240, 2021.

\bibitem[Gieres(2022)]{gieres2022}
F.~Gieres.
\newblock Improvement of a conserved current density versus adding a total
  derivative to a {L}agrangian density.
\newblock \emph{Fortschritte der Physik}, 70\penalty0 (9-10):\penalty0 2200078,
  2022.

\bibitem[Butcher(2012)]{Butcher2012PHD}
L.M. Butcher.
\newblock \emph{The localisation of gravitational energy, momentum, and spin}.
\newblock PhD thesis, University of Cambridge, 2012.

\bibitem[Kuzmin and McKeon(2001)]{kuzmin2001}
S.V. Kuzmin and D.G.C. McKeon.
\newblock Automatic derivation of improved symmetry currents in the context of
  the {W}ess-{Z}umino model.
\newblock \emph{Physical Review D}, 64\penalty0 (8):\penalty0 085009, 2001.

\bibitem[Gelfand and Fomin(2012)]{gelfand2012}
I.M. Gelfand and S.V. Fomin.
\newblock \emph{Calculus of Variations}.
\newblock Dover Books on Mathematics. Dover Publications, 2012.
\newblock ISBN 9780486135014.

\bibitem[Bak et~al.(1994)Bak, Cangemi, and Jackiw]{jackiw1994}
D.~Bak, D.~Cangemi, and R.~Jackiw.
\newblock Energy-momentum conservation in gravity theories.
\newblock \emph{Phys. Rev. D}, 49:\penalty0 5173--5181, May 1994.
\newblock \doi{10.1103/PhysRevD.49.5173}.

\bibitem[Leclerc(2006{\natexlab{a}})]{leclerc2006}
M.~Leclerc.
\newblock Canonical and gravitational stress-energy tensors.
\newblock \emph{International Journal of Modern Physics D}, 15\penalty0
  (07):\penalty0 959--989, 2006{\natexlab{a}}.

\bibitem[Leclerc(2006{\natexlab{b}})]{leclerc2006b}
M.~Leclerc.
\newblock Noether's theorem, the stress-energy tensor and {H}amiltonian
  constraints.
\newblock \emph{arXiv preprint gr-qc/0608096}, 2006{\natexlab{b}}.

\bibitem[Baker and Kuzmin(2019)]{baker2019}
M.R. Baker and S.~Kuzmin.
\newblock A connection between linearized {G}auss--{B}onnet gravity and
  classical electrodynamics.
\newblock \emph{International Journal of Modern Physics D}, 28\penalty0
  (07):\penalty0 1950092, 2019.

\bibitem[M{\o}ller(1961)]{moller1961}
C.~M{\o}ller.
\newblock Further remarks on the localization of the energy in the general
  theory of relativity.
\newblock \emph{Annals of Physics}, 12\penalty0 (1):\penalty0 118--133, 1961.

\bibitem[Favata(2001)]{favata2001}
M.~Favata.
\newblock Energy localization invariance of tidal work in general relativity.
\newblock \emph{Physical Review D}, 63\penalty0 (6):\penalty0 064013, 2001.

\bibitem[Landau and Lifshitz(1971)]{landau1971}
L.D. Landau and E.M. Lifshitz.
\newblock The classical theory of fields.
\newblock 1971.

\bibitem[Einstein(1916{\natexlab{a}})]{Einstein1916}
A~Einstein.
\newblock Hamiltonsches prinzip und allgemeine relativit{\"a}tstheorie, sitz.
\newblock \emph{Ber. Preuss. Akad. Wiss. Berlin}, 111, 1916{\natexlab{a}}.

\bibitem[Goldberg(1958)]{Goldberg1958}
J.N. Goldberg.
\newblock Conservation laws in general relativity.
\newblock \emph{Physical Review}, 111\penalty0 (1):\penalty0 315, 1958.

\bibitem[Weinberg(1972)]{weinberg1972}
S.~Weinberg.
\newblock Gravitation and cosmology: principles and applications of the general
  theory of relativity.
\newblock 1972.

\bibitem[Papapetrou(1948)]{Papapetrou1948}
A.~Papapetrou.
\newblock Einstein's theory of gravitation and flat space.
\newblock In \emph{Proceedings of the Royal Irish Academy. Section A:
  Mathematical and Physical Sciences}, volume~52, pages 11--23. JSTOR, 1948.

\bibitem[Bergmann and Thomson(1953)]{Bergmann1953}
P.G. Bergmann and R.~Thomson.
\newblock Spin and angular momentum in general relativity.
\newblock \emph{Physical Review}, 89\penalty0 (2):\penalty0 400, 1953.

\bibitem[M{\o}ller(1958)]{Moller1958}
C.~M{\o}ller.
\newblock On the localization of the energy of a physical system in the general
  theory of relativity.
\newblock \emph{Annals of Physics}, 4\penalty0 (4):\penalty0 347--371, 1958.

\bibitem[Tolman(1930)]{tolman1930}
R.C. Tolman.
\newblock On the use of the energy-momentum principle in general relativity.
\newblock \emph{Physical Review}, 35\penalty0 (8):\penalty0 875, 1930.

\bibitem[Cooperstock(2000)]{cooperstock2000}
F.I. Cooperstock.
\newblock The role of energy and a new approach to gravitational waves in
  general relativity.
\newblock \emph{Annals of Physics}, 282\penalty0 (1):\penalty0 115--137, 2000.

\bibitem[Xulu(2003)]{xulu2003phd}
S.S. Xulu.
\newblock The energy-momentum problem in general relativity.
\newblock \emph{Ph. D. Thesis, University of Zululand}, 2003.

\bibitem[Sharif and Fatima(2005)]{sharif2005}
M.~Sharif and T.~Fatima.
\newblock Energy--momentum distribution: A crucial problem in general
  relativity.
\newblock \emph{International Journal of Modern Physics A}, 20\penalty0
  (18):\penalty0 4309--4330, 2005.

\bibitem[Padmanabhan(2010)]{padmanabhan2010}
T.~Padmanabhan.
\newblock \emph{Gravitation: foundations and frontiers}.
\newblock Cambridge University Press, 2010.

\bibitem[Duerr(2019)]{duerr2019}
P.M. Duerr.
\newblock Fantastic beasts and where (not) to find them: Energy conservation
  and local gravitational energy in general relativity.
\newblock \emph{Studies in History and Philosophy of Modern Physics},
  65:\penalty0 1--14, 2019.

\bibitem[De~Haro(2022)]{deHaro2022}
S.~De~Haro.
\newblock Noether's theorems and energy in general relativity.
\newblock \emph{The Physics and Philosophy of Noether's Theorems}, 2022.

\bibitem[Hulse and Taylor(1975)]{hulse1975}
R.A. Hulse and J.H. Taylor.
\newblock Discovery of a pulsar in a binary system.
\newblock \emph{The Astrophysical Journal}, 195:\penalty0 L51--L53, 1975.

\bibitem[Chen and Nester(1999)]{chen1999}
C.-M. Chen and J.M. Nester.
\newblock Quasilocal quantities for general relativity and other gravity
  theories.
\newblock \emph{Classical and Quantum Gravity}, 16\penalty0 (4):\penalty0 1279,
  1999.

\bibitem[Chang et~al.(1999)Chang, Nester, and Chen]{chang1999}
C.-C. Chang, J.M. Nester, and C.-M. Chen.
\newblock Pseudotensors and quasilocal energy-momentum.
\newblock \emph{Physical Review Letters}, 83\penalty0 (10):\penalty0 1897,
  1999.

\bibitem[Szabados(2009)]{szabados2009}
L.B. Szabados.
\newblock Quasi-local energy-momentum and angular momentum in general
  relativity.
\newblock \emph{Living reviews in relativity}, 12:\penalty0 1--163, 2009.

\bibitem[Einstein(1916{\natexlab{b}})]{einstein1916gw}
A.~Einstein.
\newblock N{\"a}herungsweise integration der feldgleichungen der gravitation.
\newblock \emph{Sitzungsberichte der K{\"o}niglich Preu{\ss}ischen Akademie der
  Wissenschaften}, pages 688--696, 1916{\natexlab{b}}.

\bibitem[Poincar{\'e}(1893)]{poincare1893}
H.~Poincar{\'e}.
\newblock \emph{Les m{\'e}thodes nouvelles de la m{\'e}canique c{\'e}leste},
  volume~2.
\newblock Gauthier-Villars et fils, imprimeurs-libraires, 1893.

\bibitem[Freud(1939)]{Freud1939}
P.~Freud.
\newblock Uber die ausdrucke der gesamtenergie und des gesamtimpulses eines
  materiellen systems in der allgemeinen relativitatstheorie.
\newblock \emph{Annals of Mathematics}, pages 417--419, 1939.

\bibitem[Chen et~al.(2015)Chen, Nester, and Tung]{Chen2015}
C.-M. Chen, J.M. Nester, and R.-S. Tung.
\newblock Gravitational energy for {GR} and {P}oincar{\'e} gauge theories: A
  covariant {H}amiltonian approach.
\newblock \emph{International Journal of Modern Physics D}, 24\penalty0
  (11):\penalty0 1530026, 2015.

\bibitem[Einstein(1915)]{Einstein1915}
A.~Einstein.
\newblock Erkl{\"a}rung der perihelbewegung des merkur aus der allgemeinen
  relativit{\"a}tstheorie.
\newblock \emph{Sitzungsberichte der K{\"o}niglich Preu{\ss}ischen Akademie der
  Wissenschaften Berlin}, pages 831--839, 1915.

\bibitem[Antonov(2023)]{Antonov2023}
L.~Antonov.
\newblock Landau-{L}ifshitz stress-energy pseudotensor.
\newblock 2023.
\newblock \doi{10.13140/RG.2.2.27292.36483/2}.

\bibitem[Notte-Cuello and Rodrigues~Jr(2009)]{Notte2009}
E.A. Notte-Cuello and W.A. Rodrigues~Jr.
\newblock Freud’s identity of differential geometry, the {E}instein-{H}ilbert
  equations and the vexatious problem of the energy-momentum conservation in
  {GR}.
\newblock \emph{Advances in applied Clifford algebras}, 19\penalty0
  (1):\penalty0 113--145, 2009.

\bibitem[Bhattacharyya(2023)]{Bhattacharyya2021}
S.~Bhattacharyya.
\newblock Two body dynamics in a quadratic modification of general relativity.
\newblock In \emph{Proceedings of the MG16 Meeting on General Relativity},
  pages 883--906. World Scientific, 2023.

\bibitem[Kopeikin and Petrov(2013)]{Kopeikin2013}
S.M. Kopeikin and A.N. Petrov.
\newblock Post-{N}ewtonian celestial dynamics in cosmology: Field equations.
\newblock \emph{Physical Review D}, 87\penalty0 (4):\penalty0 044029, 2013.

\bibitem[Matyjasek(2008)]{matyjasek2008}
J.~Matyjasek.
\newblock Some remarks on the {E}instein and {M}{\o}ller pseudotensors for
  static and spherically-symmetric configurations.
\newblock \emph{Modern Physics Letters A}, 23\penalty0 (08):\penalty0 591--601,
  2008.

\bibitem[Rosen(1994)]{Rosen1994}
N.~Rosen.
\newblock The energy of the universe.
\newblock \emph{General Relativity and Gravitation}, 26:\penalty0 319--321,
  1994.

\end{thebibliography}

\appendix

\section{Appendix A - General set of energy-momentum tensors derived from the Noether current system of equations} \label{Fock to Noether Coefficient Comparison}

In Appendix \ref{Fock to Noether Coefficient Comparison} we will include all coefficient restrictions on the general set of energy-momentum tensors in linearized gravity which can be derived from the Noether current, given in (\ref{FockEMT}). In \ref{Restriction1} we include the system of equations that restrict the possible Lagrangian densities (\ref{genlag}) such that the linearized Einstein field equations are obtained from the Euler-Lagrange equation. In \ref{Restriction2} we include the conditions imposed on (\ref{GenSymTransform}) to preserve symmetry of the transformations ($\delta h_{\mu\nu} = \delta h_{\nu\mu}$). In \ref{Restriction3} we include the system of equations corresponding to the energy-momentum tensors which can be derived from the Noether current given the general Lagrangian (\ref{genlag}) and field transformations (\ref{GenSymTransform}). All restrictions \ref{Restriction1}-\ref{Restriction3} must simultaneously be satisfied in order to derive an energy-momentum tensor from the Noether current in linearized gravity.

\subsection{Equation of motion coefficient restrictions} \label{Restriction1}

\begin{tabular}{@{}p{.5\linewidth}@{}p{.5\linewidth}@{}}
\begin{gather}
    D_4 - C_2 = - \frac{n}{4}    \label{eomcondition1}
\\
    D_3 - C_1 = \frac{n}{4}   \label{eomcondition2}
\end{gather}
  &
\begin{gather}
    D_1 - C_3 + D_5 = \frac{n}{2}   \label{eomcondition3}
\\
    D_2 - C_4 - C_5 = - \frac{n}{2}    \label{eomcondition4}
\end{gather}
\end{tabular}

\subsection{Field transformation coefficient restrictions} \label{Restriction2}

\begin{tabular}{@{}p{.33333333\linewidth}@{}p{.33333333\linewidth}@{}p{.33333333\linewidth}@{}}
\begin{equation}
    B_2 = B_3 
\end{equation}
  &
\begin{equation}
B_4 = B_5
\end{equation}
&
\begin{equation}
 B_6 = B_7
\end{equation}
\end{tabular}

\subsection{General set of energy-momentum tensors derived from the Noether current coefficient restriction} \label{Restriction3}

\small

\begin{equation}
    b_1 = B_2 C_5 + B_3 C_5 + B_4 C_4 + B_5 C_4 - \frac{1}{4} B_2 D_2 - \frac{1}{4} B_3 D_2 - \frac{1}{4} B_4 D_2 - \frac{1}{4} B_5 D_2 
\end{equation}

\begin{equation}
    b_2 = \frac{1}{2} B_4 C_3 + \frac{1}{2} B_5 C_3 + B_6 C_5 + B_7 C_5 - \frac{1}{4} B_6 D_2 - \frac{1}{4} B_7 D_2 - \frac{1}{2} B_4 D_5 - \frac{1}{2} B_5 D_5
\end{equation}

\begin{multline}
    b_3 = B_2 C_3 + B_2 C_4 + B_3 C_3 + B_3 C_4 + B_4 C_3 + B_5 C_3 + 4 B_8 C_3 + 2 B_8 C_4 + 2 B_8 C_5
    \\
    - B_2 D_1 - B_3 D_1 - B_4 D_1 - B_5 D_1 - 4 B_8 D_1 - \frac{1}{4} B_2 D_2 - \frac{1}{4} B_3 D_2 - B_8 D_2 
\end{multline}

\begin{tabular}{@{}p{.5\linewidth}@{}p{.5\linewidth}@{}}
\begin{gather}
    b_4 = B_4 C_5 + B_5 C_5 - \frac{1}{4} B_4 D_2 - \frac{1}{4} B_5 D_2     \label{eqn:b_4 Coefficient}
\\
    b_5 = B_4 C_5 + B_5 C_5 - \frac{1}{4} B_4 D_2 - \frac{1}{4} B_5 D_2     \label{eqn:b_5 Coefficient}
\\
    b_7 = 2B_1 C_1 - B_1 D_3     \label{eqn:b_7 Coefficient}
\end{gather}
  &
\begin{gather}
b_{11_{i}} = 2 B_4 C_1 + 2 B_5 C_1 - B_4 D_3 - B_5 D_3 
\\
b_{11_{ii}} = 2 B_1 C_5 - \frac{1}{2} B_1 D_2    \label{eqn:b_11_ii Coefficient}
\end{gather}
\end{tabular}

\begin{multline}
    b_6 = 2 B_1 C_2 + 2 B_6 C_2 + \frac{1}{2} B_6 C_3 + 2 B_7 C_2 + \frac{1}{2} B_7 C_3 + 2 B_9 C_1 + 8 B_9 C_2 + B_9 C_3 
    \\
    - B_9 D_3 - B_1 D_4 - B_6 D_4 - B_7 D_4 - 4 B_9 D_4 - \frac{1}{2} B_6 D_5 - \frac{1}{2} B_7 D_5 - B_9 D_5 
\end{multline}

\begin{equation}
    b_{8_{i}}= \frac{1}{2} B_4 C_3 + \frac{1}{2} B_5 C_3 + 2 B_6 C_1 + 2 B_7 C_1 - B_6 D_3 - B_7 D_3 - \frac{1}{2} B_4 D_5 - \frac{1}{2} B_5 D_5 
\end{equation}

\begin{equation}
    b_{8_{ii}}= B_1 C_3 + B_6 C_5 + B_7 C_5 - \frac{1}{4} B_6 D_2 - \frac{1}{4} B_7 D_2 - B_1 D_5
\end{equation}

\begin{multline}
    b_{9_{i}} = 2 B_2 C_2 + 2 B_3 C_2 + 2 B_4 C_2 + 2 B_5 C_2 + B_6 C_4 + B_7 C_4 + 2 B_8 C_1 + 8 B_8 C_2 + B_8 C_3 
    \\
    - \frac{1}{4} B_6 D_2 - \frac{1}{4} B_7 D_2 - B_8 D_3 - B_2 D_4 - B_3 D_4 - B_4 D_4 - B_5 D_4 - 4 B_8 D_4 - B_8 D_5 
\end{multline}

\begin{multline}
    b_{9_{ii}} = B_1 C_3 + \frac{1}{2} B_2 C_3 + \frac{1}{2} B_3 C_3 + B_6 C_3 + B_7 C_3 + 4 B_9 C_3 + 2 B_9 C_4 + 2 B_9 C_5 
    \\
    - B_1 D_1 - B_6 D_1 - B_7 D_1 - 4 B_9 D_1 - B_9 D_2 - \frac{1}{2} B_2 D_5 - \frac{1}{2} B_3 D_5 
\end{multline}

\begin{equation}
    b_{10_{i}} = 2 B_2 C_1 + 2 B_3 C_1 + B_4 C_4 + B_5 C_4 - \frac{1}{4} B_4 D_2 - \frac{1}{4} B_5 D_2 - B_2 D_3 - B_3 D_3 
\end{equation}

\begin{equation}
    \label{eqn:b_10_ii Coefficient}
    b_{10_{ii}} = 2 B_1 C_4 + B_2 C_5 + B_3 C_5 - \frac{1}{2} B_1 D_2 - \frac{1}{4} B_2 D_2 - \frac{1}{4} B_3 D_2 
\end{equation}

\begin{tabular}{@{}p{.5\linewidth}@{}p{.5\linewidth}@{}}
\begin{gather}
d_1 = \frac{1}{4} B_6 D_2 + \frac{1}{4} B_7 D_2     \label{eqn:d_1 Coefficient}
\\
    d_2 = \frac{1}{4} B_2 D_2 + \frac{1}{4} B_3 D_2 
\\
d_3 = \frac{1}{2} B_4 D_5 + \frac{1}{2} B_5 D_5 
\\
d_4 = \frac{1}{4} B_4 D_2 + \frac{1}{4} B_5 D_2     \label{eqn:d_4 Coefficient}
\\
    d_{5_{i}} = \frac{1}{4} B_4 D_2 + \frac{1}{4} B_5 D_2     \label{eqn:d_5_i Coefficient}
\\
    d_{5_{ii}} = 0    \label{eqn:d_5_ii Coefficient}
\\
d_{6_{ii}} = \frac{1}{4} B_2 D_2 + \frac{1}{4} B_3 D_2 
\end{gather}
  &
\begin{gather}
    d_8 = B_1 D_3 
\\
    d_{9_{ii}} = \frac{1}{4} B_6 D_2 + \frac{1}{4} B_7 D_2 + B_6 D_3 + B_7 D_3     \label{eqn:d_9_ii Coefficient}
\\
    d_{10_{i}} = \frac{1}{2} B_1 D_2 + \frac{1}{4} B_2 D_2 + \frac{1}{4} B_3 D_2 
\\
    d_{10_{ii}} = B_2 D_3 + B_3 D_3 
    \\
d_{11_{ii}} = B_1 D_5 + \frac{1}{2} B_2 D_5 + \frac{1}{2} B_3 D_5 
    \\
d_{12_{i}} = \frac{1}{4} B_4 D_2 + \frac{1}{4} B_5 D_2 + B_4 D_3 + B_5 D_3 
    \\
d_{12_{ii}} = \frac{1}{2} B_1 D_2        \label{eqn:d_12_ii Coefficient}
\end{gather}
\end{tabular}

\begin{equation}
    d_{6_{i}} = B_2 D_1 + B_3 D_1 + B_4 D_1 + B_5 D_1 + 4 B_8 D_1 + \frac{1}{4} B_4 D_2 + \frac{1}{4} B_5 D_2 + B_8 D_2 
\end{equation}

\begin{equation}
    d_7 = B_9 D_3 + B_1 D_4 + B_6 D_4 + B_7 D_4 + 4 B_9 D_4 + \frac{1}{2} B_6 D_5 + \frac{1}{2} B_7 D_5 + B_9 D_5 
\end{equation}

\begin{equation}
    d_{9_{i}} = B_1 D_1 + B_6 D_1 + B_7 D_1 + 4 B_9 D_1 + \frac{1}{4} B_6 D_2 + \frac{1}{4} B_7 D_2 + B_9 D_2 
\end{equation}

\begin{equation}
    d_{11_{i}} = B_8 D_3 + B_2 D_4 + B_3 D_4 + B_4 D_4 + B_5 D_4 + 4 B_8 D_4 + \frac{1}{2} B_4 D_5 + \frac{1}{2} B_5 D_5 + B_8 D_5 
\end{equation}

\begin{tabular}{@{}p{.5\linewidth}@{}p{.5\linewidth}@{}}
\begin{gather}
a_1 = D_1 + \frac{1}{4} B_6 D_2 + \frac{1}{4} B_7 D_2 
\\
a_2 = D_5 + \frac{1}{2} B_2 D_5 + \frac{1}{2} B_3 D_5 
\\
a_3 = D_2 + \frac{1}{4} B_2 D_2  + \frac{1}{4} B_3 D_2 
\\
a_4 = D_4 + \frac{1}{2} B_6 D_5 + \frac{1}{2} B_7 D_5 
\\
a_5 = D_3     \label{eqn:a_5 Coefficient}
\end{gather}
  &
\begin{gather}
c_1 = C_2 + \frac{1}{2} B_6 C_3 + \frac{1}{2} B_7 C_3 - \frac{1}{2} B_6 D_5 - \frac{1}{2} B_7 D_5 
\\
c_2 = C_1     \label{eqn:c_2 Coefficient}
\\
    c_3 = C_4 + B_2 C_4 + B_3 C_4 - \frac{1}{4} B_2 D_2 - \frac{1}{4} B_3 D_2 
\\
  c_4 = C_5     \label{eqn:c_4 Coefficient}
\end{gather}
\end{tabular}

\begin{equation}
    c_5 = C_3 + \frac{1}{2} B_2 C_3 + \frac{1}{2} B_3 C_3 + B_6 C_4 + B_7 C_4 - \frac{1}{4} B_6 D_2 - \frac{1}{4} B_7 D_2 - \frac{1}{2} B_2 D_5 - \frac{1}{2} B_3 D_5 
\end{equation}

\large

\section{Appendix B - Restricted energy-momentum conditions}

In this Appendix we will include all restricted coefficient conditions related to results in Section \ref{ResultsRestriction}.

\subsection{Symmetric energy-momentum tensor conditions}\label{Symmetry conditions}

The modified (symmetric) coefficient conditions required for a symmetric energy-momentum tensor, corresponding to discussion in Section \ref{SectionSymmetric}, are included below. The coefficients in Appendix \ref{Fock to Noether Coefficient Comparison} still need to be satisfied, so only the symmetric coefficient equations in \ref{Restriction3} need to be replaced by the following:

\small

\begin{multline}
    b_8 
    = \frac{1}{2} B_4 C_3 + \frac{1}{2} B_5 C_3 + 2 B_6 C_1 + 2 B_7 C_1 - B_6 D_3 - B_7 D_3 - \frac{1}{2} B_4 D_5 - \frac{1}{2} B_5 D_5 
    \\
    = B_1 C_3 + B_6 C_5 + B_7 C_5 - \frac{1}{4} B_6 D_2 - \frac{1}{4} B_7 D_2 - B_1 D_5 
\end{multline}

\begin{multline}
    b_9 
    = 2 B_2 C_2 + 2 B_3 C_2 + 2 B_4 C_2 + 2 B_5 C_2 + B_6 C_4 + B_7 C_4 + 2 B_8 C_1 + 8 B_8 C_2 + B_8 C_3 
    \\
    - \frac{1}{4} B_6 D_2 - \frac{1}{4} B_7 D_2 - B_8 D_3 - B_2 D_4 - B_3 D_4 - B_4 D_4 - B_5 D_4 - 4 B_8 D_4 - B_8 D_5 
    \\
    = B_1 C_3 + \frac{1}{2} B_2 C_3 + \frac{1}{2} B_3 C_3 + B_6 C_3 + B_7 C_3 + 4 B_9 C_3 + 2 B_9 C_4 + 2 B_9 C_5 
    \\
    - B_1 D_1 - B_6 D_1 - B_7 D_1 - 4 B_9 D_1 - B_9 D_2 - \frac{1}{2} B_2 D_5 - \frac{1}{2} B_3 D_5 
\end{multline}

\begin{multline}
    b_{10} 
    = 2 B_2 C_1 + 2 B_3 C_1 + B_4 C_4 + B_5 C_4 - \frac{1}{4} B_4 D_2 - \frac{1}{4} B_5 D_2 - B_2 D_3 - B_3 D_3 
    \\
    = 2 B_1 C_4 + B_2 C_5 + B_3 C_5 - \frac{1}{2} B_1 D_2 - \frac{1}{4} B_2 D_2 - \frac{1}{4} B_3 D_2 
\end{multline}

\begin{equation}
    b_{11} 
    = 2 B_4 C_1 + 2 B_5 C_1 - B_4 D_3 - B_5 D_3 
    = 2 B_1 C_5 - \frac{1}{2} B_1 D_2 
\end{equation}

\begin{equation}
    d_5 
    = \frac{1}{4} B_4 D_2 + \frac{1}{4} B_5 D_2 
    = 0
\end{equation}

\begin{equation}
    d_6 
    = B_2 D_1 + B_3 D_1 + B_4 D_1 + B_5 D_1 + 4 B_8 D_1 + \frac{1}{4} B_4 D_2 + \frac{1}{4} B_5 D_2 + B_8 D_2 
    = \frac{1}{4} B_2 D_2 + \frac{1}{4} B_3 D_2 
\end{equation}

\begin{equation}
    d_9 
    = B_1 D_1 + B_6 D_1 + B_7 D_1 + 4 B_9 D_1 + \frac{1}{4} B_6 D_2 + \frac{1}{4} B_7 D_2 + B_9 D_2 
    = \frac{1}{4} B_6 D_2 + \frac{1}{4} B_7 D_2 + B_6 D_3 + B_7 D_3 
\end{equation}

\begin{equation}
    d_{10} 
    = \frac{1}{2} B_1 D_2 + \frac{1}{4} B_2 D_2 + \frac{1}{4} B_3 D_2 
    = B_2 D_3 + B_3 D_3 
\end{equation}

\begin{equation}
    d_{11} 
    = B_8 D_3 + B_2 D_4 + B_3 D_4 + B_4 D_4 + B_5 D_4 + 4 B_8 D_4 + \frac{1}{2} B_4 D_5 + \frac{1}{2} B_5 D_5 + B_8 D_5 
    = B_1 D_5 + \frac{1}{2} B_2 D_5 + \frac{1}{2} B_3 D_5 
\end{equation}

\begin{equation}
    d_{12} 
    = \frac{1}{4} B_4 D_2 + \frac{1}{4} B_5 D_2 + B_4 D_3 + B_5 D_3 
    = \frac{1}{2} B_1 D_2 
\end{equation}

\large

\subsection{Traceless energy-momentum tensor conditions}\label{Traceless conditions}

The following tracelessness conditions correspond to traceless energy-momentum tensor solutions discussed in Section \ref{SectionTraceless}.

\begin{tabular}{@{}p{.5\linewidth}@{}p{.5\linewidth}@{}}
\begin{gather}
 b_1 + b_{8_i} + b_{8_{ii}} + b_{9_i} + b_{9_{ii}} + 4 c_5 = 0
\\
 b_2 + b_6 + 4 c_1 = 0
\\
b_3 + b_{10_i} + b_{10_{ii}} + 4 c_3 = 0
\\
b_4 + b_7 + 4 c_2 = 0
\\
b_5 + b_{11_i} + b_{11_{ii}} + 4 c_4  = 0
\end{gather}
  &
\begin{gather}
d_1 + d_3 + d_7 + 4 a_4 = 0
\\
d_2 + d_{11_{i}} + d_{11_{ii}} + 4 a_2  = 0
\\
d_4 + d_{9_i} + d_{9_{ii}} + 4 a_1  = 0
\\
d_{5_i} + d_{5_{ii}} + d_8 + 4 a_5  = 0
\\
d_{6_i} + d_{6_{ii}} + d_{10_{i}} + d_{10_{ii}} + d_{12_{i}} + d_{12_{ii}} + 4 a_3 = 0
\end{gather}
\end{tabular}

\vspace{-0.5cm}

\subsection{Christoffel symmetry conditions}\label{Christoffel symmetry solution}

The following conditions are the non-zero coefficients in (\ref{FockEMT}) required for the linearized Christoffel symbol energy-momentum tensors in Section \ref{LinChrisTransSec}: $b_1 = 2 C_4 - \frac{1}{2} D_2 $, $ b_2 = C_3 - D_5$, $b_3 = 2 C_3 - 2 D_1$, $b_4 = 2 C_5 - \frac{1}{2} D_2 $, $b_5 = 2 C_5 - \frac{1}{2} D_2 $, $b_6 = - 2 C_2 + D_4 $, $b_7 = - 2 C_1 + D_3 $, $ b_{8_{i}}= C_3 - D_5 $, $b_{8_{ii}}= - C_3 + D_5$, $b_{9_{i}} = 4 C_2 - 2 D_4 $, $b_{9_{ii}} = - C_3 + D_1$, $b_{10_{i}} = 2 C_4 - \frac{1}{2} D_2 $, $b_{10_{ii}} = - 2 C_4 + \frac{1}{2} D_2 $, $b_{11_{i}} = 4 C_1 - 2 D_3 $, $b_{11_{ii}} = - 2 C_5 + \frac{1}{2} D_2 $, $c_1 = C_2 $, $c_2 = C_1 $, $ c_3 = C_4 $, $c_4 = C_5 $, $c_5 = C_3 $, $ d_3 = D_5 $, $d_4 = \frac{1}{2} D_2 $, $d_{5_{i}} = \frac{1}{2} D_2 $, $ d_{6_{i}} = 2 D_1 + \frac{1}{2} D_2 $, $ d_7 = - D_4 $, $d_8 = - D_3 $, $ d_{9_{i}} = - D_1 $, $d_{10_{i}} = - \frac{1}{2} D_2 $, $d_{11_{i}} = 2 D_4 + D_5$, $d_{11_{ii}} = - D_5 $, $ d_{12_{i}} = \frac{1}{2} D_2 + 2 D_3 $, $d_{12_{ii}} = - \frac{1}{2} D_2 $, $a_1 = D_1 $, $a_2 = D_5 $, $a_3 = D_2 $, $a_4 = D_4 $ and $ a_5 = D_3 $.

\subsection{Canonical symmetry conditions}\label{Canonical symmetry solution}

The following conditions are the non-zero coefficients in (\ref{FockEMT}) required for the canonical energy-momentum tensors in Section \ref{SectionCanonical}: $    b_6 = - 2 C_2 + D_4 $, $   b_7 = - 2 C_1 + D_3 $, $    b_{8_{ii}} = - C_3 + D_5$, $  b_{9_{ii}} = - C_3 + D_1 $, $ b_{10_{ii}} = - 2 C_4 + \frac{1}{2} D_2 $, $ b_{11_{ii}} = - 2 C_5 + \frac{1}{2} D_2 $, $c_1 = C_2 $, $c_2 = C_1 $, $c_3 = C_4 $, $c_4 = C_5 $, $ c_5 = C_3 $, $ d_7 = - D_4$, $d_8 = - D_3 $, $d_{9_{i}} = - D_1 $, $d_{10_{i}} = - \frac{1}{2} D_2 $, $d_{11_{ii}} = - D_5 $, $d_{12_{ii}} = - \frac{1}{2} D_2 $, $a_1 = D_1 $, $a_2 = D_5$, $a_3 = D_2 $, $a_4 = D_4 $, and $ a_5 = D_3 $.

\subsection{Simultaneously symmetric and traceless energy-momentum tensor conditions} \label{Symmetric Traceless Conditions}

The following conditions correspond to energy-momentum tensor solutions which are simultaneously symmetric and traceless, discussed in Section \ref{symmtracecombo}.

\begin{tabular}{@{}p{.5\linewidth}@{}p{.5\linewidth}@{}}
\begin{gather}
b_1 + 2 b_8 + 2 b_9 + 4 c_5 = 0
\\
b_2 + b_6 + 4 c_1 = 0
\\
b_3 + 2 b_{10} + 4 c_3 = 0
\\
b_4 + b_7 + 4 c_2 = 0
\\
b_5 + 2 b_{11} + 4 c_4 = 0
\end{gather}
  &
\begin{gather}
d_1 + d_3 + d_7 + 4 a_4 = 0
\\
d_2 + 2 d_{11} + 4 a_2 = 0
\\
d_4 + 2 d_9 + 4 a_1 = 0
\\
2 d_5 + d_8 + 4 a_5 = 0
\\
2 d_6 + 2 d_{10} + 2 d_{12} + 4 a_3 = 0
\end{gather}
\end{tabular}

\large

\section{Appendix C - Table of coefficients vs. linearized gravity energy-momentum tensors}\label{Table of Fock Coefficients vs. Pseudotensors}

In this Appendix we will include all of the linearized gravity energy-momentum tensors considered in this article in tables which outline which terms in (\ref{FockEMT}) are non-zero. 

\small

\begin{tabular}{l*{20}{c}r}
\rotatebox{90}{$\partial h \partial h$ Terms}
& \rotatebox{90}{Einstein Left (\ref{EinsteinLeftRaise})} 
& \rotatebox{90}{Einstein Right (\ref{EinsteinRightRaise})} 
& \rotatebox{90}{Landau-Lifshitz (\ref{LLLin})} 
& \rotatebox{90}{Goldberg Mixed (\ref{GoldbergMixedLin})} 
& \rotatebox{90}{Goldberg Symmetric (\ref{GoldbergSymLin})}  
& \rotatebox{90}{Weinberg (\ref{LinWein})} 
& \rotatebox{90}{Papapetrou (\ref{PapaLin})} 
& \rotatebox{90}{Bergmann-Thomson (\ref{BTPseudo})} 
& \rotatebox{90}{Imitation Einstein (\ref{LinImEin})} 
& \rotatebox{90}{M{\o}ller (\ref{LinMoller})} 
& \rotatebox{90}{Hilbert (\ref{HilbertEMT})} 
& \rotatebox{90}{Fierz (\ref{FierzEMTExpand})}
& \rotatebox{90}{Butcher $\tau$ (\ref{Butcher1})} 
& \rotatebox{90}{Butcher t (\ref{Butcher3})} 
& \rotatebox{90}{Padmanabhan (\ref{PaddyEMT})} 
& \rotatebox{90}{Rose-Xavier (\ref{RoseXavier})}\\
\hline
$b_1$            &   &   & X &   & X & X &   & X & X & X & X & X &   & X & X & X \\
\rowcolor{gray!20}
$b_2$            &   &   & X &   & X & X & X & X & X & X & X & X &   & X & X & X \\
$b_3$            &   &   & X &   & X &   &   &   &   &   &   & X &   &   &   & X \\
\rowcolor{gray!20}
$b_4$            &   &   & X &   & X & X &   & X & X & X & X & X &   &   & X & X \\
$b_5$            &   &   &   &   &   & X &   & X & X & X & X &   &   & X & X & X \\
\rowcolor{gray!20}
$b_6$            & X & X & X & X & X &   &   & X & X &   & X & X & X & X & X & X \\
$b_7$            & X & X & X & X & X & X &   & X &   &   & X & X & X & X & X & X \\
\rowcolor{gray!20}
$b_{8_i}$        & X &   & X & X & X & X & X & X & X &   &   &   &   & X & X & X \\
$b_{8_{ii}}$     &   & X & X &   & X & X & X & X & X & X &   & X &   & X & X & X \\
\rowcolor{gray!20}
$b_{9_i}$        & X &   & X & X & X &   & X & X & X &   & X & X &   &   & X & X \\
$b_{9_{ii}}$     &   & X & X &   & X &   & X &   &   &   & X & X &   &   & X & X \\
\rowcolor{gray!20}
$b_{10_i}$       &   &   &   &   &   & X &   & X &   &   &   &   &   & X &   & X \\
$b_{10_{ii}}$    &   &   &   &   &   & X &   & X &   & X &   & X &   & X &   & X \\
\rowcolor{gray!20}
$b_{11_i}$       & X &   & X & X & X &   &   & X &   &   & X & X &   & X & X & X \\
$b_{11_{ii}}$    &   & X & X &   & X &   &   &   & X &   & X & X &   & X & X & X \\
\rowcolor{gray!20}
$c_1$            & X & X & X & X & X & X &   & X & X &   & X & X & X & X & X & X \\
$c_2$            & X & X & X & X & X & X &   & X &   &   & X & X & X & X & X & X \\
\rowcolor{gray!20}
$c_3$            &   &   &   &   &   & X &   & X &   &   &   & X &   &   &   & X \\
$c_4$            & X & X & X & X & X & X &   & X &   &   & X & X &   &   & X & X \\
\rowcolor{gray!20}
$c_5$            & X & X & X & X & X & X & X & X & X &   &   & X &   & X & X & X \\
\end{tabular}

\vspace{0.2cm}

\begin{tabular}{l*{20}{c}r}
\rotatebox{90}{$h \partial \partial h$ Terms}
& \rotatebox{90}{Einstein Left (\ref{EinsteinLeftRaise})} 
& \rotatebox{90}{Einstein Right (\ref{EinsteinRightRaise})} 
& \rotatebox{90}{Landau-Lifshitz (\ref{LLLin})} 
& \rotatebox{90}{Goldberg Mixed (\ref{GoldbergMixedLin})} 
& \rotatebox{90}{Goldberg Symmetric (\ref{GoldbergSymLin})}  
& \rotatebox{90}{Weinberg (\ref{LinWein})} 
& \rotatebox{90}{Papapetrou (\ref{PapaLin})} 
& \rotatebox{90}{Bergmann-Thomson (\ref{BTPseudo})} 
& \rotatebox{90}{Imitation Einstein (\ref{LinImEin})} 
& \rotatebox{90}{M{\o}ller (\ref{LinMoller})} 
& \rotatebox{90}{Hilbert (\ref{HilbertEMT})} 
& \rotatebox{90}{Fierz (\ref{FierzEMTExpand})}
& \rotatebox{90}{Butcher $\tau$ (\ref{Butcher1})} 
& \rotatebox{90}{Butcher t (\ref{Butcher3})} 
& \rotatebox{90}{Padmanabhan (\ref{PaddyEMT})} 
& \rotatebox{90}{Rose-Xavier (\ref{RoseXavier})}\\
\hline
$d_1$            &   &   &   &   &   & X & X &   &   &   & X &   &   &   &   &   \\
\rowcolor{gray!20}
$d_2$            &   &   &   &   &   & X &   &   &   &   & X &   &   &   &   &   \\
$d_3$            &   &   &   &   &   &   & X & X & X & X &   &   &   & X &   &   \\
\rowcolor{gray!20}
$d_4$            &   &   &   &   &   & X &   & X & X & X & X &   &   & X &   &   \\
$d_{5_i}$        &   &   &   &   &   &   &   &   & X &   &   &   &   & X &   &   \\
\rowcolor{gray!20}
$d_{5_{ii}}$     &   &   &   &   &   &   &   & X &   & X &   &   &   & X &   &   \\
$d_{6_i}$        &   &   &   &   &   &   &   &   &   &   & X &   &   &   &   &   \\
\rowcolor{gray!20}
$d_{6_{ii}}$     &   &   &   &   &   &   &   & X & X & X & X &   &   & X &   &   \\
$d_7$            &   &   &   &   &   &   &   & X & X &   &   &   &   & X &   &   \\
\rowcolor{gray!20}
$d_8$            &   &   &   &   &   & X &   & X &   &   &   &   &   &   &   &   \\
$d_{9_i}$        &   &   &   &   &   &   & X &   &   &   & X &   &   &   &   &   \\
\rowcolor{gray!20}
$d_{9_{ii}}$     &   &   &   &   &   &   & X & X & X &   & X &   &   & X &   &   \\
$d_{10_i}$       &   &   &   &   &   &   &   &   & X &   &   &   &   & X &   &   \\
\rowcolor{gray!20}
$d_{10_{ii}}$    &   &   &   &   &   &   &   & X &   &   &   &   &   & X &   &   \\
$d_{11_i}$       &   &   &   &   &   &   & X & X & X &   &   &   &   &   &   &   \\
\rowcolor{gray!20}
$d_{11_{ii}}$    &   &   &   &   &   &   & X & X & X & X &   &   &   & X &   &   \\
$d_{12_i}$       &   &   &   &   &   & X &   & X &   &   &   &   &   & X &   &   \\
\rowcolor{gray!20}
$d_{12_{ii}}$    &   &   &   &   &   & X &   & X &   & X &   &   &   & X &   &   \\
$a_1$            &   &   &   &   &   & X & X & X & X &   & X &   &   & X &   &   \\
\rowcolor{gray!20}
$a_2$            &   &   &   &   &   &   & X & X & X &   &   &   &   &   &   &   \\
$a_3$            &   &   &   &   &   & X &   & X &   &   &   &   &   &   &   &   \\
\rowcolor{gray!20}
$a_4$            &   &   &   &   &   &   &   & X & X &   &   &   &   & X &   &   \\
$a_5$            &   &   &   &   &   & X &   & X &   &   &   &   &   &   &   &   \\
\end{tabular}

\large

\section{Appendix D - All four terms in the Noether current} \label{AppendixFourTerms}

In this Appendix we include all four terms in the Noether current (\ref{NoetherCurrent2}) for the general Lagrangian density (\ref{genlag}) and general field transformations (\ref{GenSymTransform}),

\small

\begin{multline}
    \eta^{\rho \lambda} \mathcal{L} \delta x_\lambda = 
    \eta^{\rho \lambda} (C_1 \partial_\beta h_{\mu \nu} \partial^\beta h^{\mu \nu} 
    + C_2 \partial_\beta h^\mu_\mu \partial^\beta h^\nu_\nu 
    + C_3 \partial_\beta h^{\beta \mu} \partial_\mu h^\nu_\nu 
    + C_4 \partial_\beta h^{\beta \mu} \partial^\nu h_{\mu \nu} 
    + C_5 \partial_\mu h_{\nu \beta} \partial^\nu h^{\mu \beta} 
    \\ 
    + D_1 h_{\mu \nu} \partial^\mu \partial^\nu h^\beta_\beta 
    + D_2 h_{\mu \nu} \partial^\mu \partial_\beta h^{\nu \beta} 
    + D_3 h_{\mu \nu} \partial^\beta \partial_\beta h^{\mu \nu} 
    + D_4 h^\mu_\mu \partial^\beta \partial_\beta h^\nu_\nu 
    + D_5 h^\beta_\beta \partial^\mu \partial^\nu h_{\mu \nu}) a_\lambda ,
\end{multline}

\begin{multline}
    \frac{\partial \mathcal{L}}{\partial(\partial_\rho h_{\omega \sigma})} \delta h_{\omega \sigma} = 
    (2 B_9 C_1 + 2 B_1 C_2 + 2 B_6 C_2 + 2 B_7 C_2 + 8 B_9 C_2 + \frac{1}{2} B_6 C_3 + \frac{1}{2} B_7 C_3 + B_9 C_3)\partial^\rho h^\mu_\mu \partial^\lambda h^\beta_\beta a_\lambda
    + 2 B_1 C_1 \partial^\rho h^{\mu \nu} \partial^\lambda h_{\mu \nu} a_\lambda
    \\
    + (2 B_8 C_1 + 2 B_2 C_2 + 2 B_3 C_2 + 2 B_4 C_2 + 2 B_5 C_2 + 8 B_8 C_2 + B_8 C_3 + B_6 C_4 + B_7 C_4) \partial^\rho h^\nu_\nu \partial_\mu h^{\lambda \mu} a_\lambda
    + (2 B_1 C_4 + B_2 C_5 + B_3 C_5) \partial_\mu h^{\mu \nu} \partial^\lambda h^\rho_{\ \nu} a_\lambda 
    \\
    + (\frac{1}{2} B_2 C_3 + \frac{1}{2} B_3 C_3 + B_1 C_3 + 4 B_9 C_3 + B_6 C_3 + B_7 C_3 + 2 B_9 C_4 + 2 B_9 C_5) \partial^\lambda h^\nu_\nu \partial^\mu h^\rho_{\ \mu} a_\lambda 
    + (B_1 C_3 + B_6 C_5 + B_7 C_5) \partial^\mu h^\nu_\nu \partial^\lambda h^\rho_{\ \mu} a_\lambda 
    \\
    + (B_2 C_3 + B_3 C_3 + B_4 C_3 + B_5 C_3 + 4 B_8 C_3 + B_2 C_4 + B_3 C_4 + 2 B_8 C_4 + 2 B_8 C_5) \partial_\mu h^{\mu \rho} \partial_\nu h^{\nu \lambda} a_\lambda
    + 2 B_1 C_5 \partial^\mu h^{\rho \nu} \partial^\lambda h_{\mu \nu} a_\lambda
    \\
    + (\frac{1}{2} B_4 C_3 + \frac{1}{2} B_5 C_3 + B_6 C_5 + B_7 C_5) \partial^\mu h^\nu_\nu \partial_\mu h^{\rho \lambda} a_\lambda
    + (B_4 C_4 + B_5 C_4 + B_2 C_5 + B_3 C_5) \partial_\mu h^{\mu \nu} \partial_\nu h^{\rho \lambda} a_\lambda
    + (B_4 C_5 + B_5 C_5) \partial^\mu h^{\rho \nu} \partial_\nu h^\lambda_{\ \mu} a_\lambda
    \\
    + (2 B_2 C_1 + 2 B_3 C_1 + B_4 C_4 + B_5 C_4) \partial^\rho h^{\lambda \nu} \partial^\mu h_{\mu \nu} a_\lambda
    + (2 B_6 C_1 + 2 B_7 C_1 + \frac{1}{2} B_4 C_3 + \frac{1}{2} B_5 C_3) \partial^\rho h^{\mu \lambda} \partial_\mu h^\nu_\nu a_\lambda
    \\
    + (B_4 C_5 + B_5 C_5) \partial^\mu h^{\rho \nu} \partial_\mu h^\lambda_{\ \nu} a_\lambda
    + (2 B_4 C_1 + 2 B_5 C_1) \partial^\rho h^{\mu \nu} \partial_\mu h^\lambda_{\ \nu} a_\lambda 
    + (\frac{1}{2} B_2 C_3 + \frac{1}{2} B_3 C_3 + B_6 C_4 + B_7 C_4) \eta^{\rho \lambda} \partial^\mu h^\nu_\nu \partial^\beta h_{\mu \beta} a_\lambda
    \\
    + (\frac{1}{2} B_6 C_3 + \frac{1}{2} B_7 C_3) \eta^{\rho \lambda} \partial^\mu h^\nu_\nu \partial_\mu h^\beta_\beta a_\lambda
    + (B_2 C_4 + B_3 C_4) \eta^{\rho \lambda} \partial_\beta h^{\beta \mu} \partial^\nu h_{\mu \nu} a_\lambda ,
\end{multline}

\begin{multline}
    \frac{\partial \mathcal{L}}{\partial(\partial_\rho \partial_\zeta h_{\omega \sigma})} \partial_\zeta \delta h_{\omega \sigma} = 
    (B_1 D_1 + B_6 D_1 + B_7 D_1 + 4 B_9 D_1 + \frac{1}{4} B_6 D_2 + \frac{1}{4} B_7 D_2 + B_9 D_2) h^{\rho \mu} \partial_\mu \partial^\lambda h^\nu_\nu a_\lambda
    +B_1 D_3 h_{\mu \nu} \partial^\rho \partial^\lambda h^{\mu \nu} a_\lambda  
    \\
    + (B_2 D_1 + B_3 D_1 + B_4 D_1 + B_5 D_1 + 4 B_8 D_1 + \frac{1}{4} B_4 D_2 + \frac{1}{4} B_5 D_2 + B_8 D_2) h^{\rho \mu} \partial_\mu \partial_\nu h^{\lambda \nu} a_\lambda 
    + (B_2 D_3 + B_3 D_3) h^{\lambda \mu} \partial^\rho \partial^\nu h_{\mu \nu} a_\lambda  
    \\
    + (\frac{1}{4} B_4 D_2 + \frac{1}{4} B_5 D_2 + B_4 D_3 + B_5 D_3) h_{\mu \nu} \partial^\rho \partial^\mu h^{\lambda \nu} a_\lambda  
    + (\frac{1}{4} B_6 D_2 + \frac{1}{4} B_7 D_2 + B_6 D_3 + B_7 D_3) h^{\lambda \mu} \partial^\rho \partial_\mu h^\nu_\nu a_\lambda
    \\
    + (B_8 D_3 + B_2 D_4 + B_3 D_4 + B_4 D_4 + B_5 D_4 + 4 B_8 D_4 + \frac{1}{2} B_4 D_5 + \frac{1}{2} B_5 D_5 + B_8 D_5) 
    h^\mu_\mu \partial^\rho \partial_\nu h^{\lambda \nu} a_\lambda 
    + (\frac{1}{2} B_2 D_5 + \frac{1}{2} B_3 D_5) \eta^{\rho \lambda} h^\mu_\mu \partial^\nu \partial^\alpha h_{\nu \alpha} a_\lambda     
    \\
    + (B_9 D_3 + B_1 D_4 + B_6 D_4 + B_7 D_4 + 4 B_9 D_4 + \frac{1}{2} B_6 D_5 + \frac{1}{2} B_7 D_5 + B_9 D_5) 
    h^\mu_\mu \partial^\rho \partial^\lambda h^\nu_\nu a_\lambda
    + (B_1 D_5 + \frac{1}{2} B_2 D_5 + \frac{1}{2} B_3 D_5) h^\mu_\mu \partial_\nu \partial^\lambda h^{\rho \nu} a_\lambda     
    \\
    + (\frac{1}{2} B_4 D_5 + \frac{1}{2} B_5 D_5) h^\mu_\mu \partial_\nu \partial^\nu h^{\rho \lambda} a_\lambda
    + (\frac{1}{2} B_6 D_5 + \frac{1}{2} B_7 D_5) \eta^{\rho \lambda} h^\mu_\mu \partial_\omega \partial^\omega h^\nu_\nu a_\lambda 
    + (\frac{1}{2} B_1 D_2 + \frac{1}{4} B_2 D_2 + \frac{1}{4} B_3 D_2) h^{\rho \mu} \partial^\lambda \partial^\nu h_{\mu \nu} a_\lambda 
    \\
    + \frac{1}{2} B_1 D_2 h_{\mu \nu} \partial^\mu \partial^\lambda h^{\nu \rho} a_\lambda 
    + (\frac{1}{4} B_2 D_2 + \frac{1}{4} B_3 D_2) h^{\rho \lambda} \partial_\mu \partial_\nu h^{\mu \nu} a_\lambda 
    + (\frac{1}{4} B_2 D_2 + \frac{1}{4} B_3 D_2) h^{\mu \lambda} \partial_\mu \partial_\nu h^{\nu \rho} a_\lambda 
    \\
    + (\frac{1}{4} B_2 D_2  + \frac{1}{4} B_3 D_2) \eta^{\rho \lambda} h^{\mu \nu} \partial_\mu \partial^\alpha h_{\nu \alpha} a_\lambda  
    + (\frac{1}{4} B_4 D_2 + \frac{1}{4} B_5 D_2) h_{\mu \nu} \partial^\mu \partial^\nu h^{\rho \lambda} a_\lambda 
    + (\frac{1}{4} B_4 D_2 + \frac{1}{4} B_5 D_2) h^{\rho \mu} \partial_\nu \partial^\nu h^\lambda_\mu a_\lambda 
    \\
    + (\frac{1}{4} B_6 D_2 + \frac{1}{4} B_7 D_2) \eta^{\rho \lambda} h^{\mu \nu} \partial_\mu \partial_\nu h^\alpha_\alpha a_\lambda 
    + (\frac{1}{4} B_6 D_2 + \frac{1}{4} B_7 D_2) h^{\rho \lambda} \partial_\mu \partial^\mu h^\nu_\nu a_\lambda ,
\end{multline}

\begin{multline}
    [\partial_\zeta \frac{\partial \mathcal{L}}{\partial(\partial_\rho \partial_\zeta h_{\omega \sigma})}]\delta h_{\omega \sigma} = 
    (B_1 D_1 + B_6 D_1 + B_7 D_1 + 4 B_9 D_1 + B_9 D_2 + \frac{1}{2} B_2 D_5 + \frac{1}{2} B_3 D_5) \partial_\mu h^{\rho \mu} \partial^\lambda h^\nu_\nu a_\lambda 
    \\
    + (B_2 D_1 + B_3 D_1 + B_4 D_1 + B_5 D_1 + 4 B_8 D_1 + \frac{1}{4} B_2 D_2 + \frac{1}{4} B_3 D_2 + B_8 D_2) \partial_\mu h^{\rho \mu} \partial_\nu h^{\lambda \nu} a_\lambda
    \\
    + \frac{1}{2} B_1 D_2 \partial^\mu h^{\rho \nu} \partial^\lambda h_{\mu \nu} a_\lambda 
    + (\frac{1}{2} B_1 D_2 + \frac{1}{4} B_2 D_2 + \frac{1}{4} B_3 D_2)\partial^\mu h_{\mu \nu} \partial^\lambda h^{\rho \nu} a_\lambda 
    + (\frac{1}{4} B_2 D_2 + \frac{1}{4} B_3 D_2 + \frac{1}{4} B_4 D_2 + \frac{1}{4} B_5 D_2) \partial_\mu h^{\rho \lambda} \partial_\nu h^{\mu \nu} a_\lambda 
    \\
    + (\frac{1}{4} B_2 D_2 + \frac{1}{4} B_3 D_2) \eta^{\rho \lambda} \partial_\mu h^{\mu \nu} \partial^\alpha h_{\alpha \nu} a_\lambda 
    + (\frac{1}{4} B_4 D_2 + \frac{1}{4} B_5 D_2) \partial_\mu h^{\rho \nu} \partial_\nu h^{\lambda \mu} a_\lambda 
    + (\frac{1}{4} B_4 D_2 + \frac{1}{4} B_5 D_2) \partial^\mu h^{\rho \nu} \partial_\mu h^\lambda_\nu a_\lambda 
    \\
    + (\frac{1}{4} B_4 D_2 + \frac{1}{4} B_5 D_2 + B_2 D_3 + B_3 D_3) \partial^\mu h_{\mu \nu} \partial^\rho h^{\lambda \nu} a_\lambda 
    + (\frac{1}{4} B_6 D_2 + \frac{1}{4} B_7 D_2 + B_1 D_5) \partial^\lambda h^{\rho \mu} \partial_\mu h^\nu_\nu a_\lambda 
    \\
    + (\frac{1}{4} B_6 D_2 + \frac{1}{4} B_7 D_2 + \frac{1}{2} B_2 D_5 + \frac{1}{2} B_3 D_5) \eta^{\rho \lambda} \partial_\mu h^{\mu \nu} \partial_\nu h^\alpha_\alpha a_\lambda 
    + (\frac{1}{4} B_6 D_2 + \frac{1}{4} B_7 D_2 + \frac{1}{2} B_4 D_5 + \frac{1}{2} B_5 D_5) \partial_\mu h^{\rho \lambda} \partial^\mu h^\nu_\nu a_\lambda 
    \\
    + (\frac{1}{4} B_6 D_2 + \frac{1}{4} B_7 D_2 + B_8 D_3 + B_2 D_4 + B_3 D_4 + B_4 D_4 + B_5 D_4 + 4 B_8 D_4 + B_8 D_5) \partial_\mu h^{\lambda \mu} \partial^\rho h^\nu_\nu a_\lambda 
    + B_1 D_3 \partial^\rho h_{\mu \nu} \partial^\lambda h^{\mu \nu} a_\lambda 
    \\
    + (B_4 D_3 + B_5 D_3)\partial^\rho h_{\mu \nu} \partial^\mu h^{\lambda \nu} a_\lambda
    + (B_6 D_3 + B_7 D_3 + \frac{1}{2} B_4 D_5 + \frac{1}{2} B_5 D_5) \partial^\rho h^{\mu \lambda} \partial_\mu h^\nu_\nu a_\lambda    + (\frac{1}{2} B_6 D_5 + \frac{1}{2} B_7 D_5) \eta^{\rho \lambda} \partial_\mu h^\alpha_\alpha \partial^\mu h^\nu_\nu a_\lambda
    \\
    + (B_9 D_3 + B_1 D_4 + B_6 D_4 + B_7 D_4 + 4 B_9 D_4 + \frac{1}{2} B_6 D_5 + \frac{1}{2} B_7 D_5 + B_9 D_5) \partial^\lambda h^\mu_\mu \partial^\rho h^\nu_\nu a_\lambda .
\end{multline} 

\end{document}